\documentclass[traditabstract]{aa}
\usepackage{amsmath}
\usepackage{txfonts}
\usepackage{natbib}
\usepackage{graphicx}
\usepackage{multirow}
\bibpunct{(}{)}{;}{a}{}{,}

\def\ie{{\it i.e.,}\,}
\def\eg{{\it e.g.,}\,}

\begin{document}
\title{Parametrized 3D models of neutrino-driven supernova explosions}
\subtitle{Neutrino emission asymmetries and gravitational-wave signals}
\abstract {Time-dependent and direction-dependent neutrino and
  gravitational-wave (GW) signatures are presented for a set of
  three-dimensional (3D) hydrodynamic models of parametrized,
  neutrino-driven supernova explosions of non-rotating 15 and
  20\,$M_\odot$ stars. We emploed an approximate treatment of neutrino
  transport based on a gray spectral description and a ray-by-ray
  treatment of multi-dimensional effects.  Owing to the excision of
  the high-density core of the proto-neutron star (PNS) and the use of
  an axis-free (Yin-Yang) overset grid, the models can be followed
  from the post-bounce accretion phase through the onset of the
  explosion into more than one second of the early cooling evolution
  of the PNS without imposing any symmetry restrictions and covering a
  full sphere. Gravitational waves and neutrino emission exhibit the
  generic time-dependent features already known from 2D
  (axi-symmetric) models. Violent non-radial hydrodynamic mass motions
  in the accretion layer and their interaction with the outer layers
  of the proto-neutron star together with anisotropic neutrino
  emission give rise to a GW signal with an amplitude of
  $\sim\,5\,-\,20\,$cm in the frequency range of 100--500\,Hz. The GW
  emission from mass motions usually reaches a maximum before the
  explosion sets in. After the onset of the explosion the GW signal
  exhibits a low-frequency modulation, in some cases describing a
  quasi-monotonic growth, associated with the non-spherical expansion
  of the explosion shock wave and the large-scale anisotropy of the
  escaping neutrino flow. Variations of the mass-quadrupole moment
  caused by convective activity inside the nascent neutron star add a
  high-frequency component to the GW signal during the post-explosion
  phase. The GW signals exhibit strong variability between the two
  polarizations, different explosion simulations and different
  observer directions, and besides common basic features do not
  possess any template character. The neutrino emission properties
  (fluxes and effective spectral temperatures) show fluctuations over
  the neutron star surface on spatial and temporal scales that reflect
  the different types of non-spherical mass motions in the supernova
  core, \ie post-shock overturn flows and proto-neutron star
  convection. However, because very prominent, quasi-periodic sloshing
  motions of the shock caused by the standing accretion-shock
  instability are absent and the emission from different surface areas
  facing an observer adds up incoherently, the modulation amplitudes
  of the measurable neutrino luminosities and mean energies are
  significantly lower than predicted by 2D simulations.  }
\author{E. M\"uller \and H.-Th. Janka \and A. Wongwathanarat }
\institute{Max-Planck-Institut f\"{u}r Astrophysik,
           Karl-Schwarzschild-Stra{\ss}e 1, 
           D-85748 Garching, Germany}
\keywords{Supernovae: general --- neutron stars: general} 
\maketitle

\section{Introduction}
The electromagnetic signature of core-collapse supernovae has been
exploited comprehensively and thoroughly by countless observations
during the past decades, providing only indirect information about the
explosion mechanism, however.  The up to now only recorded neutrino
signal of a core-collapse supernova (SN1987A) confirmed the idea that
the collapse of the core of a massive star to neutron star densities
provides the necessary energy for the explosion \citep{BaadeZwicky34}.
Because gravitational waves (GW), the only other means to probe the
supernova engine besides neutrinos, are yet to be detected, supernova
modelers are preparing for this prospective measurement by predicting
the gravitational wave signature of core-collapse supernovae with ever
increasing realism \citep[for reviews, see \eg][]{Kotake_etal06,
  Ott09, FryerNew11}.

According the Einstein's theory of general relativity (GR),
gravitational waves will be generated by any mass-energy distribution
whose quadrupole (or higher) moment varies non-linearly with time.  In
core-collapse supernovae this criterion is satisfied by time-dependent
rotational flattening particularly during collapse and bounce, prompt
post-shock convection, non-radial flow inside the proto-neutron star
and in the neutrino-heated hot bubble, the activity of the standing
accretion shock instability (SASI), asymmetric emission of neutrinos,
and by asymmetries associated with the effects of magnetic fields
\citep[for a recent review see, \eg][and references therein]{Ott09}.
While significant rotational deformation and dynamically relevant
magnetic fields require particular progenitors that possess some
(considerable) amount of rotational and magnetic energy or that must
rotate fast and differentially (additional differential rotation
develops during collapse) to amplify an initially weak magnetic field,
all other processes are genuinely operative in any core-collapse
supernova.

Obviously, an adequate modeling of these effects and an accurate
prediction of the GW signal ultimately requires three dimensional (3D)
hydrodynamic simulations covering the collapse, bounce, and
post-bounce evolution (at least) until a successful launch of the
explosion and including a proper treatment of neutrino transport and
the relevant microphysics.  However, most studies of the past thirty
years were either concerned with the collapse and bounce signal only
\citep{Mueller82, FinnEvans90, Moenchmeyer_etal91, YamadaSato94,
  ZwergerMueller97, Rampp_etal98, Dimmelmeier_etal01,
  Dimmelmeier_etal02, Kotake_etal03, Shibata03, ShibataSekiguchi04,
  Ott_etal04, Cerda-Duran_etal05, Saijo05, ShibataSekiguchi05,
  Kotake_etal06, Dimmelmeier_etal07, Ott_etal07, Dimmelmeier_etal08},
or were restricted to axisymmetric (2D) models \citep{Mueller_etal04,
  Ott_etal06, Kotake_etal07, Marek_etal09, Murphy_etal09,
  Yakunin_etal10}.  Several authors also investigated the influence of
magnetic fields on the GW signal during the collapse and early
post-bounce evolution assuming axisymmetry \citep{Kotake_etal04,
  YamadaSawai04, Kotake_etal05, Obergaulinger_etal06a,
  Obergaulinger_etal06b} and no symmetry restriction at all
\citep{Scheidegger_etal08, Scheidegger_etal10}. The GW signal caused
by aspherical neutrino emission was first studied by \citet{Epstein78}
and subsequently by \citet{BurrowsHayes96}, \citet{MuellerJanka97},
and \citet{Kotake_etal07, Kotake_etal09a, Kotake_etal09b,
  Kotake_etal11}, where the investigations by \citet{ MuellerJanka97}
and \citet{Kotake_etal09b, Kotake_etal11} considered also 3D, \ie
non-axisymmetric models.

Concerning 3D post-bounce models, the topic of the study presented
here, \citet{MuellerJanka97} were the first to analyze the GW
signature of 3D non-radial flow and anisotropic neutrino emission from
prompt post-bounce convection in the outer layers of a proto-neutron
star during the first 30\,msec after supernova-shock formation.
Because of smaller convective structures and slower overturn
velocities, the GW wave amplitudes of their 3D models are more than a
factor of 10 lower than those of the corresponding 2D ones, and the
wave amplitudes associated with anisotropic neutrino emission are a
factor of 10 higher than those caused by non-radial gas flow.
\citet{Fryer04} and \citet{Fryer_etal04} presented gravitational wave
signals obtained from 3D core-collapse simulations that extend up to
150\,msec past bounce and were performed with a Newtonian smoothed
particle hydrodynamics code coupled to a three-flavor gray
flux-limited diffusion neutrino transport scheme. Gravitational wave
emission occurs owing to matter asymmetries that arise from
perturbations caused by precollapse convection, core rotation, and
low-mode convection in the explosion engine itself, and owing to
anisotropic neutrino emission.  \citet{Kotake_etal09b} simulated 3D
mock-up models that mimic neutrino-driven explosions aided by the
SASI, and computed the GW signal resulting from anisotropic neutrino
emission by means of a ray-tracing method in a post-processing
step. They pointed out that the gravitational waveforms of 3D models
vary much more stochastically than those of axisymmetric ones, \ie in
3D the GW signals do not possess any template character.  However,
when considering rotating models, \citet{Kotake_etal11} argue that the
GW waveforms resulting from neutrino emission exhibit a common
feature, which results from an excess of neutrino emission along the
spin axis due to the growth of spiral SASI modes.
\citet{Scheidegger_etal08} simulated the collapse of two rotating and
magnetized cores in 3D until several 10\,msec past bounce, applying a
parametrized deleptonization scheme \citep{Liebendoerfer05} that is a
good approximation until a few milliseconds past bounce.
\citet{Scheidegger_etal10} extended their study by systematically
investigating the effects of the equation of state, the initial
rotation rate, and both the initial magnetic field strength and
configuration on the GW signature. They also simulated three
representative models until $\sim\,$200\,msec of post-bounce accretion
(no development of explosions) incorporating a treatment for neutrino
transport based on a partial implementation of the isotropic diffusion
source approximation \citep{Liebendoerfer_etal09}.

In the following we present the GW analysis of a set of parametrized
3D models of neutrino-powered supernova explosions covering the
evolution from core bounce until $\sim\,$1.4\,s later, where the
high-density inner core of the proto-neutron star (PNS) is excised and
replaced by a time-dependent boundary condition and a central point
mass. The neutrino transport is treated by an approximate solver based
on a ray-by-ray treatment of the multi-dimensional effects
\citep{Scheck_etal06}.  Bceause we analyze the GW signal arising from
both non-radial mass flow and anisotropic neutrino emission, we
discuss the neutrino emission of these 3D models as well, and
particularly address its multidimensional properties, some of which
have previously been investigated in 2D by \citet{JankaMoenchmeyer89a,
  JankaMoenchmeyer89b}, \citet{Ott_etal08}, \citet{Kotake_etal09a},
\citet{MarekJanka09}, \citet{Marek_etal09}, and \citet{Brandt_etal11}.

Based on 2D simulations of rotational core-collapse,
\citet{JankaMoenchmeyer89a, JankaMoenchmeyer89b} pointed out that the
energy output in neutrinos seen by an observer may vary as much as a
factor of 3 with his inclination angle relative to the rotation axis,
while for the neutrino energy much smaller angular variations are to
be expected.  \citet{Marek_etal09} and \cite{MarekJanka09} found that
neutrino luminosities and mean energies show quasi-periodic time
variability in their 2D simulations of non-rotating and slowly
rotating 15\,$M_\odot$ progenitors covering up to $\sim$700\,ms past
bounce.  Caused by the expansion and contraction of the shock in the
course of SASI oscillations, the level of the fluctuations is
$\la$50\% for the luminosities and roughly 1\,MeV for the mean
neutrino energies. The luminosity fluctuations are somewhat bigger for
$\nu_{\rm e}$ and $\bar\nu_{\rm e}$ than for heavy-lepton neutrinos.
The neutrino quantities also depend on polar angle as a consequence of
the preference of the SASI motions along the symmetry axis of the 2D
models. Additional short-wavelength spatial variations of the average
radiated energies and of the neutrino fluxes are caused by local
downdrafts of accreted matter. This is in accordance with the findings
of \citet{MuellerJanka97}, who inferred from a post-processing
analysis of the neutrino emission anisotropy that features in the GW
signal of their 2D models of convection in the hot-bubble region are
well-correlated with structures in the neutrino signal. The features
are associated with sinking and rising lumps of matter and with
temporal variations of aspherical accretion flows toward the
proto-neutron star. \citet{Kotake_etal09a} computed neutrino
anisotropies with a ray-tracing scheme by post-processing their 2D
SASI models and derived analytical expressions for evaluating GW
signals for neutrinos in 3D models, too.  A generalization of these
expressions will be presented below.  \citet{Brandt_etal11} performed
2D multi-group, multi-angle neutrino transport simulations for both a
non-rotating and rapidly rotating 20\,$M_\odot$ model extending
$\sim$400\,ms beyond bounce.  Their simulations predict that the
neutrino radiation fields vary much less with angle than the matter
quantities in the region of net neutrino heating because most
neutrinos are emitted from deeper radiative regions and because the
neutrino energy density combines the specific intensities as integrals
over sources at many angles and depths. The rapidly rotating model
exhibits strong, flavor-dependent asymmetries in both peak neutrino
flux and light curves, the peak flux and decline rate having
pole-equator ratios $\la$3 and $\la$2, respectively.
\citet{Brandt_etal11} also provide estimates of the detectability of
neutrino fluctuations in IceCube and Super-Kamiokande as previously
done by \citet{Lund_etal10} on the basis of the \citet{Marek_etal09}
non-rotating models.

The paper is organized as follows: in Section\,2 we discuss the
numerical methods, the input physics, and the properties of the
progenitor models and the set of 3D simulations that we analyzed.
Section\,3 contains a description of the formalism we used to extract
the observable neutrino signal of our 3D models, and a discussion of
some of its properties relevant for the corresponding GW signal. In
Section\,4 we give the formalism necessary to calculate the GW
signature of 3D non-radial flow and anisotropic neutrino emission, and
discuss the GW signature of the investigated 3D models. Finally, in
Section\,5 we summarize our results and discuss shortcomings and
possible implications of our study.

\section{Model setup}
\subsection{Code and computational grid}
\label{subsec:codegrid}

The 3D supernova models we analyzed for their neutrino and GW
signature have been simulated with the explicit finite-volume,
Eulerian, multi-fluid hydrodynamics code {\sc Prometheus}
\citep{PROMET1,PROMET2,PROMET3}. This code integrates the
multidimensional hydrodynamic equations using the dimensional
splitting method of \citet{Strang68}, the piecewise parabolic method
of \citet{PPM}, and a Riemann solver for real gases proposed by
\citet{CollelaGlaz85}. Inside grid cells with strong grid-aligned
shocks fluxes computed from the Riemann solver are replaced by the
AUSM+ fluxes of \citet{Liou96} in order to prevent odd-even decoupling
\citep{Quirk94}. Nuclear species are advected using the consistent
multi-fluid advection (CMA) scheme of \citet{CMA}.

The simulation code employs an axis-free overlapping ``Yin-Yang'' grid
\citep{YinYang} in spherical polar coordinates, which was recently
implemented into {\sc Prometheus}, for spatial discretization
\citep{Wong_etal10a}. The Yin-Yang grid relaxes the CFL-timestep
condition and avoids numerical artifacts near the polar axis.
Concerning the ray-by-ray neutrino transport no special procedure
needs to be applied for the Yin-Yang grid. The (scalar) quantities
involved in the transport algorithm are computed on both the Yin and
Yang grid and are then linearly interpolated as any other scalar
quantity.

The grid consists of $400(r) \times 47(\theta) \times 137(\phi)
\times2$ cells corresponding to an angular resolution of $2^\circ$ and
covers the full $4\pi$ solid angle.  The radial grid has an
equidistant spacing of 0.3\,km from the inner grid boundary out to $r
= 80\,$km (models W15 and N20; see Table\,1) or 115\,km (model L15;
see Table\,1), respectively. Beyond this radius the radial grid is
logarithmically spaced. The outer radial grid boundary $R_\mathrm{ob}$
is at 18000\,km, which is sufficiently far out to prevent the
supernova shock from leaving the computational domain during the
simulated epoch.  This radial resolution suffices the requirement that
there are always more than 15 radial zones per decade in density.

A central region, the dense inner core of the proto-neutron star (PNS)
at $\rho \ga 10^{12\ldots13}\,$gcm$^{-3}$, is excised from the
computational domain and replaced by an inner time-dependent radial
boundary condition and a point mass at the coordinate origin.  The
radius of the inner radial boundary shrinks according to Eq.(13) of
\citet{Scheck_etal08}. For the W15 and N20 models the initial and
final (asymptotic) boundary radii are $R^\mathrm{i}_\mathrm{ib} =
65\,$km and $R^\mathrm{f}_\mathrm{ib} = 15\,$km, respectively.  For
the L15 models the corresponding radii are 82\,km and 25\,km,
accounting for a less extreme contraction of the neutron star within
the simulation time.  The timescale for the contraction is
$t_\mathrm{ib} = 1\,$sec for all models. This choice of parameters
implies $R_\mathrm{ib} \approx 19\,$km at 1.3\,s for models W15 and
N20, and $R_\mathrm{ib} \approx 30\,$km at 1.4\,s for the L15 models.
Hydrostatic equilibrium is assumed at the inner radial grid boundary
$R_\mathrm{ib}$, which is thus a Lagrangian (co-moving) position,
while a free outflow boundary condition is employed at the outer
radial grid boundary \citep[for more details, see][]{Wong11,
  Wong_etal10a}.

\subsection{Input physics}
\label{subsec:physics}

Self-gravity is fully taken into account by solving Poisson's equation
in integral form using an expansion into spherical harmonics as in
\citet{MuellerSteinmetz95}.  The monopole term of the potential is
corrected for general relativistic effects as described in
\citet{Scheck_etal06} and \citet{Arcones_etal07}.  The cooling of the
PNS is prescribed by neutrino emission properties (luminosities and
mean neutrino energies) as boundary condition at the inner radial grid
boundary \citep[for details, see][]{Scheck_etal06}. The contraction of
the PNS is mimicked by the movement of the inner radial grid boundary
(see Sect.\,\ref{subsec:codegrid}).  ``Ray-by-ray'' neutrino transport
and neutrino-matter interactions are approximated as in
\citet{Scheck_etal06} by radial integration of the one-dimensional
(spherical), gray transport equation for all angular grid directions
($\theta$,\,$\phi$) independently. This approach allows for angular
variations of the neutrino fluxes. The tabulated equation of state
(EoS) of \citet{JankaMueller96} is used to describe the stellar
fluid. It includes arbitrarily degenerate and arbitrarily relativistic
electrons and positrons, photons, and four predefined nuclear species
(n, p, $\alpha$, and a representative Fe-group nucleus) in nuclear
statistical equilibrium.

%
\begin{table}
\caption{Some properties of the analyzed 3D models including the mass
  of the progenitor star $M_\mathrm{PS}$, the mass of the neutron star
  $M_\mathrm{NS}$, the time of explosion $t_\mathrm{exp}$, and the
  explosion energy $E_\mathrm{exp}$ at the time $t_\mathrm{f}$ after
  bounce when we stopped the simulation. Note that $1\,\mathrm{B} =
  1\,\mathrm{bethe} = 10^{51}\,$erg.}
\centering
\begin{tabular}{lccccc}
\hline\hline\multirow{2}{*}{Model} 
 & $M_\mathrm{PS}$  & $M_\mathrm{NS}$ & $t_\mathrm{exp}$
 & $E_\mathrm{exp}$ & $t_\mathrm{f}$  \\
 & [$M_\odot$]     & [$M_\odot$]     & [ms]           
 & [B]            & [s]\\
\hline
W15-2 & 15 & 1.37 & 248 & 1.13 & 1.3 \\
W15-4 & 15 & 1.38 & 272 & 0.94 & 1.3 \\
L15-2 & 15 & 1.51 & 381 & 1.74 & 1.4 \\
L15-3 & 15 & 1.62 & 477 & 0.84 & 1.4 \\
N20-2 & 20 & 1.28 & 265 & 3.12 & 1.3 \\
\hline
\end{tabular}
\label{tab:models}
\end{table}

\subsection{Models}
\label{subsec:models}

We have analyzed a set of 3D simulations \citep{Wong11, Wong_etal10b,
  Wong_etal11} based on two 15\,$M_\odot$ progenitor models (W15 and
L15), and a 20\,$M_\odot$ progenitor model (N20).  The W15 model is
obtained from the non-rotating 15\,$M_\odot$ progenitor s15s7b2 of
\citet{WoosleyWeaver95}, the L15 model from a star evolved by
\citet{Limongi_etal00}, and the N20 model from a SN\,1987A progenitor
of \citet{ShigeyamaNomoto90}.  The progenitor models were evolved
through collapse to 15\,ms after bounce with the {\sc
  Prometheus-Vertex} code in one dimension (A.\,Marek and R.\,Buras,
private communication) providing the initial models for the 3D
simulations. To break the spherical symmetry of the initial models,
random seed perturbations of 0.1\% amplitude are imposed on the radial
velocity ($v_r$) field at the beginning of the 3D
simulations. Explosions are initiated by neutrino heating at a rate
that depends on suitable values of the neutrino luminosities imposed
at the lower boundary chosen such that the desired value of the
explosion energy is obtained.  The evolution is followed until 1.3\.s
after bounce for the W15 and N20 progenitor models, while the L15
models are simulated until 1.4\,s post-bounce.  The GW analysis
presented below comprises five models (see Table\,1), where models W15-2
and W15-4 differ only by the initial seed perturbations.  The
explosion energies, $E_\mathrm{exp}$, given in Table\,1 are
instantaneous values at the end of the simulations ($t =
t_\mathrm{f}$), adding up the total energies (kinetic + internal +
gravitational) in all zones where the sum of these energies is
positive. The explosion time, $t_\mathrm{exp}$, is defined as the time
when this sum exceeds a value of $10^{48}\,$erg, roughly corresponding
to the time when the average shock radius is 400 to 500\,km (see,
however, \citet{PejchaThompson11} for an alternative definition of the
time of the onset of the explosion).

\section{Neutrino signal}
\label{sec:neutrino-signal}

The non-radial motions caused by the SASI and convection in the
neutrino-heated hot-bubble as well as by convection inside the
proto-neutron star (driven by Ledoux unstable lepton gradients) give
rise to a time-dependent, anisotropic emission of neutrinos of all
flavors, and thus to the emission of gravitational waves
\citep{Epstein78, BurrowsHayes96, MuellerJanka97, Kotake_etal07,
  Kotake_etal09a, Kotake_etal09b}, as discussed in
Sect.\,\ref{sec:gw-signal}. We have analyzed this emission for the 3D
models discussed above (see Sect.\,\ref{subsec:models}), particularly
addressing its multidimensional properties.

\subsection{Formalism}
\label{subsec:nuform}

\begin{figure}
\centering
\resizebox{\hsize}{!}{\includegraphics{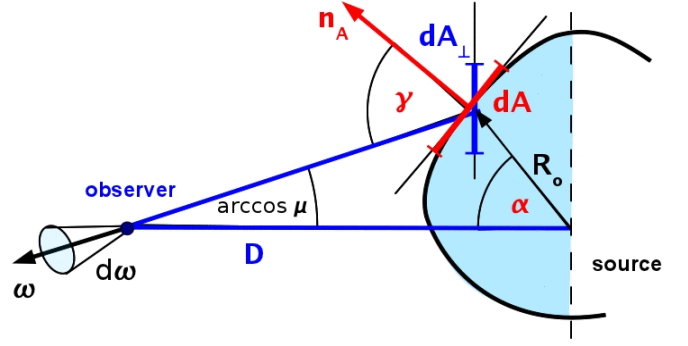}}
\caption{Sketch illustrating the various quantities involved when
  deriving the observable luminosity of a radiating source, whose
  visible surface is shaded in blue. When the observer is at infinity,
  the emitting surface is a sphere, and $\mathbf{R}_o$ is measured
  from the origin of this sphere, one obtains $\alpha = \gamma$.}
\label{fig:applum}
\end{figure}

To derive \emph{observable luminosities} of an emitting source we
consider an observer located at far distance $D$ from that source
(see Fig.\,\ref{fig:applum}).  According to definition the flux
measured by the observer is given by the following integral at the
position of the observer:
\begin{equation}
  F(D,t) = \oint {\mathrm d} \omega\, \mu\, I(D, {\vec \omega}, t) \, , 
\label{eq:obsflux}
\end{equation}
where $\mu$ is the cosine of the angle between the direction of the
radiation and the line of sight (between the observer and the center
of the source), $\vec \omega$ denotes the radiation direction at the
observer's location (defined by a pair of angles), and ${\mathrm d}
\omega$ is the solid angle element around the radiation direction
$\vec \omega$.  The intensity $I$ adopts non-zero values within the
opening angle subtended by the emitting surface (not necessarily a
sphere).
We note that here and in the following we suppress the
dependence of the intensity on the neutrino energy and assume that
energy-integrated quantities are considered (the outlined formalism,
however, is valid also for an energy-dependent treatment).  The
integration over $\vec \omega$ at the observer's location can be
substituted by an integration over the emitting surface of the source,
because the radiation intensity is constant along rays, \ie
\begin{equation}
  I(D, {\vec \omega}, t) = I_o({\vec R}_o, {\vec \omega}_o, t) 
\label{eq:radintens}
\end{equation}
for any ray arriving at the observer from the source (and zero
otherwise), where ${\vec R}_o$ denotes the position of a surface
element of the emitting surface in the coordinate frame of the source
and ${\vec \omega}_o$ the direction of the radiation field at that
position toward the observer.  Note that we ignore in
Eq.\,(\ref{eq:radintens}) the trivial effect that the time $t$ for
$I_o$ relative to the time for $I$ is subject to a retardation.

Moreover, in the following we disregard spectral and angular
corrections that may be relevant when the emitting surface is in
relative motion to the observer or sitting deep in the gravitational
potential of a compact star (in which case general relativistic (GR)
effects like redshifting and ray bending would be important).
Considering the source to be at rest is a good assumption for the
neutrinospheric region in the supernova core after bounce (the
velocities of mass motions in this layer are unlikely to be higher
than some 1000\,km/s, \ie at most around one percent of the speed of
light), while GR energy redshift is certainly of relevance on the
$\sim$10--20\% level during the proto-neutron star cooling phase ($t
\gtrsim 1$\,s after bounce), but much lower in the accretion and
shock revival phases, when the forming neutron star is still
considerably less compact than the final remnant.

%
%

For a distant observer $D \gg \mathrm{max} \lbrace |{\vec R}_o|
\rbrace $ holds, \ie the value of $\mu$ is very close to one for the
whole emitting surface. Denoting the solid angle subtended by a
surface element of the emitting surface by ${\rm d} \omega$, we have
${\rm d} \omega = {\rm d} A_{\perp} / D^2$, where ${\rm d} A_{\perp} =
\cos\gamma\, {\rm d} A$ is the projected area of this surface element
perpendicular to the line of sight, when $\gamma$ is the angle between
the normal unit vector ${\vec n}_A$ of the emitting surface element
${\rm d} A$ and the line of sight (see Fig.\,\ref{fig:applum}), but
specified to the case $D \gg |{\vec R}_o|$.  Hence, we obtain for the
observable luminosity the expression
\begin{equation}
  L_o(t) = 4\pi D^2 F(D,t) = 4\pi \int_\mathrm{vis. surf.} 
           {\rm d} A\, \cos\gamma\, 
           I_o ({\vec R}_o, {\vec \omega}_o, t) \, . 
\label{eq:app-lum0}
\end{equation}

In order to evaluate the integral in Eq.\,(\ref{eq:app-lum0}), one
needs to know the intensity $I_o$ as a function of energy, emission
direction, and time at every point of the radiating surface of the
source.  Determining $I_o({\vec R}_o, {\vec \omega}_o, t)$ in general
requires calculating full-scale neutrino transport. With this quantity
as the solution of the transport problem at hand,
Eq.\,(\ref{eq:app-lum0}) can be evaluated directly (in general with
energy dependence) by performing the integration over {\em any}
surface that encloses the radiation (neutrino) emitting source and
that lies outside the volume where radiation interacts with matter
(\ie the intensity $I_o$ in all points on the chosen surface must be
given in the reference frame of the observer and must be preserved on
the way from the emission point to the observer).

Our ray-by-ray transport approximation, however, yields only the local
neutrino energy density $E({\vec R}_o, t)$ and the neutrino flux
density $F({\vec R}_o, t)$. To estimate the neutrinos radiated from
every point of the neutrinosphere to the observer, we therefore have to
develop an approach that yields a reasonable representation of the
direction-dependent intensity as function of the quantities delivered
by our transport approximation
\footnote{Note that the procedure described in the following does not
  depend on whether the transport is performed in the gray
  approximation or is energy dependent. We therefore suppress the
  energy variable in all transport quantities and introduce mean
  energies in our gray treatment instead of considering neutrino
  energies as a function of spectral frequencies.}.
To this end, we assume that the neutrino distribution is axisymmetric
around the normal vector ${\vec n}_A$ at all points ${\vec R}_o$. This
implies that the direction dependence of the intensity $I_o$ is
described by the direction angle $\gamma$ only (see
Fig.\,\ref{fig:applum}), \ie $I_o = I_o({\vec R}_o, \gamma, t)$, and
that the flux direction is given by ${\vec n}_A$. Assuming further
that $I_o(\gamma)$ can be approximated by the lowest two terms of an
expansion in spherical harmonics (as in the diffusion approximation),
one can write
\begin{equation} 
  I_o ({\vec R}_o, \gamma, t) = \frac{F_o ({\vec R}_o, t)}{2\pi} 
       \left( 1 + \frac{3}{2} \cos\gamma \right) \, .
\label{eq:app-int}
\end{equation}

Because the radiation flux density is defined as the first angular
moment of the intensity, one can easily verify that the numerical
coefficient 3/2 of the dipole term ensures that the flux density
$F_o({\vec R}_o,t)$ (normal to the emitting surface element
$\mathrm{d}A$) is given by $\int_0^{2\pi}\mathrm{d}\varphi \int_0^1
\mathrm{d} \cos\gamma\, \cos\gamma I_o({\vec R}_o,\gamma,t)$, if the
radiating surface does not receive any incident neutrinos from outside
(i.e., $I_o({\vec R}_o, \gamma, t) = 0$ for $\cos\gamma < 0$)
\footnote{Note that the requirement $I_o\ge 0$ implies that
  Eq.\,(\ref{eq:app-int}) is valid in the whole range of $\cos\gamma
  \ge -\frac{2}{3}$, which includes inward going radiation for
  $\cos\gamma < 0$.  Extending the integration over all directions of
  validity, one obtains $\frac{25}{27}F_o = \int_0^{2\pi} \mathrm{d}
  \varphi \int_{-2/3}^1 \mathrm{d} \cos\gamma\, \cos\gamma I_o \approx
  F_o$ and $E_o = \int_0^{2\pi} \mathrm{d} \varphi\int_{-2/3}^1
  \mathrm{d} \cos\gamma\, I_o = \frac{35}{12} F_o/c \approx 3
  F_o/c$. This means that we have $F_o \approx \frac{1}{3} cE_o$,
  which is a reasonably good approximation of the relation between
  flux density and energy density at the neutrinosphere, where one
  typically obtains $f_o \equiv F_o/(E_oc) \approx \frac{1}{4} \dots
  \frac{1}{3}$ for the flux factor at an optical depth between unity
  and about 2/3 in detailed neutrino transport calculations in
  spherical symmetry \citep{Janka91}.  The ansatz of
  Eq.\,(\ref{eq:app-int}) is therefore consistent with basic
  properties of the neutrinospheric emission. Moreover, we note that
  the expression corresponds to the limb-darkening law $I_E
  (\cos\gamma) / I_E (1) = (2/5) (1 + 3/2\times \cos\gamma)$ that can
  be derived on grounds of the Eddington approximation (see, \eg,
  \citet{Mihalas78}, page 61, or \citet{MorseFeshbach53}, page 187).}.

Inserting Eq.\,(\ref{eq:app-int}) into Eq.\,(\ref{eq:app-lum0}), we
find for the observable neutrino luminosity the expression
\begin{equation}
  L_o(t) = 2\, \int_\mathrm{vis. surf.} {\rm d} A\,
           \cos\gamma\, F_o ({\vec R}_o,t)\, 
           \left (1 + \frac{3}{2} \cos\gamma \right) \, . 
\label{eq:app-lum}
\end{equation}
We further define an \emph{observable mean neutrino energy} according
to
\begin{equation}
  \langle E \rangle_o\, (t) \,=\, \frac{L_o (t)}{L_{n,o}(t)} \, ,
\label{eq:enue_mean}
\end{equation}
where
\begin{equation}
  L_{n,o} (t) = 2\, \int_\mathrm{vis. surf.} {\rm d} A\, \cos\gamma\, 
               \frac{F_o ({\vec R}_o, t)}{\epsilon ({\vec R}_o,t)} \, 
              \left (1 + \frac{3}{2} \cos\gamma \right) 
\label{eq:lnum}
\end{equation}
is the observable \emph{neutrino number flux} with $\epsilon$ being
the mean energy of the neutrino energy spectrum radiated from point
${\vec R}_o$.

Our 3D radiation hydrodynamics code computes the time-dependent
\emph{neutrino energy flux density}, $F_o ({\vec R}_o, t)$, and
\emph{neutrino number flux density}, $F_{n,o} ({\vec R}_o,t)$, through
a sphere of radius $R_o = |{\vec R}_o|$ in dependence of the angular
position ${\vec \Omega} \equiv (\theta, \phi)$, but actually stores
the related quantities
\begin{equation}
  \Lambda ({\vec \Omega},t) \equiv 4\pi R^2_o\, 
                                   F_o (R_o, {\vec \Omega},t) \, ,
\label{eq:lambdae}
\end{equation}
and
\begin{equation}
  \Lambda_n ({\vec \Omega},t) 
     \equiv \frac{\Lambda ({\vec \Omega},t) }{ \epsilon ({\vec \Omega},t)} 
     =      4\pi R^2_o\, F_{n,o} (R_o, {\vec \Omega},t) \, ,
\label{eq:lambdan}
\end{equation} 
because these quantities are constant in the ray-by-ray approximation
of the free streaming region. In this approximation, both the neutrino
energy flux and the neutrino number flux are purely radial.

Using Eqs.\,(\ref{eq:lambdae}) and (\ref{eq:lambdan}), and the fact
that ${\rm d} A = R^2_o {\rm d} \Omega$ with ${\rm d}\Omega =
\sin\theta {\rm d}\theta {\rm d} \phi$ for the special case of an
emitting sphere of radius $R_o$, we can rewrite the general expression
for the observable neutrino luminosity given in
Eq.\,(\ref{eq:app-lum}) in the form
\begin{equation}
  L_o(t) = \frac{1}{2\pi}\, \int_\mathrm{vis. hem.} {\rm d} \Omega\,
             \cos\gamma\, \Lambda ({\vec \Omega},t)\, 
             \left (1 + \frac{3}{2} \cos\gamma \right) 
\label{eq:app-lum_code}
\end{equation} 
and that of the observable neutrino number flux given in
Eq.\,(\ref{eq:lnum}) in the form
\begin{equation}
  L_{n,o} (t) = \frac{1}{2\pi}\, \int_\mathrm{vis. hem.}  {\rm d} \Omega\,
               \cos\gamma\, \Lambda_n ({\vec \Omega}, t)\, 
               \left (1 + \frac{3}{2} \cos\gamma \right) \, , 
\label{eq:lnum_code}
\end{equation}
where in both cases the integration is performed over the visible
hemisphere.
\footnote{Using of the ray-by-ray approximation has the advantage that
  the evaluation of the integrals on the rhs of Eqs.\,\ref{eq:app-lum}
  and \ref{eq:lnum} does not require the specification of a suitable
  surface, but can be done on any sphere outside the
  neutrino-decoupling region, as Eqs.\,\ref{eq:app-lum_code} and
  \ref{eq:lnum_code} are independent of ${\vec R}_o$. }

For the evaluation of the gravitational wave amplitude in
Sect.\,\ref{subsec:gw-form-nu} we will also need the quantity
\begin{equation}
 \frac{ {\rm d}\Lambda }{ {\rm d}\Omega} ({\vec \Omega},t)
    \equiv F_o (R_0, {\vec \Omega},t)\, R^2_o 
\label{eq:dlamdom}
\end{equation}
and the corresponding angle-integrated quantity
\begin{equation}
 \Lambda (t) \equiv \oint_\mathrm{surf.} {\rm d}\Omega\, 
              \frac{ {\rm d}\Lambda }{ {\rm d}\Omega} ({\vec \Omega},t)
              = \frac{1}{4\pi} 
                \oint_\mathrm{surf.} {\rm d} \Omega\, 
                 \Lambda ({\vec \Omega}, t)  \, .
\label{eq:lamt}
\end{equation}

For the later discussion of the results we finally define the
\emph{surface-averaged neutrino flux density}
\begin{equation}
 \langle F_o \rangle\, (t) 
   \equiv \frac{1}{4\pi} \oint_\mathrm{surf.}  
          {\rm d}\Omega\, F_0 (R_0, {\vec \Omega},t) 
   \equiv \frac{1}{4\pi R^2_o}\, \frac{{\rm d}{\mathcal E}(t)}{{\rm d} t} 
   \, ,
\label{eq:flxave}
\end{equation}
where
\begin{equation}
 \frac{{\rm d}{\mathcal E}(t)}{{\rm d} t} = \oint_\mathrm{surf.} 
       {\rm d}\Omega\, \frac{ {\rm d}\Lambda }{ {\rm d}\Omega} ({\vec \Omega},t) 
 =  \Lambda (t) 
\label{eq:eloss}
\end{equation}
is the total energy loss rate at time $t$ from the supernova core to
all directions, which (because of the flux variations over the sphere)
is no directly observable quantity.

We have also analyzed the evolution of the neutrino flux asymmetry by
calculating the angular pseudo-power spectrum of the neutrino energy
flux variation 
\begin{equation}
  \delta_\Lambda ({\vec \Omega},t) \equiv 
    \frac{ \Lambda ({\vec \Omega},t) - \Lambda(t) }{ \Lambda(t) } \, ,
\label{eq:dlam}
\end{equation}
where $\Lambda ({\vec \Omega},t)$ and $\Lambda(t)$ are defined in
Eqs.\,(\ref{eq:lambdae}) and (\ref{eq:lamt}), respectively.  The
pseudo-power spectrum is given by the decomposition of $\delta_\Lambda
({\vec \Omega},t)$ in spherical harmonic coefficients
\begin{equation}
  \Lambda_{lm}(t) = \oint {\rm d} \Omega\, \delta_\Lambda ({\vec \Omega},t)
                    Y^\ast_{lm} (\Omega)  \, ,
\label{eq:lamlm}
\end{equation}
where $Y^\ast_{lm} (\Omega)$ is the respective (complex conjugate)
spherical harmonics. For our mode analysis we actually used the
pseudo-power coefficients $C_0 \equiv |\Lambda_{00}|^2$ and
\begin{equation}
 C_l \equiv \frac{1}{2l+1} \left( |\Lambda_{l0}|^2 + 
                                  2 \sum_{m=1}^{m=l} |\Lambda_{lm}|^2 \right)
\label{eq:cl}
\end{equation}
for $l>0$, respectively.

%
\begin{figure*}[!]
\centering
\resizebox{0.47\hsize}{!}{\includegraphics*{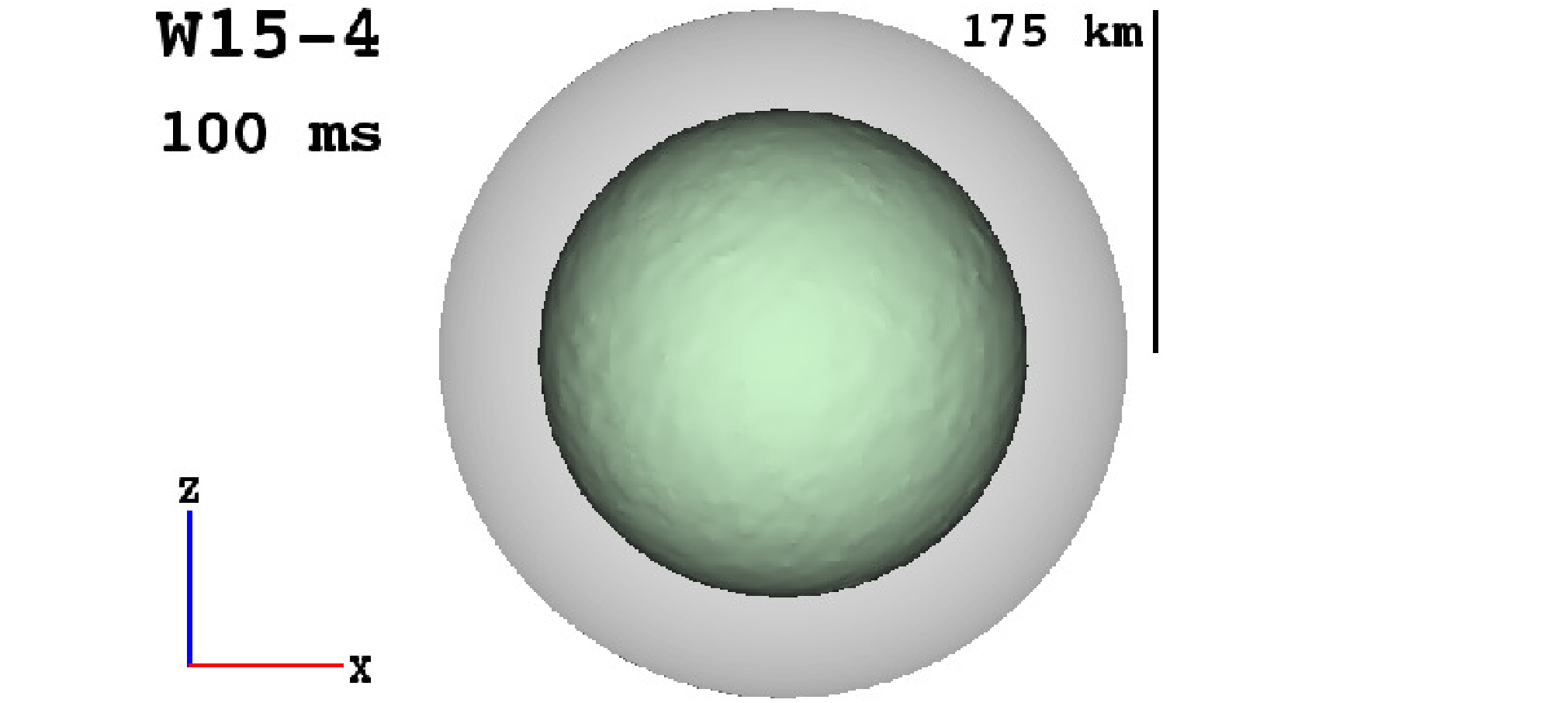}}\hspace{1cm}
\resizebox{0.47\hsize}{!}{\includegraphics*{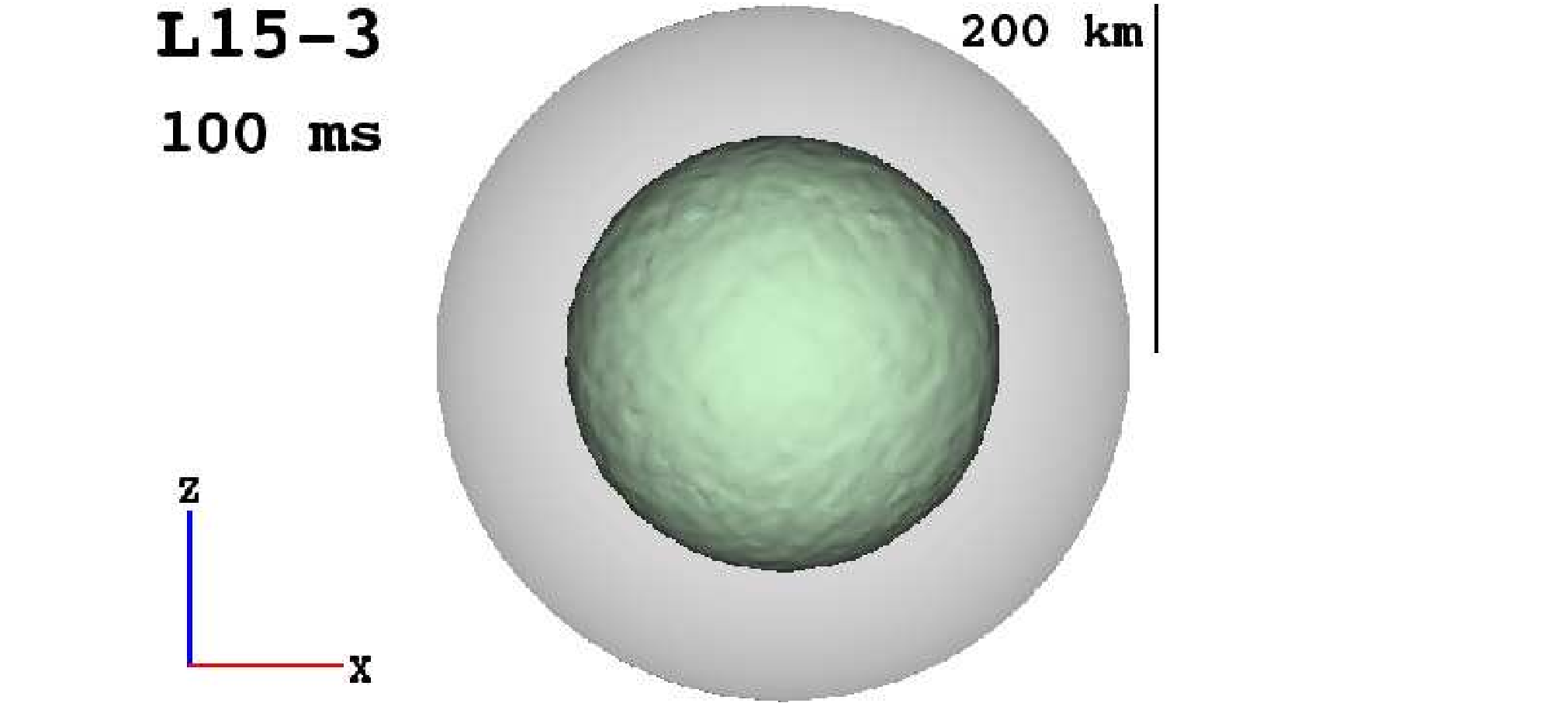}}\\
\resizebox{0.47\hsize}{!}{\includegraphics*{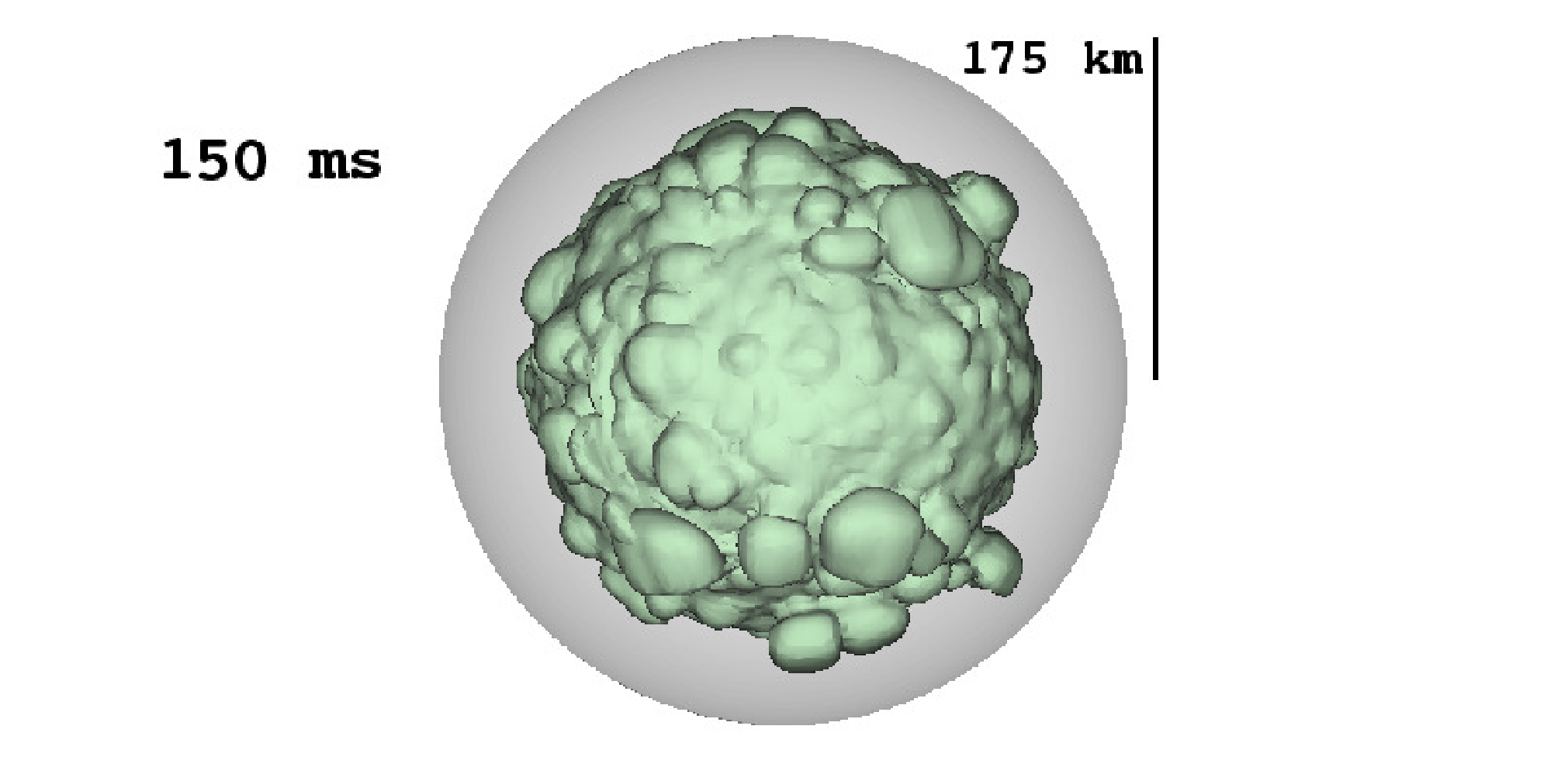}}\hspace{1cm}
\resizebox{0.47\hsize}{!}{\includegraphics*{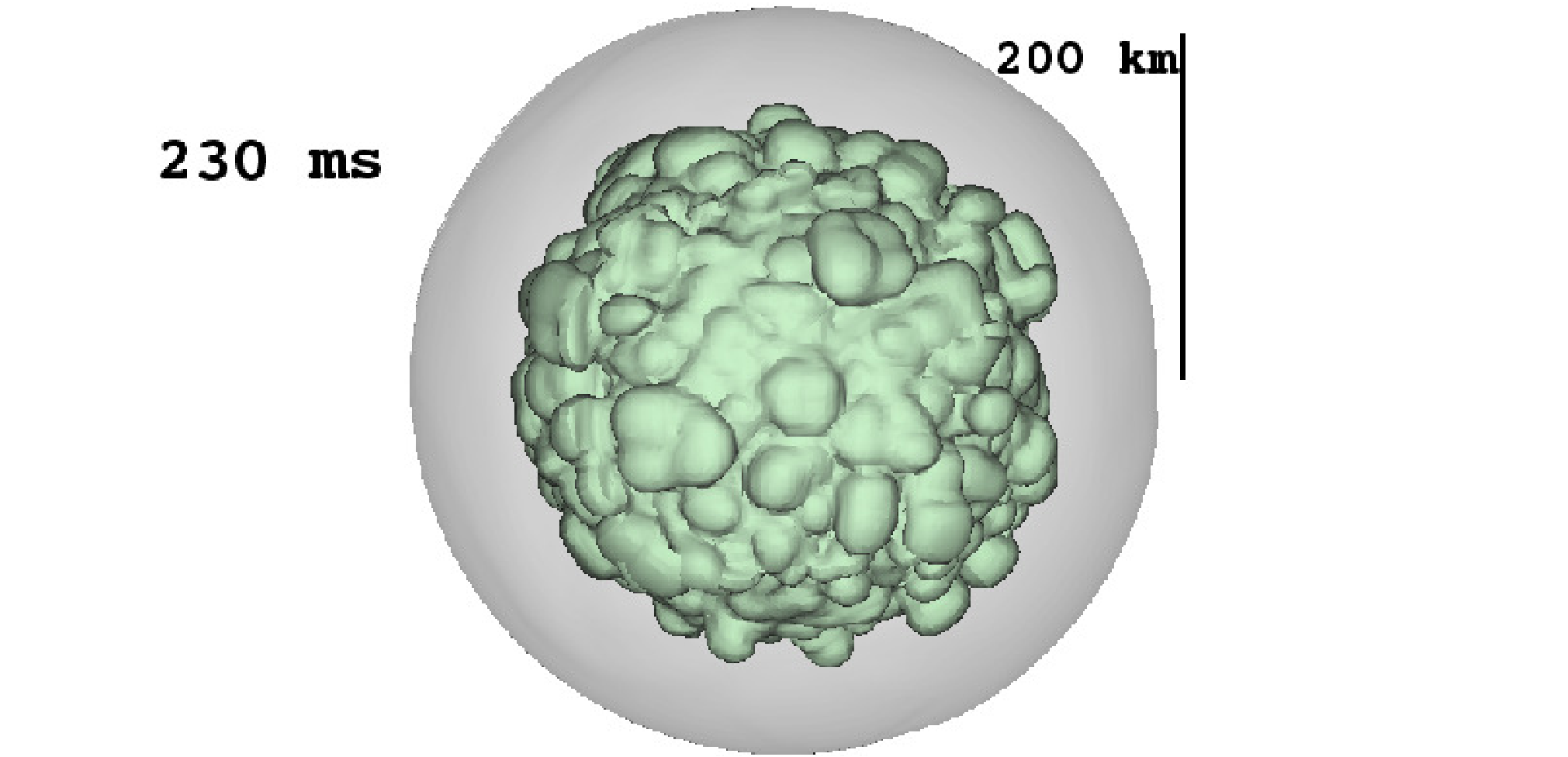}}\\
\resizebox{0.47\hsize}{!}{\includegraphics*{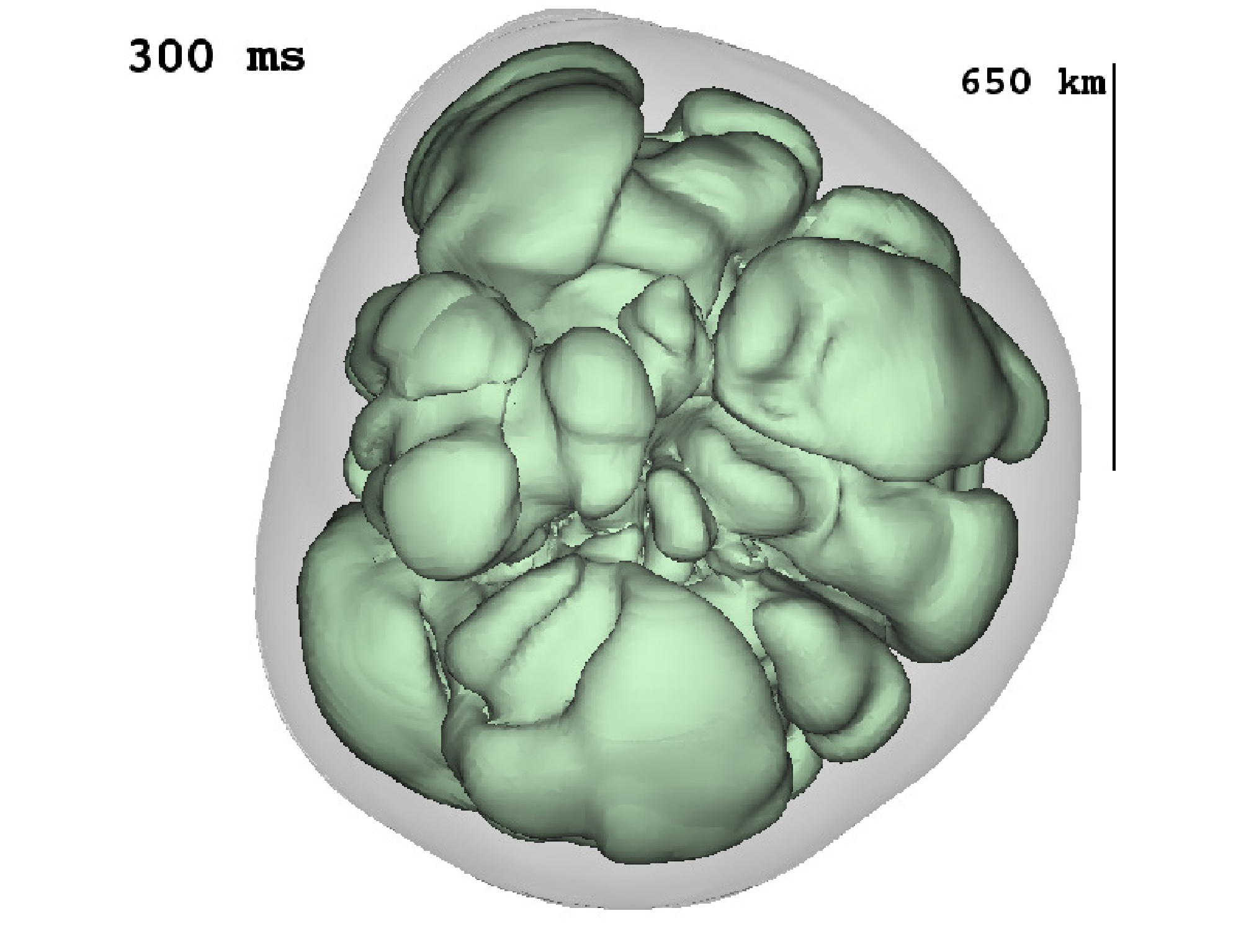}}\hspace{1cm}
\resizebox{0.47\hsize}{!}{\includegraphics*{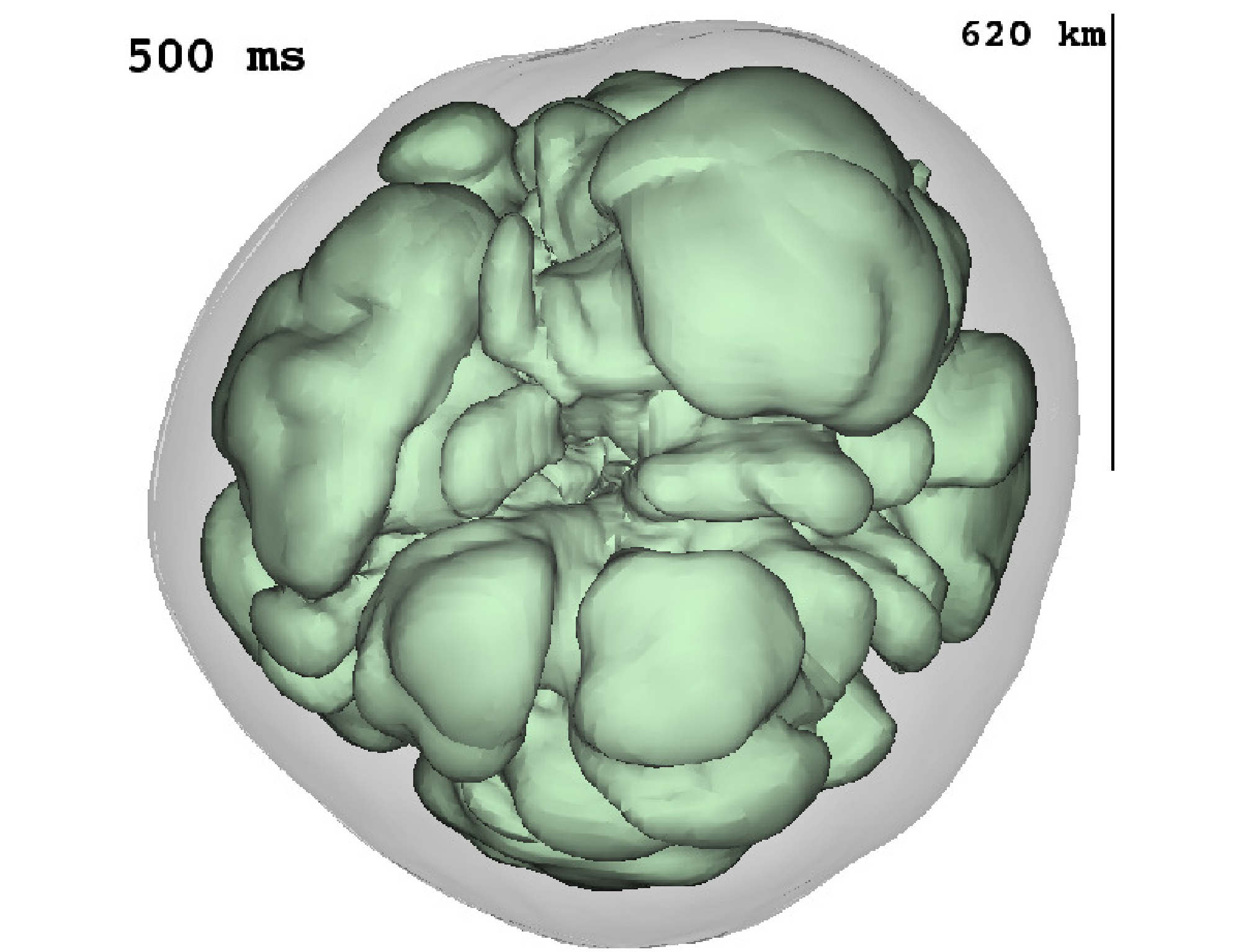}}\\
\resizebox{0.47\hsize}{!}{\includegraphics*{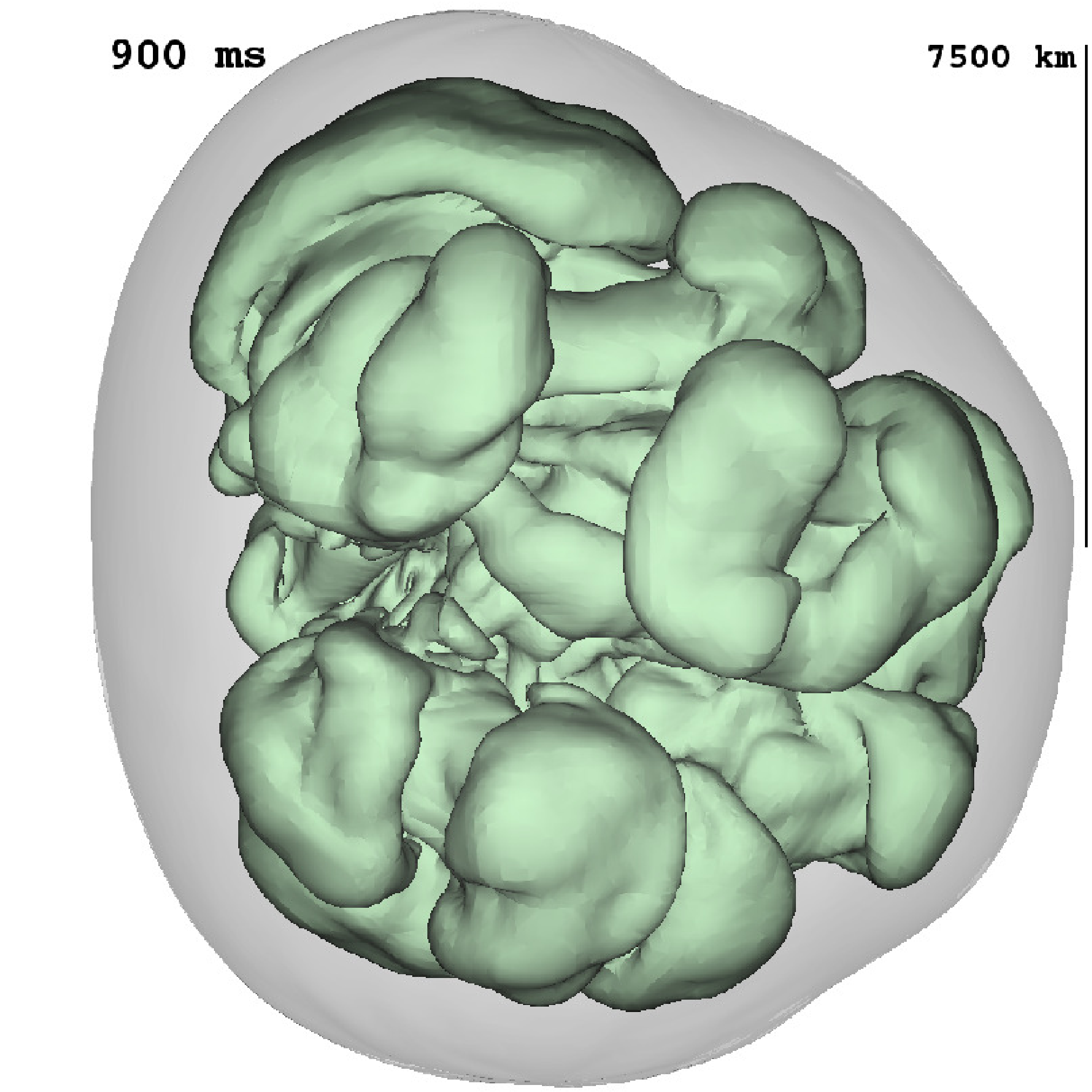}}\hspace{1cm}
\resizebox{0.47\hsize}{!}{\includegraphics*{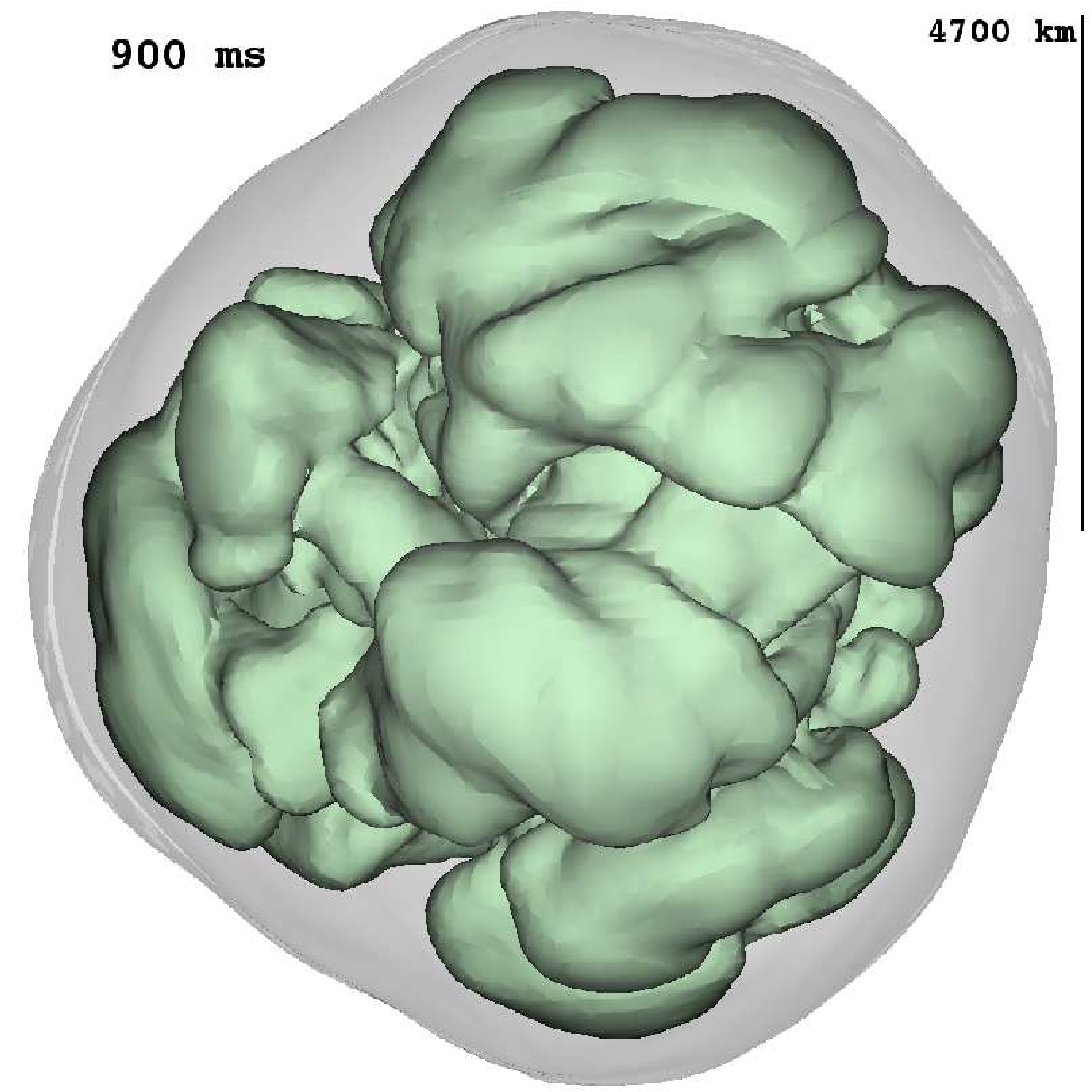}}\\
\caption{Snapshots of models W15-4 (left) and L15-3 (right)
  illustrating the four phases characterizing the evolution of our 3D
  models (see text for details). Each snapshot shows two surfaces of
  constant entropy marking the position of the shock wave (gray) and
  depicting the growth of non-radial structures (greenish). The time
  and linear scale are indicated for each snapshot. }
\label{fig:phases}
\end{figure*}
%

%
\begin{figure}[htp!]
\centering
\resizebox{0.95\hsize}{!}{\includegraphics*{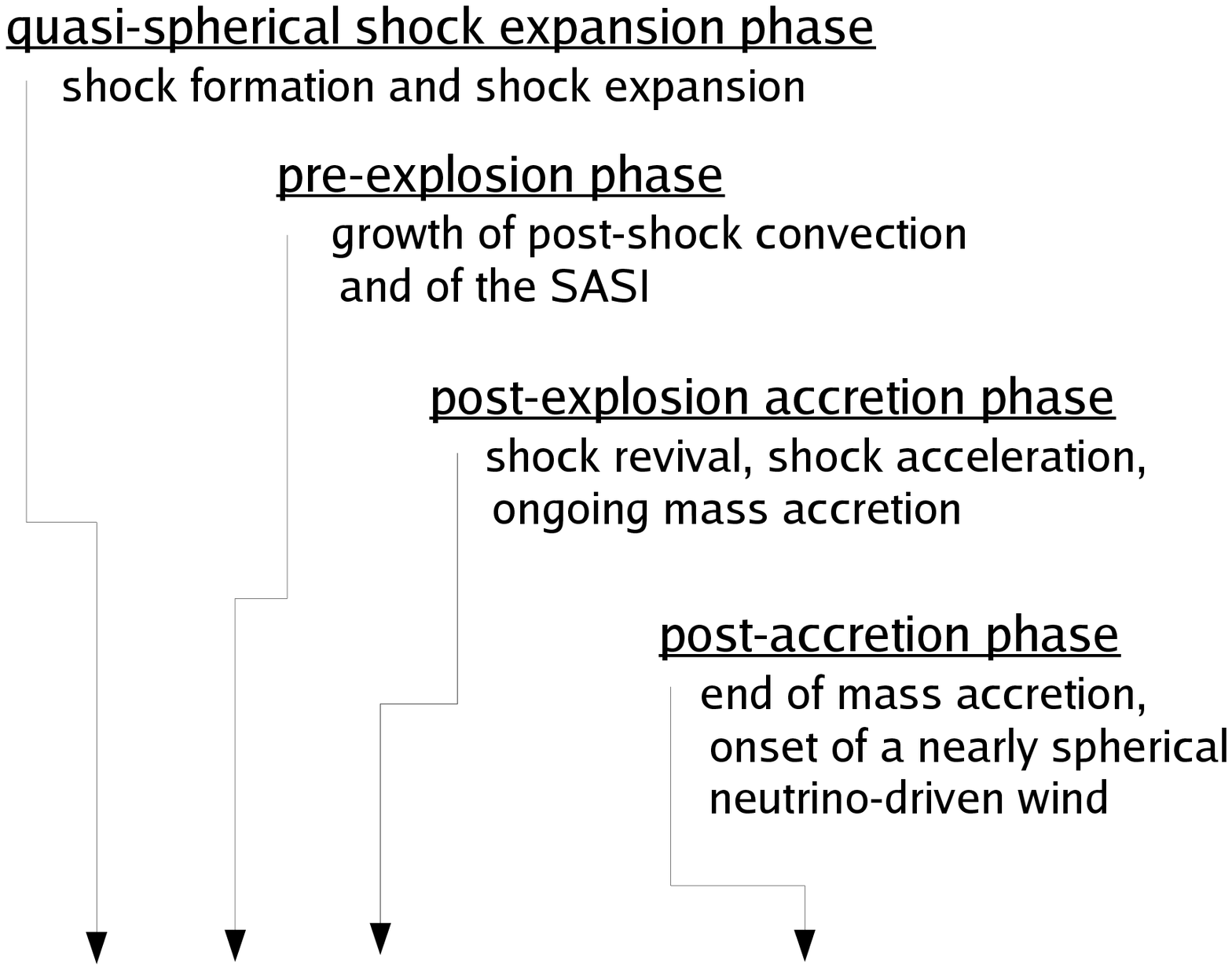}}\\
\resizebox{0.95\hsize}{!}{\includegraphics*{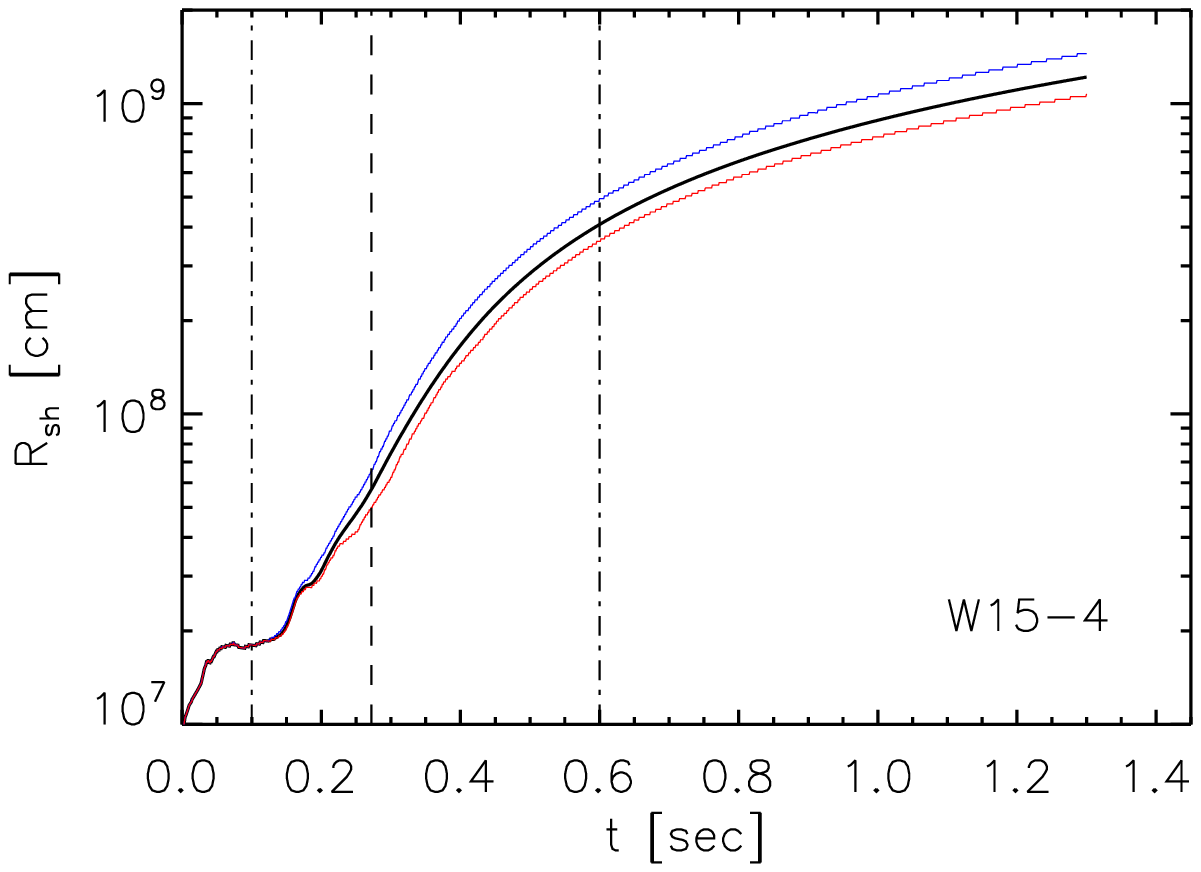}}\\
\resizebox{0.95\hsize}{!}{\includegraphics*{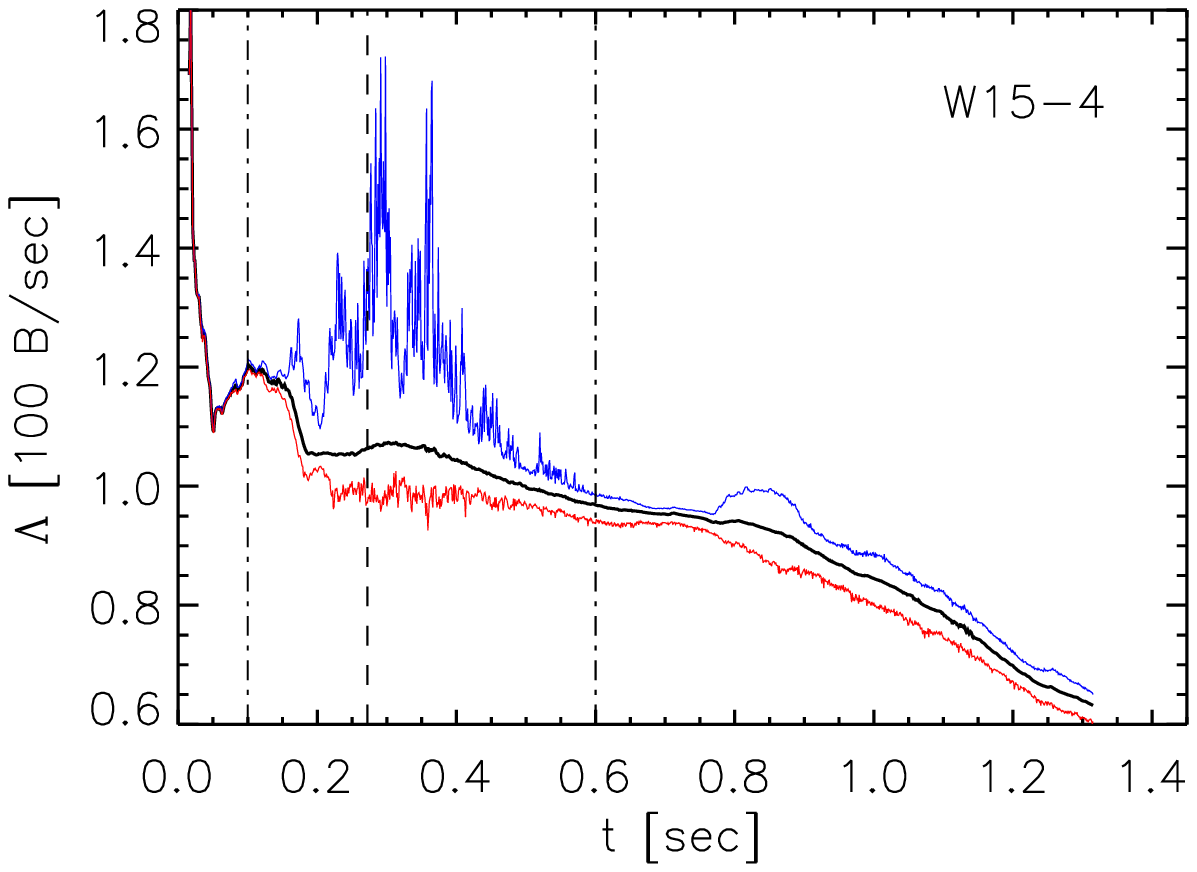}}
\caption{Shock radius (top) and total (\ie summed over all flavors)
  energy loss rate due to neutrinos (bottom) as functions of time for
  model W15-4.  In the upper panel, the black curve shows the
  angle-averaged mean shock radius, the blue (red) curve gives the
  maximum (minimum) shock radius, and the vertical dashed line marks
  the time of the onset of the explosion as defined in
  Sect.\,\ref{subsec:models}. In the lower panel, the blue and red
  curves show the time evolution of $\Lambda_\mathrm{max}(\Omega,t)$
  and $\Lambda_\mathrm{min}(\Omega,t)$, the maximum and minimum value
  of $\Lambda(\Omega,t)$ (Eq.\,\ref{eq:lambdae}) on a sphere of
  500\,km radius, respectively. The black line gives the corresponding
  surface-averaged value $\Lambda(t)$ (Eq.\,\ref{eq:lamt}). Note that
  the luminosities imposed at the inner radial grid boundary are kept
  constant during the first second and later are assumed to decay like
  $t^{-2/3}$.}
\label{fig:rsh+lnu_vs_t}
\end{figure}

\begin{figure}[!]
\centering
\resizebox{0.95\hsize}{!}{\includegraphics*{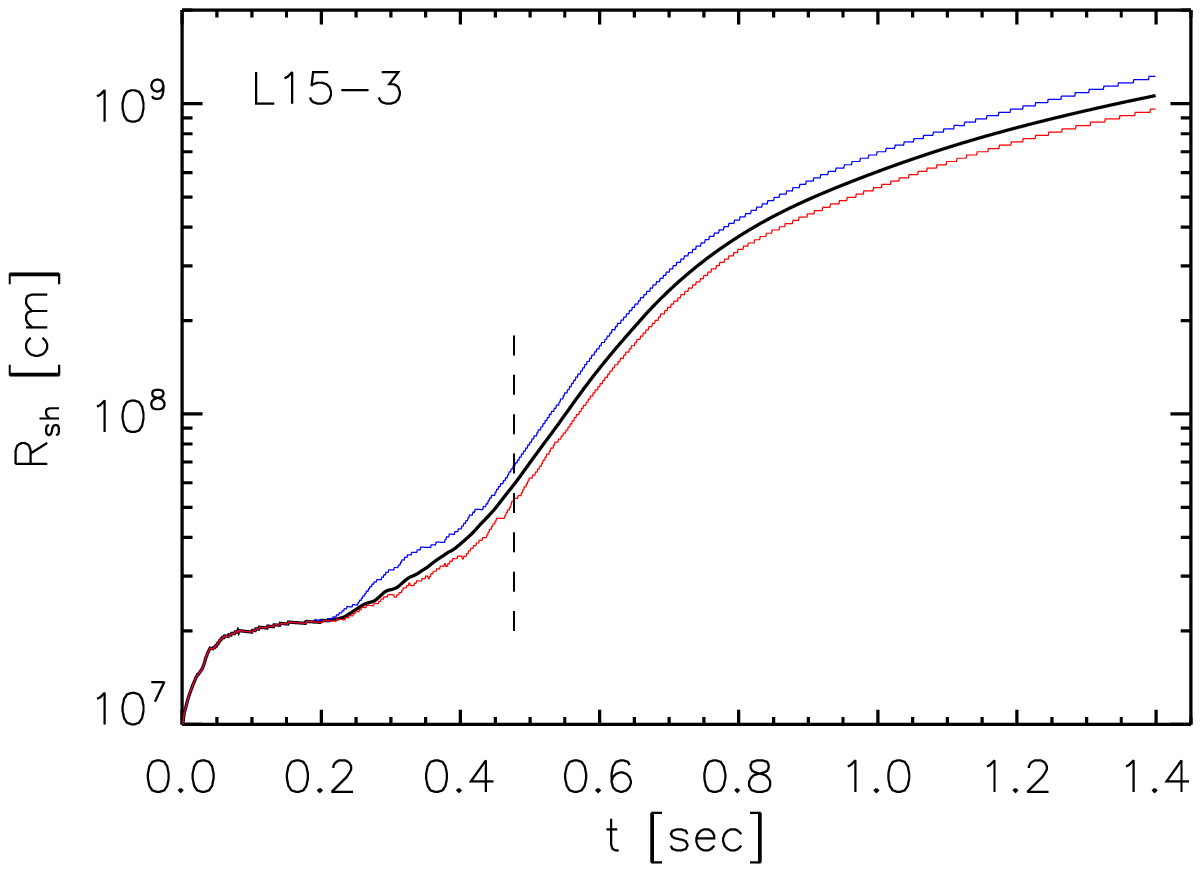}}\\
\resizebox{0.95\hsize}{!}{\includegraphics*{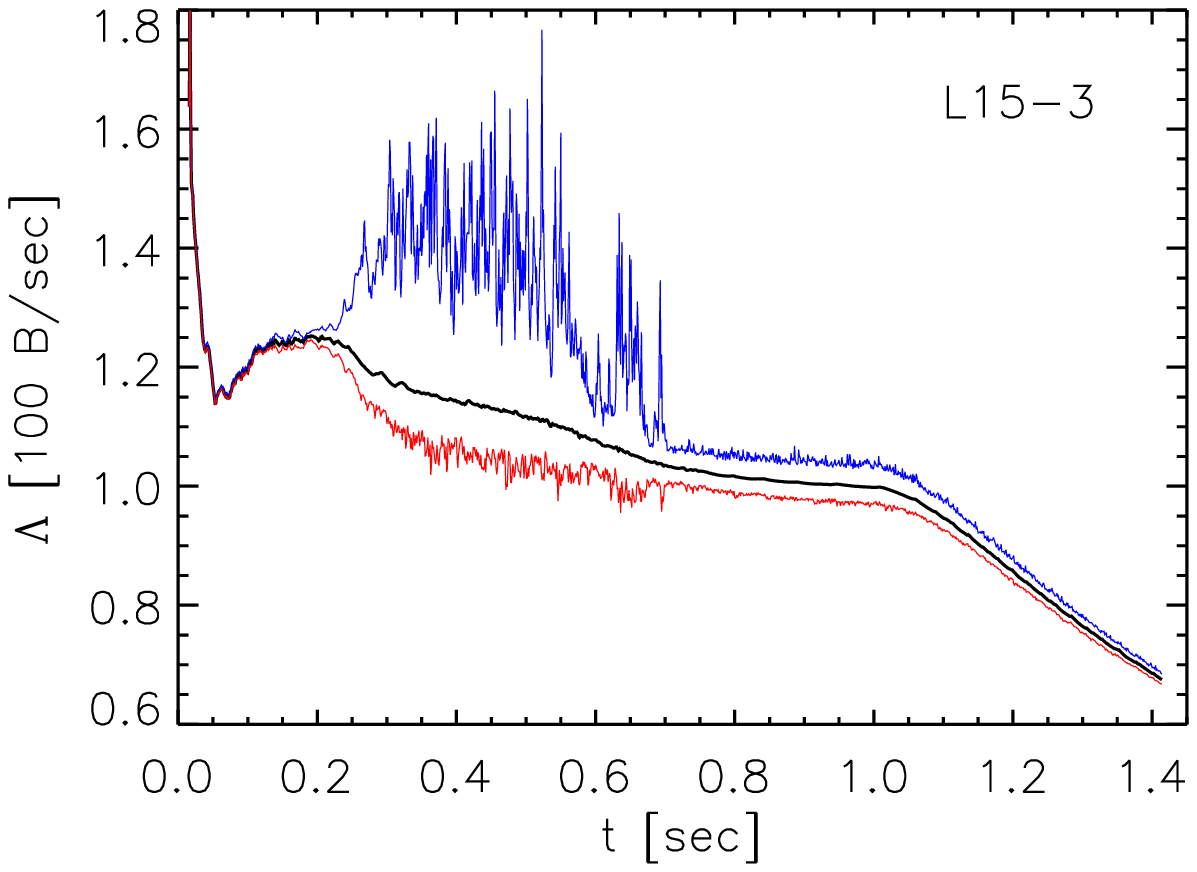}}\\
\resizebox{0.95\hsize}{!}{\includegraphics*{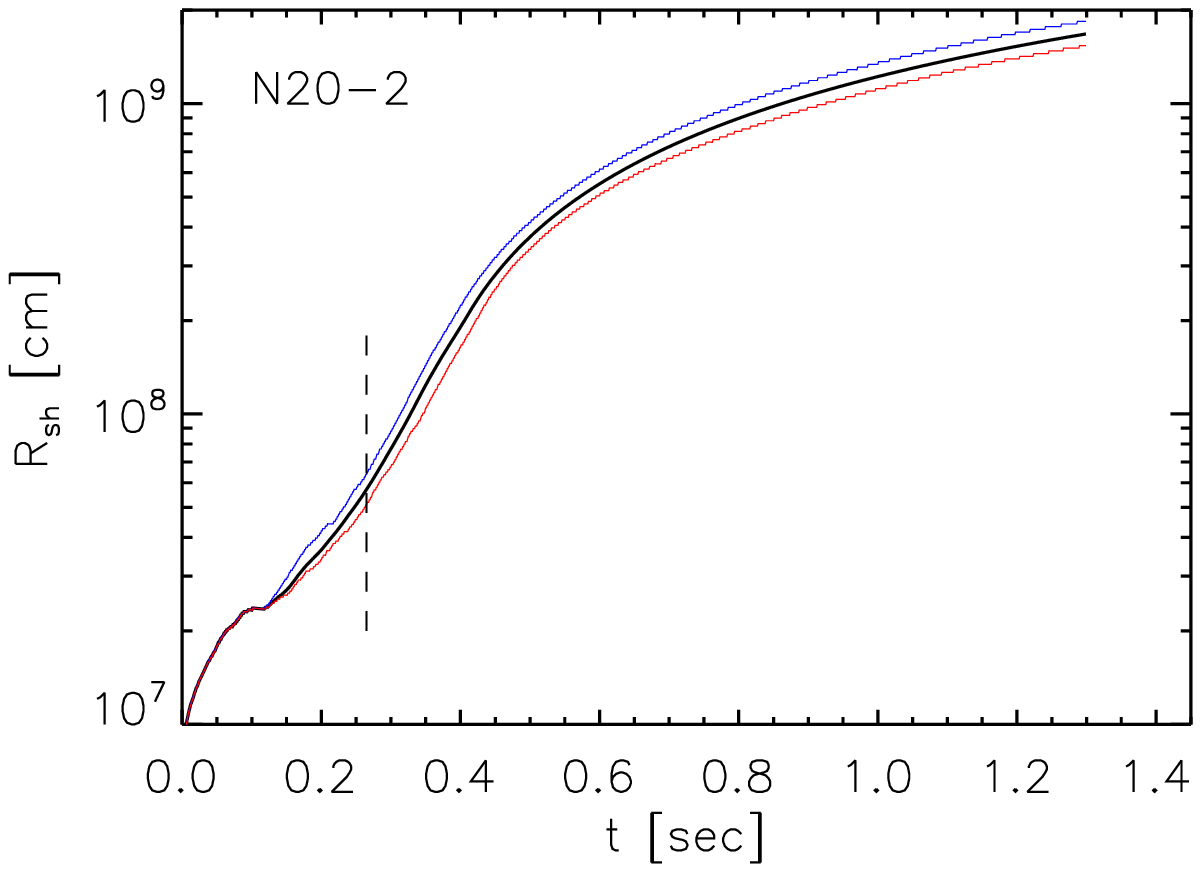}}\\
\resizebox{0.95\hsize}{!}{\includegraphics*{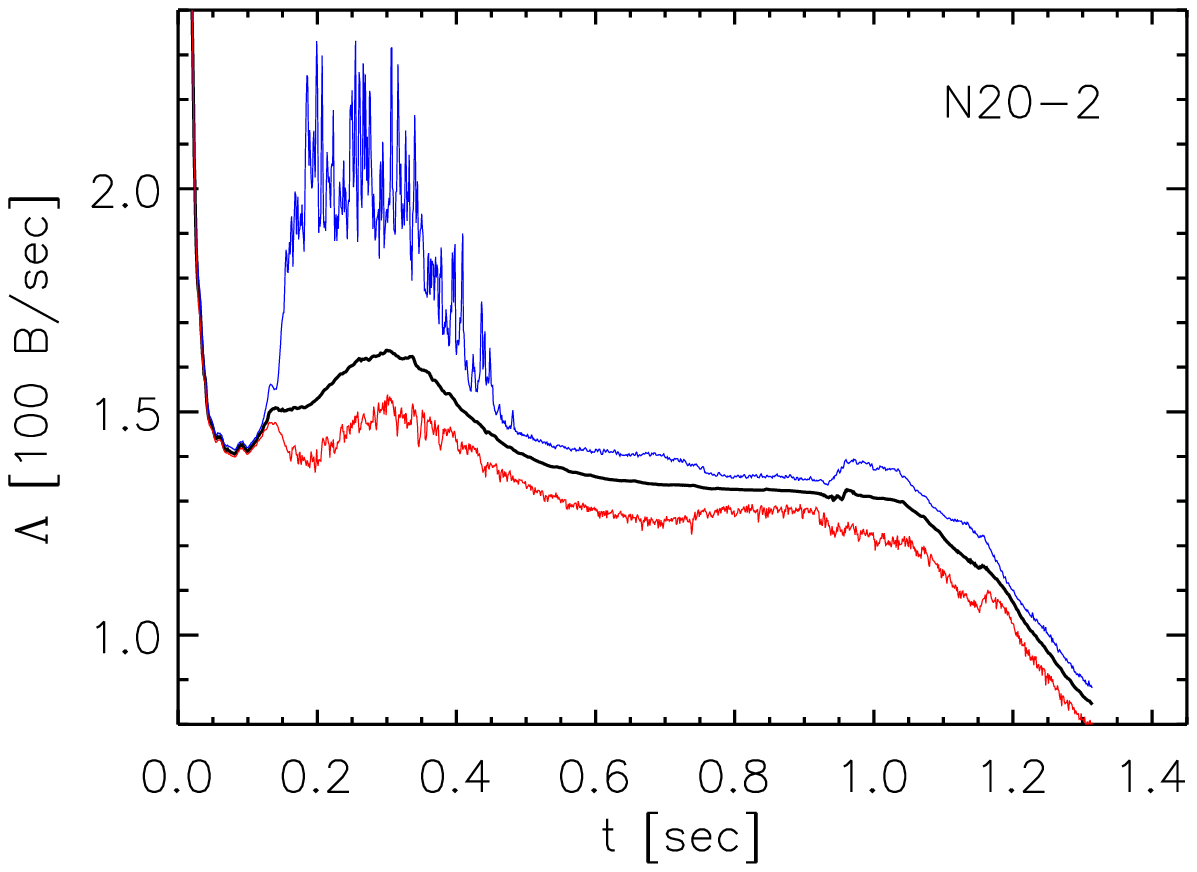}}
\caption{Same as Fig.\,\ref{fig:rsh+lnu_vs_t} but for models L15-3
  (uppermost two panels) and N20-2 (lowermost two panels),
  respectively.}
\label{fig:rsh+lnu_vs_t_a}
\end{figure}

\subsection{Results}
\label{subsec:nuresults}

\begin{figure}[!]
\centering
\resizebox{0.88\hsize}{!}{\includegraphics*{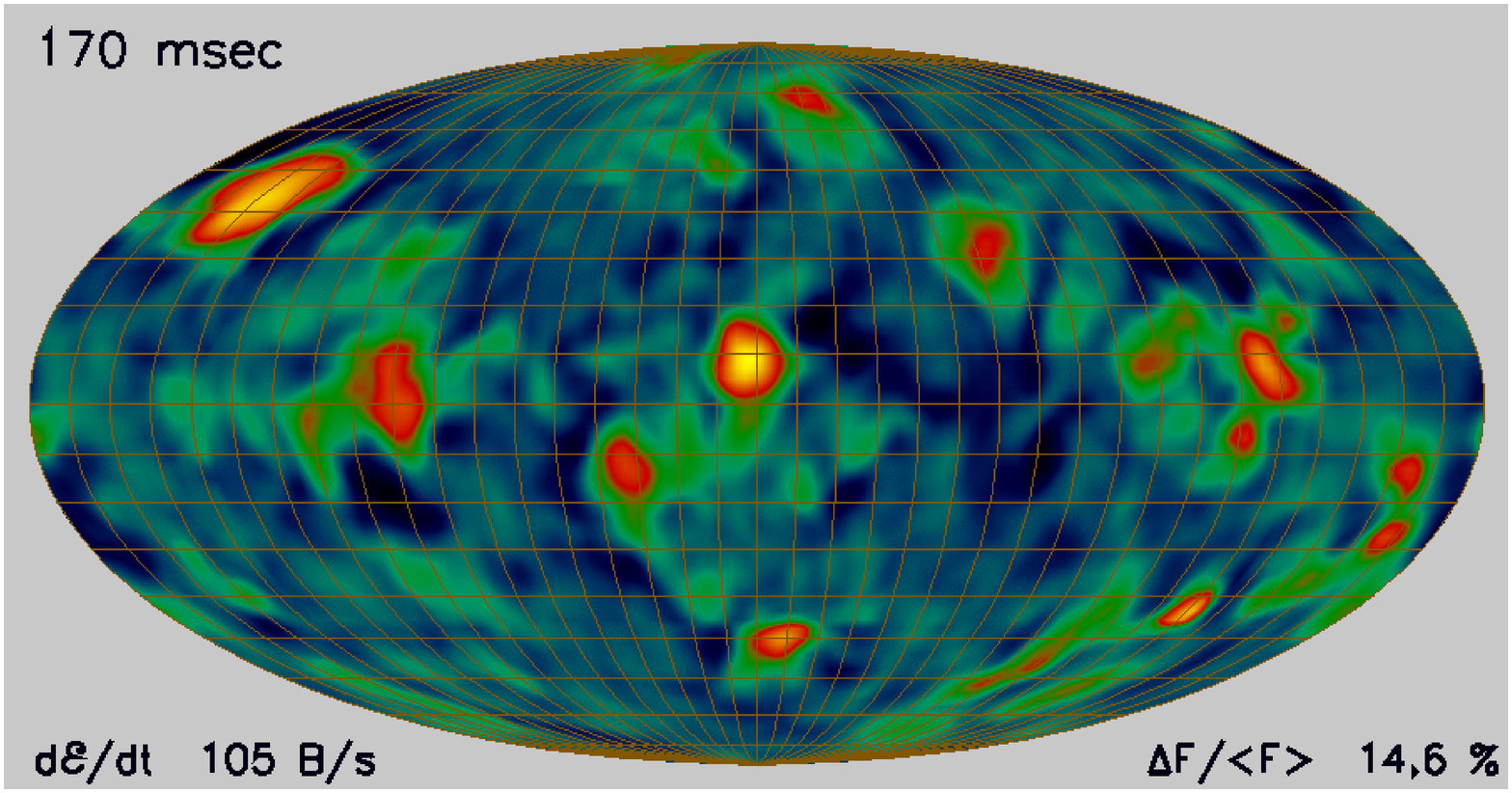}}\\
\resizebox{0.88\hsize}{!}{\includegraphics*{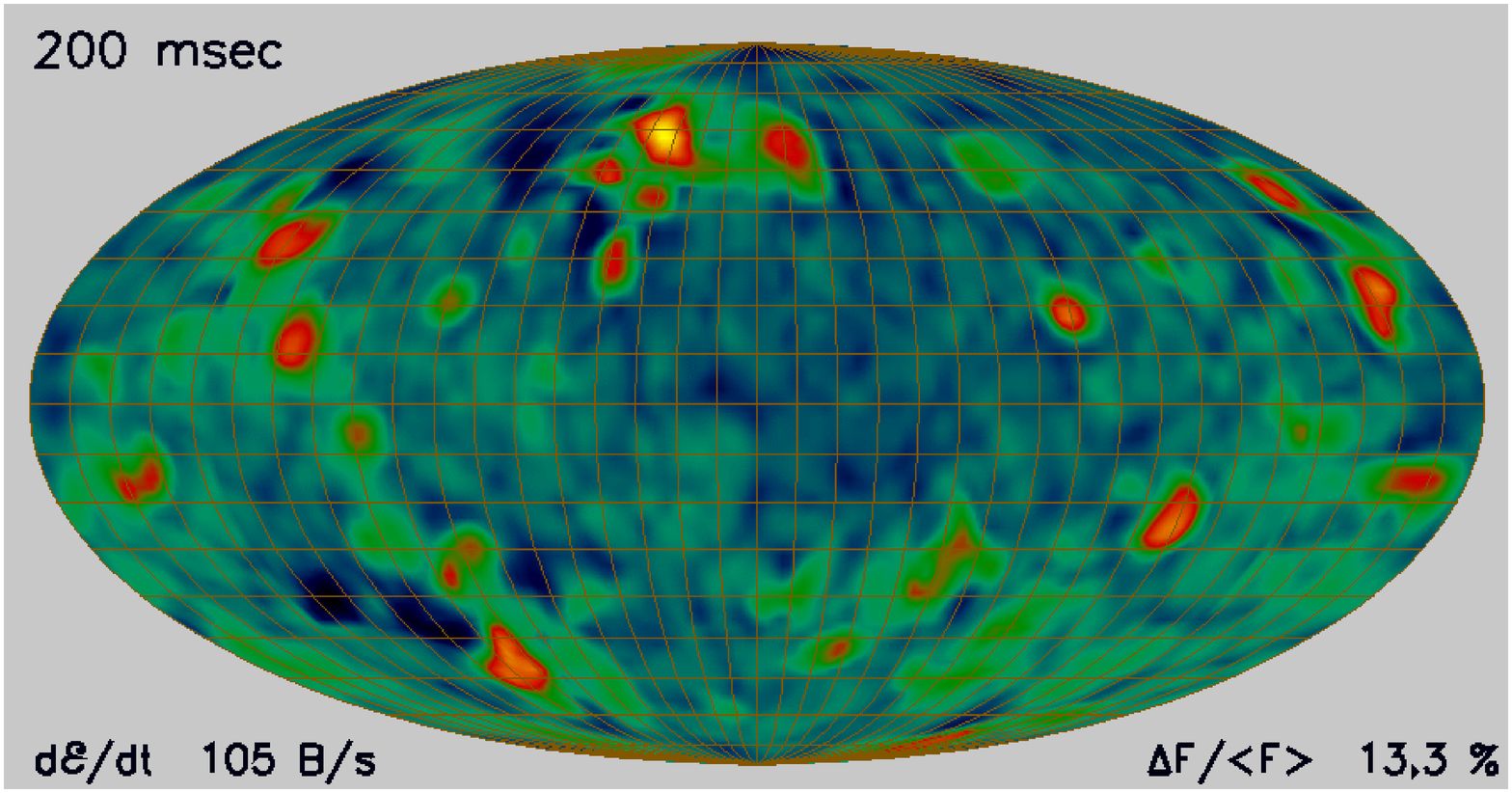}}\\
\resizebox{0.88\hsize}{!}{\includegraphics*{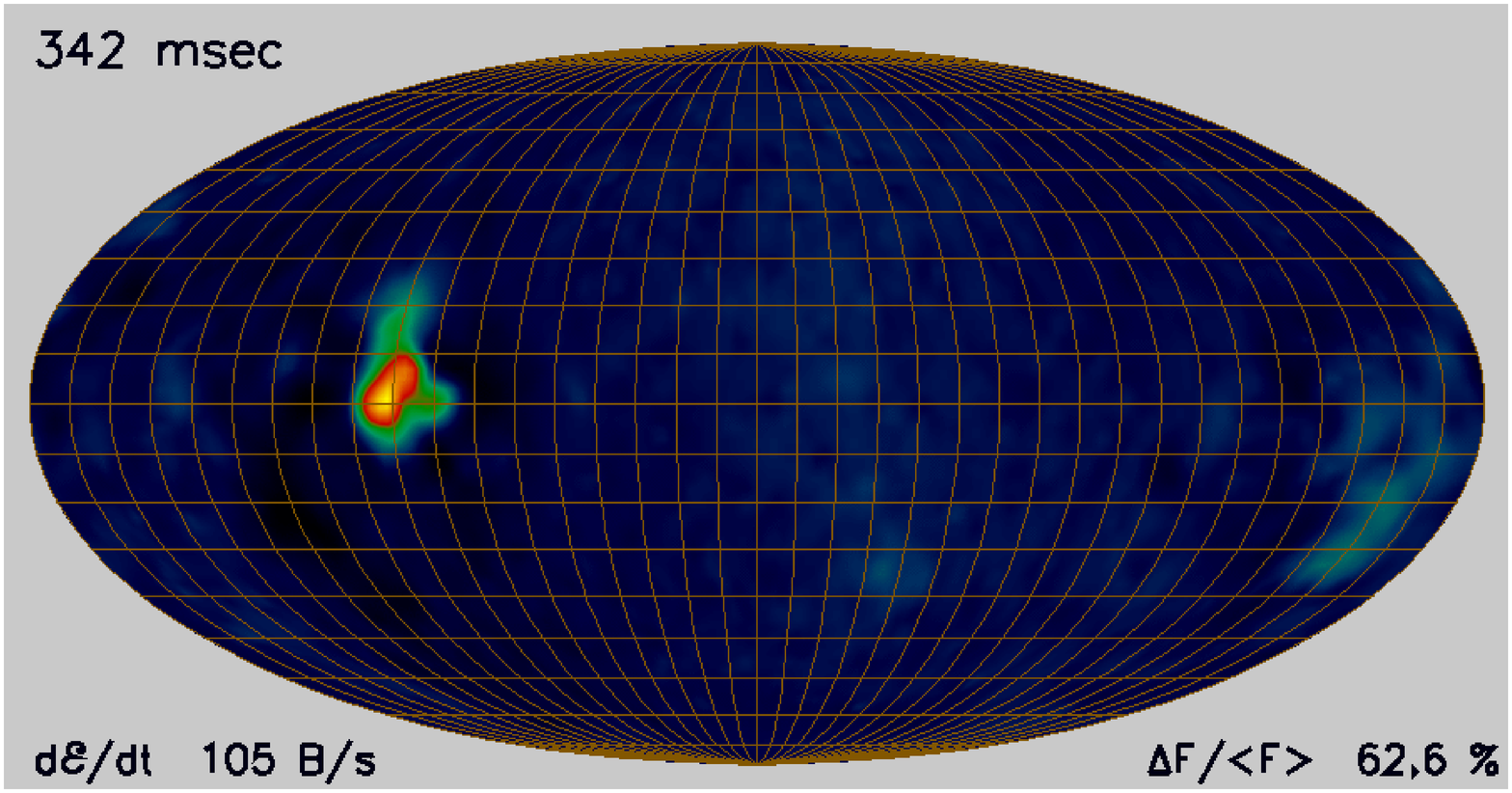}}\\
\resizebox{0.88\hsize}{!}{\includegraphics*{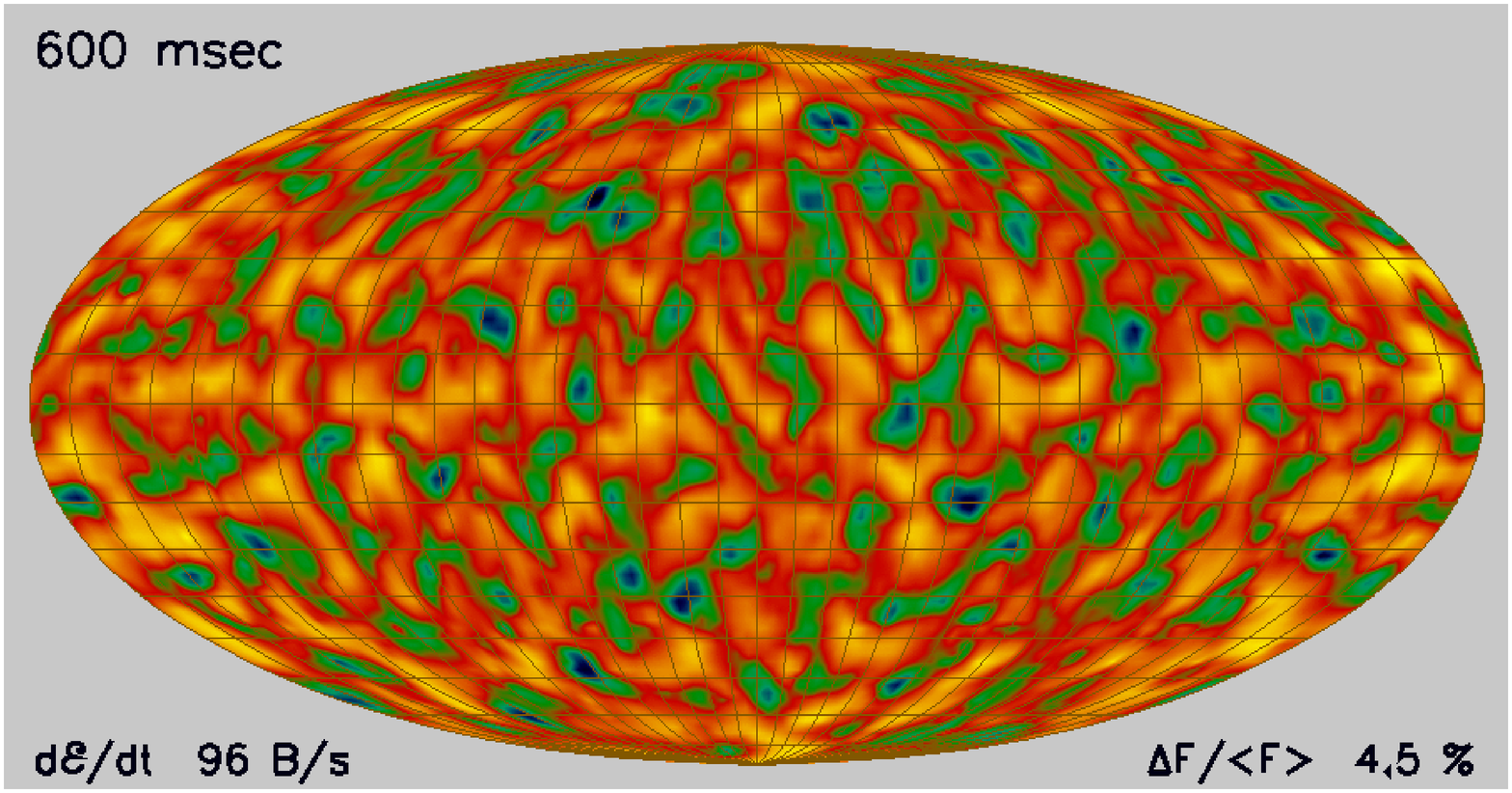}}\\
\resizebox{0.88\hsize}{!}{\includegraphics*{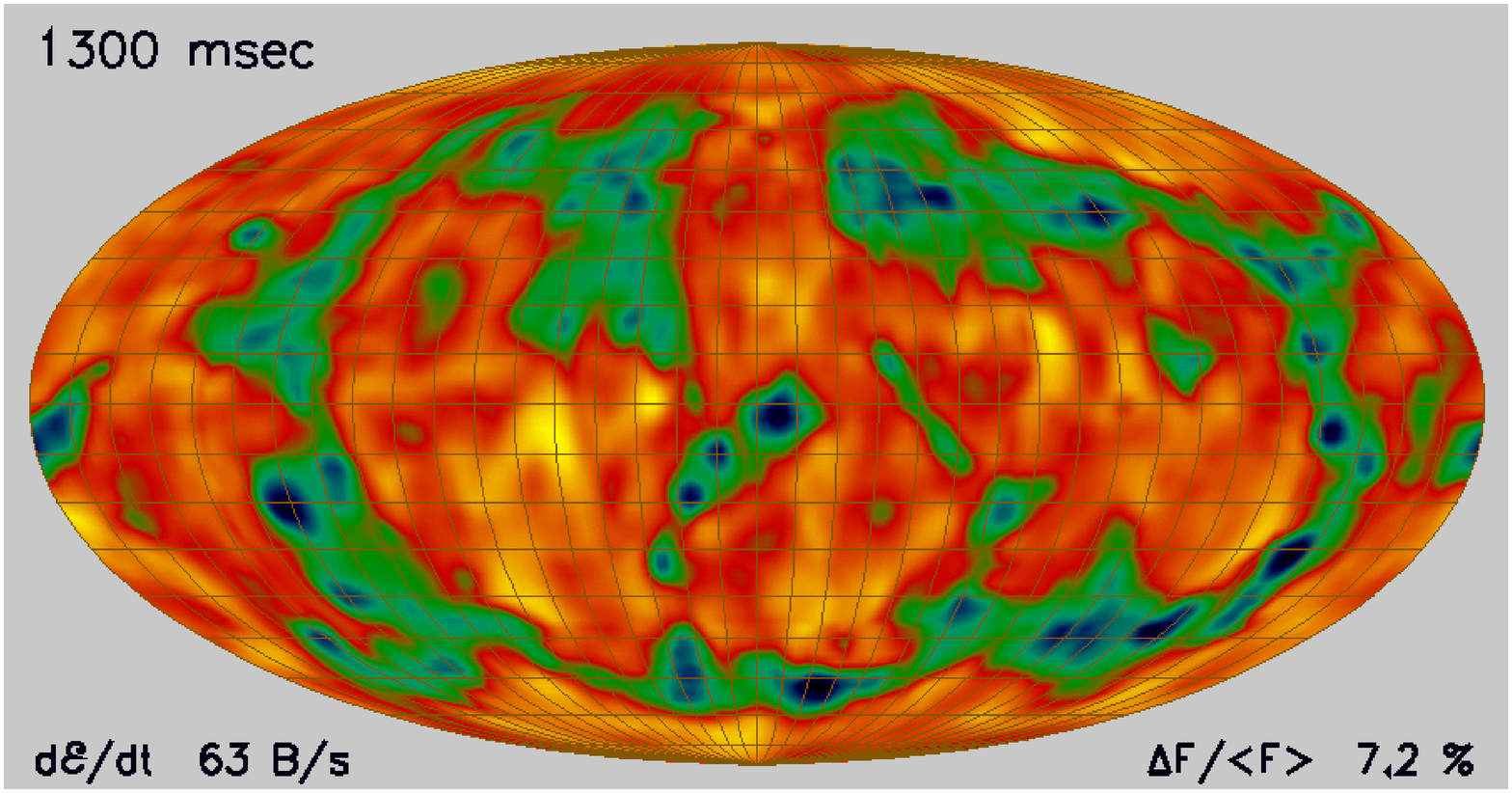}}
\caption{Neutrino flux asymmetry at 170\,msec, 200\,msec, 342\,msec,
  600\,msec, and 1.3\,sec (from top to bottom), respectively. The
  $4\pi$-maps show the relative angular variation $\Delta F_o /
  \left\langle F_o \right\rangle$ of the total (\ie sum of all
  neutrino flavors) neutrino energy flux density over a sphere
  (normalized to its angular average) for model W15-4. The maximum
  value is given in the lower right corner of each panel.  Regions of
  higher emission are shown in bright yellow, while orange, red,
  green, and blue colors indicate successively less emission.  Note
  that the color scale of each panel is adjusted to the maximum and
  minimum values at the corresponding time.  The total energy loss
  rate due to neutrinos is given in the lower left corner.}
\label{fig:W15-4_lnue}
\end{figure}

The evolution of our models can be divided into four distinct phases
(Figs.\,\ref{fig:phases}, \ref{fig:rsh+lnu_vs_t}).  
\begin{itemize}

\item[(1)] The first phase, the \emph{quasi-spherical shock-expansion
  phase} (Fig.\,\ref{fig:phases}, top row), lasts from shock formation
  shortly after core bounce to $80\,-\,150$\,msec, when convection
  sets in. During this phase the shock rapidly propagates out to a
  radius of $\sim\,$200\,km, where its expansion comes to a halt.

\item[(2)] The second phase, the hydrodynamically vigorous
  \emph{pre-explosion phase}, comprises the growth of post-shock
  convection and of the standing accretion shock instability, SASI
  (Fig.\,\ref{fig:phases}, second row from top).

\item[(3)] The \emph{post-explosion accretion phase} begins when
  energy deposition by $\nu$-heating in the post-shock layers becomes
  sufficiently strong to launch the explosion, and the total energy in
  the post-shock region ultimately becomes positive (see
  Sect.\,\ref{subsec:models} for a definition). During this phase the
  shock accelerates outward while gas is still accreted onto the PNS.
  This process is commonly called "shock revival"
  (Fig.\,\ref{fig:phases}, third row from top).

  Non-radial instabilities during the latter two stages cause
  considerable temporal and angular fluctuations of the neutrino
  energy flux density as illustrated in Figs.\,\ref{fig:rsh+lnu_vs_t}
  - \ref{fig:W15-4_lnue}. Besides the evolution of the shock radius,
  the figures show the surface-averaged neutrino light curve
  $\Lambda(t)$, \ie the total energy loss due to neutrinos versus time
  (Eqs.\,\ref{eq:lamt}, \ref{eq:eloss}), together with the time
  evolution of the maximum and minimum values of $\Lambda(\Omega,t)$
  (Eq.\,(\ref{eq:lambdae}); the numerical evaluation is performed on
  an arbitrarily chosen sphere of 500\,km radius). Distinct and
  high-amplitude spikes in $\Lambda_\mathrm{max} (\Omega,t)$ are
  visible for several 100\,msec and reflect violent post-shock
  convection, possible SASI activity, and anisotropic accretion
  fluctuations after the onset of the explosion.  We have marked the
  explosion time $t_\mathrm{exp}$ (see Section\,\ref{subsec:models},
  and Table\,1) by a vertical dashed line in
  Figs.\,\ref{fig:rsh+lnu_vs_t} and \ref{fig:rsh+lnu_vs_t_a}.  The
  post-explosion accretion phase lasts until $\sim\,$500\,msec (models
  W15-4 and N20-2) or $\sim\,$700\,msec (model L15-3) depending on the
  progenitor.

\item[(4)] During the \emph{post-accretion phase}, the fourth and
  final phase characterizing the evolution of our models
  (Fig.\,\ref{fig:phases}, bottom row), gas infall to the
  proto-neutron star has come to an end and the newly formed neutron
  star looses mass at a low rate in a nearly spherical neutrino-driven
  wind.  We find considerably less temporal variability and a lower
  level of angular variation ($\la 10\%$) of the neutrino emission
  during this fourth phase (Figs.\,\ref{fig:rsh+lnu_vs_t} -
  \ref{fig:W15-4_lnue}).

  While in model L15-3 the amplitudes of the neutrino emission
  fluctuations decrease continuously, the other two models exhibit
  growing temporal emission variations (though at a lower level than
  the earlier variability) during a later stage (notice the
  decrease/increase in $\Lambda_\mathrm{max} - \Lambda_\mathrm{min}$
  in Figs.\,\ref{fig:rsh+lnu_vs_t} and \ref{fig:rsh+lnu_vs_t_a}),
  which might be considered as a fifth evolutionary phase. This phase
  is associated with growing convective activity below the
  neutrinosphere.  This PNS convection develops more or less strongly
  in the different models depending on the location of the
  convectively unstable region relative to the inner radial boundary
  of our computational domain.
\end{itemize}

%
%

%
\begin{figure*}
\centering
\resizebox{0.33\hsize}{!}{\includegraphics*{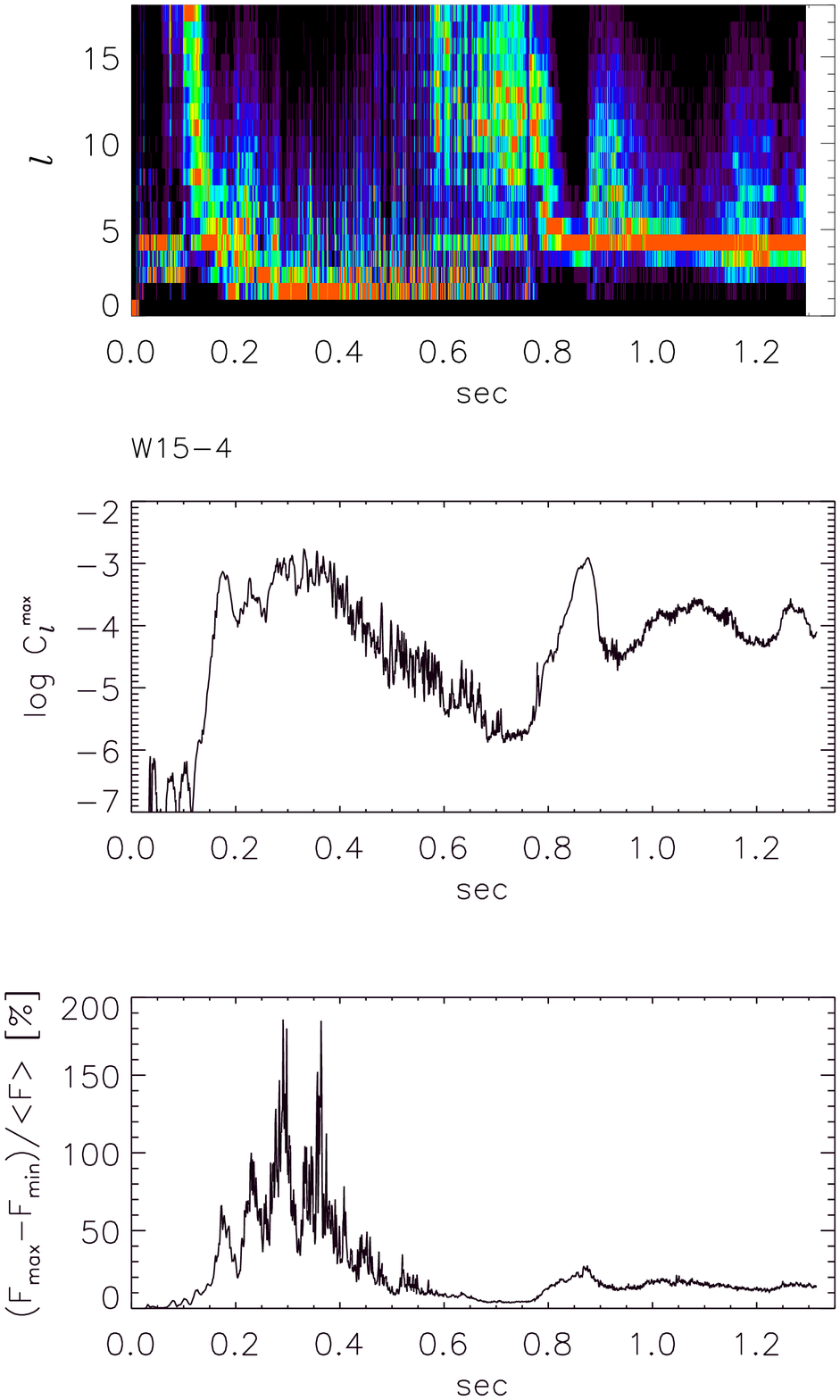}}
\resizebox{0.33\hsize}{!}{\includegraphics*{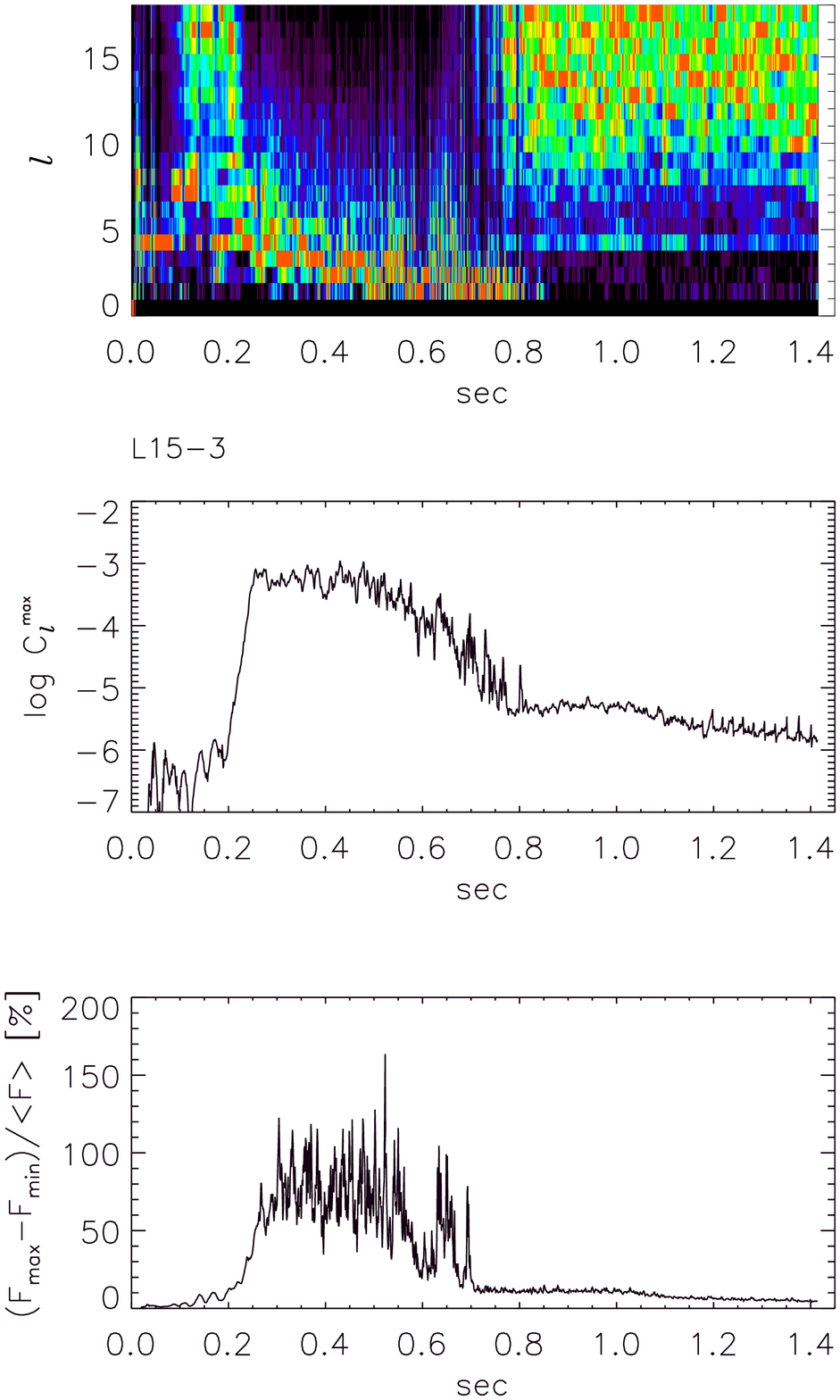}}
\resizebox{0.33\hsize}{!}{\includegraphics*{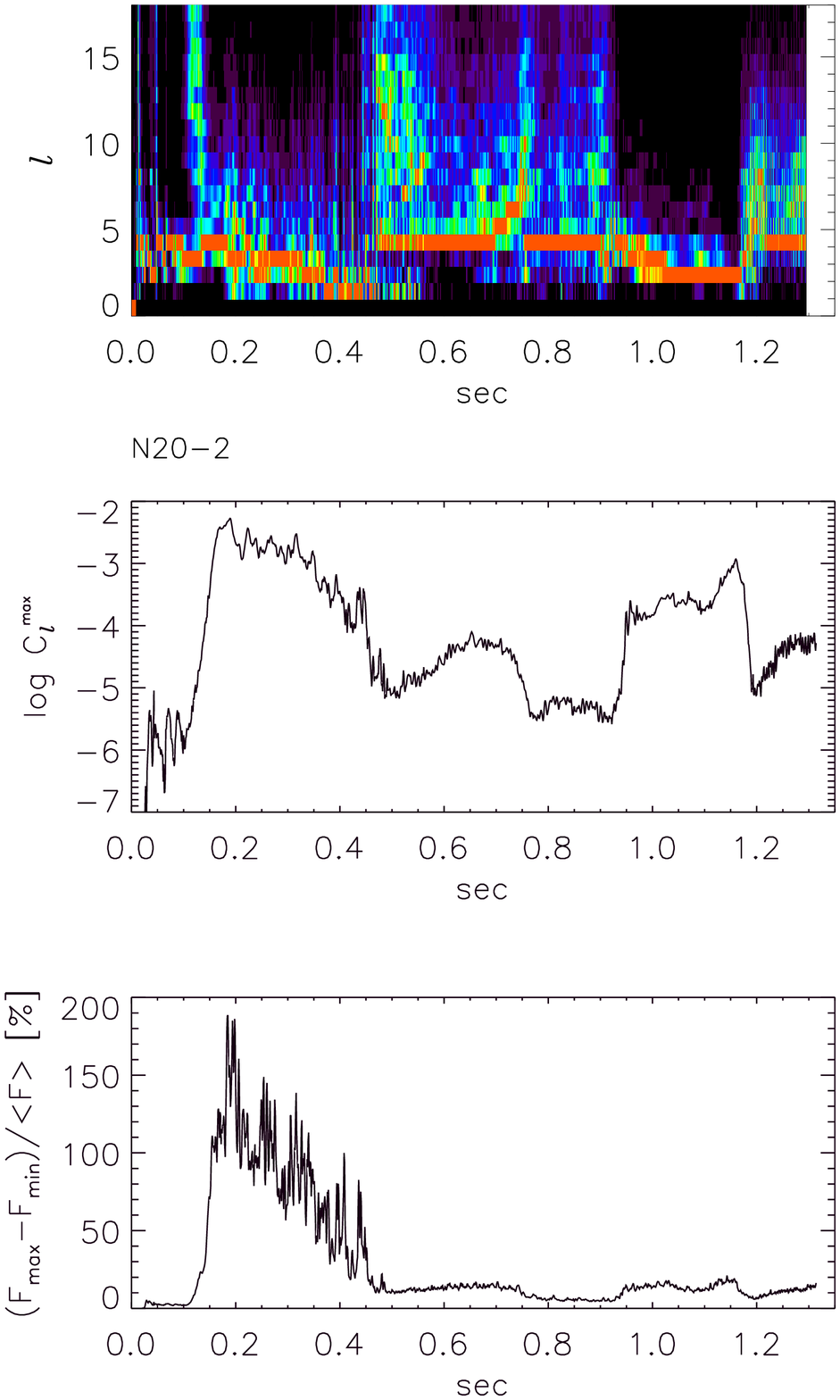}}
\caption{Pseudo-power spectrogram of the electron-neutrino energy flux
  density (top row) for models W15-4 (left), L15-3 (middle), and N20-2
  (right), respectively. The panels in the middle row show the
  corresponding maximum pseudo-power coefficient $C^\mathrm{max}_l$ as
  a function of time, and the panels in the lower row give the
  relative angular variation of the electron neutrino flux density
  (maximum minus minimum flux density on the sphere divided by the
  angle-averaged flux density in percent) with time. }
\label{fig:cl_spectrogram}
\end{figure*}
%

\begin{figure}
\centering
\resizebox{0.9\hsize}{!}{\includegraphics*{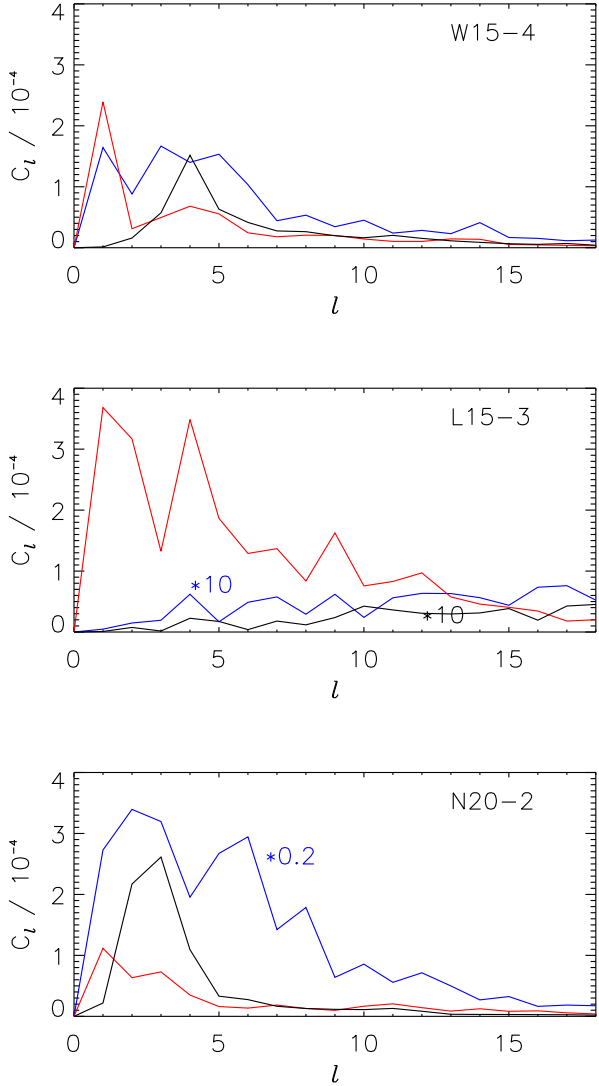}}
\caption{Pseudo-power coefficients $C^\mathrm{max}_l$ of the
  electron-neutrino flux density as functions of angular mode number
  $l$ at 200\,ms (blue), 400\,ms (red), and 1000\,ms (black) for
  models W15-4 (top), L15-3 (middle), and N20-3 (bottom),
  respectively. }
\label{fig:cl_vs_l}
\end{figure}

We have evaluated the time evolution of the neutrino energy flux
asymmetry by producing $4\pi$-maps that show the relative angular
variation $\Delta F_o / \left\langle F_o \right\rangle$ of the total
(\ie sum of all neutrino flavors) neutrino energy flux density across
a sphere (normalized to the surface-averaged flux density;
Eq.\,\ref{eq:flxave}). Several snapshots of this evolution are shown
for model W15-4 in Fig.\,\ref{fig:W15-4_lnue}.  The evolution of the
typical angular scales of the fluctuations is reflected by the
pseudo-power spectrogram of the electron neutrino energy flux
variation (Eq.\,\ref{eq:dlam}) in Fig.\,\ref{fig:cl_spectrogram}, top
panels, which give the color-coded pseudo-power coefficient
distribution normalized to the maximum value versus time. The
variation of the pseudo-power coefficients with angular mode number is
shown in Fig.\,\ref{fig:cl_vs_l} at selected times of 200\,ms (blue),
400\,ms (red), and 1000\,ms (black).

During the quasi-spherical shock expansion phase the level of angular
fluctuations of $F$ is low ($\la 10^{-2}$), while the fluctuation
amplitudes of the total neutrino energy flux density reach a level of
several 10\% during the hydrodynamically vigorous second phase and the
post-explosion accretion phase, where a few distinct regions or even
single spots with an angular size of 10$^\circ$ to 20$^\circ$ dominate
the emission (Fig.\,\ref{fig:W15-4_lnue}, panels 2 and 3). The mode
number $l$ of the dominant angular perturbation scale is of no
relevance during the first phase, as the maximum pseudo-power
coefficient $C^\mathrm{max}_l$ (see Eq.\,\ref{eq:cl}) is tiny $\la
10^{-6}$ (Fig.\,\ref{fig:cl_spectrogram}, middle panels), \ie the
dominating $l=2$ and $l=4$ modes visible in the upper panels of
Fig.\,\ref{fig:cl_spectrogram} only reflect tiny angular perturbations
imprinted presumably by the computational grid.  When neutrino heating
eventually causes significant non-radial flow during the second and
third phases, $C^\mathrm{max}_l$ rises sharply to a level of $\sim
10^{-3}$ (Fig.\,\ref{fig:cl_spectrogram}, middle panels), and the
relative angular variations of the electron neutrino flux density grow
to the several ten percent level (Fig.\,\ref{fig:cl_spectrogram},
bottom panels).  The latter quantity gives the maximum minus the
minimum flux density on the sphere divided by the angle-averaged flux
density in percent.  Compared to the \emph{total} neutrino emission in
Figs.\,\ref{fig:rsh+lnu_vs_t} - \ref{fig:W15-4_lnue}, the temporal and
angular variations in different directions are even more pronounced
when considering the energy flux of the electron neutrinos or electron
anti-neutrinos alone (Fig.\,\ref{fig:cl_spectrogram}), where angular
variations can exceed 100\% in all models during the pre-explosion and
accretion phases, and peak values are close to 200\% during short
episodes (Fig.\,\ref{fig:cl_spectrogram}, lower panels).

During the vigorous pre-explosion phase including the post-explosion
accretion stage, electron neutrinos and antineutrinos dominate the
angular flux variations, while muon and tau neutrinos (accounting for
roughly 50\% of the total luminosity) exhibit essentially isotropic
emission in all directions. The reason of this finding is that $\nu_e$
and $\bar \nu_e$ are produced almost exclusively by efficient
charged-current reactions in the accretion region perturbed by
non-radial fluid flows.  The spectrogram of the two phases is
characterized by initially very small-scale angular variations with $l
\ga 12$, which are associated with the onset of the Rayleigh-Taylor
overturn activity, and which merge to continuously larger angular
structures that correspond to $l \approx 1 \ldots 4$ modes toward the
end of the accretion period at $0.4\,-\,0.6\,$s (depending on the
model). This evolution is accompanied by a steady decrease of
$C^\mathrm{max}_l$ to a level of $\sim 10^{-5}$ and a reduction of the
electron neutrino flux density variations from values well beyond
100\% to a level of $\sim\,10\%$, only (see
Fig.\,\ref{fig:cl_spectrogram}, left panels).

When neutrino-energy deposition in the post-shock layers becomes
sufficiently strong and the explosion is eventually launched at about
250 to 500\,msec (depending on the model; Table\,1), subsequent radial
shock expansion rapidly diminishes the activity of the SASI and
freezes post-shock convection.  Single, longer lasting downdrafts of
accretion flows are associated with isolated hot spots, where the
variations of the total flux density can reach peak amplitudes of up
to $\sim\,70\,$\% (Fig.\,\ref{fig:W15-4_lnue}, panel 3). When
accretion has ended, the amplitude of the angular variations of the
total neutrino energy flux reduces to a level of a few percent
(Figs.\,\ref{fig:rsh+lnu_vs_t}, \ref{fig:rsh+lnu_vs_t_a}), and the
angular pattern of the emission becomes more uniform over the sphere,
consisting of many spots with an angular size of $\sim\,$30$^\circ$
(Fig.\,\ref{fig:W15-4_lnue}, panel 4).

In the early post-accretion phase of model W15-4, $0.6\,\mathrm{s} \la
t \la 0.8\,$s, the spectrogram indicates the presence of low-amplitude
($C^\mathrm{max}_l \la 10^{-4}$), small-angular size ($l \ga 10$)
perturbations in the electron neutrino energy flux caused by some
low-amplitude turbulent flow in and below the neutrinospheric
region. When strong convection inside the PNS is encountered for $t
\ga 0.8\,$s the spectrogram drastically changes, being dominated by
angular modes with $l=4$, but still with $C^\mathrm{max}_l \sim
10^{-4}$. The electron neutrino flux density variations rise somewhat
to a level of 10\% to 20\%, and become manifest in the total energy
loss rate, too (Figs.\,\ref{fig:rsh+lnu_vs_t},
\ref{fig:rsh+lnu_vs_t_a}).

Model N20-2 exhibits quite a similar behavior as model W15-4 except
for the appearance of even larger ($l \sim 3$) angular structures
clearly recognizable in the pseudo-power spectrogram between 1.0\,s
and 1.2\,s (Figs.\,\ref{fig:cl_spectrogram}, \ref{fig:cl_vs_l}).  This
differs from the behavior of model L15-3, where the amplitudes and
angular size of the energy flux density variations remain small and
even decrease in the post-explosion phase
(Figs.\,\ref{fig:rsh+lnu_vs_t_a}, \ref{fig:cl_spectrogram}).

%
%

%
\begin{figure}[t]
\centering
\resizebox{0.90\hsize}{!}{\includegraphics{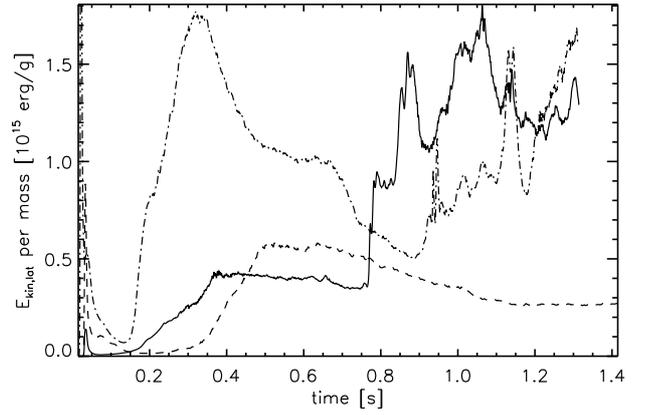}}
\caption{Evolution of the non-radial specific kinetic energy
  $(v_\theta^2 + v_\phi^2)/2$ volume averaged over the computational
  domain inside the neutrinosphere for models W15-4 (solid), L15-3
  (dashed), and N20-2 (dashed-dotted), respectively. }
\label{fig:ekin_vs_t}
\end{figure}
%

%
\begin{figure}
\centering \resizebox{1.0\hsize}{!}{\includegraphics*{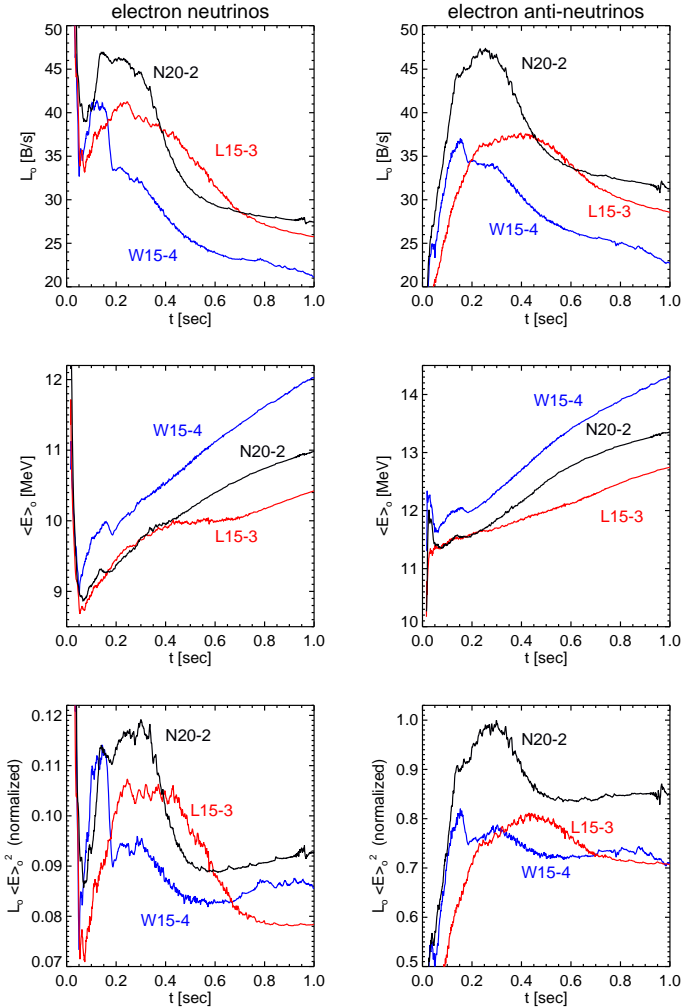}}
\caption{Observable luminosity $L_o$ (top row), observable mean energy
  $\langle E \rangle_o$ (middle row), and normalized quantity $L_o
  \langle E \rangle_o^2$ (bottom row) of electron neutrinos (left
  column) and electron anti-neutrinos (right column) as a function of
  time for three of our models. Although we only present the results
  for one particular observer direction here, the global behavior and
  characteristics are very similar for all viewing directions.}
\label{fig:lobs_vs_t}
\end{figure}
%

The reason for the fluctuation behavior of the neutrino emission
during the vigorous pre-explosion and post-explosion accretion phases
has been discussed, but what causes the spatial and temporal
variations during the post-accretion phase? Because the explosion is
well on its way at this time, neither post-shock convection nor the
SASI nor accretion can be responsible. Hence, there only remains
non-radial gas flow in the outer layers of the proto-neutron star.
Ledoux convection in the proto-neutron star thus may become visible
eventually, \ie its presence in the inner parts of the computational
domain may become dominant in observable signals. This happens in
models W15-4 and N20-2, where the level of the non-radial specific
kinetic energy $(v_\theta^2 + v_\phi^2)/2$, volume-averaged over the
computational domain, inside the neutrinosphere shows a steep rise at
$\sim\,$0.8\,s and $\sim\,$0.9\,s, respectively
(Fig.\,\ref{fig:ekin_vs_t}). These non-radial flows that develop in
models W15-4 and N20-2 at late times also become manifest in all
discussed quantities: $\Lambda_\mathrm{max}(\Omega,t)$,
$\Lambda_\mathrm{min}(\Omega,t)$, $C^\mathrm{max}_l$, the dominant low
$l$-modes ($2 \la l \la 4$), and relative angular flux-density
variations.  In contrast, no such effect is present in model L15-3
(see Fig.\,\ref{fig:ekin_vs_t}), where we find a steady decrease of
$\Lambda_\mathrm{max}(\Omega,t) - \Lambda_\mathrm{min}(\Omega,t)$,
higher $l$-modes ($l \ga 10$), smaller $C^\mathrm{max}_l$, and lower
flux-density variation amplitudes than in models W15-4 and N20-2 (see
Figs.\,\ref{fig:rsh+lnu_vs_t}, \ref{fig:rsh+lnu_vs_t_a}, and
\ref{fig:cl_spectrogram}).

Simulations with fully self-consistent treatment of the PNS interior
show the presence of convection inside the PNS, \ie below the
neutrinosphere \citep[see][]{Keil_etal96, Buras_etal06,
  Dessart_etal06} more or less from the early post-bounce phase on.
With the use of our inner radial grid boundary excising the inner
parts of the PNS, and imposing neutrino luminosities at this boundary,
convective activity is triggered only when the neutrino energy (or
lepton number) inflow into the layers close to the grid boundary is
faster than neutrino transport can carry away this energy (or lepton
number). Then convectively unstable gradients develop and convective
flows begin to carry the energy and lepton-number outward. Whether
this happens or not depends on the boundary luminosities as well as on
the location of the grid boundary within the density and temperature
profiles of the PNS layers below the neutrinosphere. That location
determines the efficiency of the neutrino transport and varies with
the stellar progenitor, whose mass-infall rate decides how much mass
accumulates in the near-surface layers of the PNS outside the inner
grid boundary. The relative strength of the artificially imposed
inflow of neutrino energy and lepton number compared to the efficiency
of the neutrino transport on the grid, both sensitive to the location
and contraction of the grid boundary on the one hand and the chosen
values of the boundary luminosities on the other, therefore decides
about when, where, and how strongly convective activity develops below
the neutrinosphere.

Because the position of and the conditions imposed at the inner
boundary can thus influence the neutrino emission properties, in
particular during the post-accretion phase, our respective model
predictions must be considered with care. While they do not allow us
to make any definite statements concerning the neutrino signal of a
particular progenitor model because of the neglected treatment of the
inner parts of the proto-neutron star, the models nevertheless show
that convective flows below the neutrinosphere are likely to imprint
themselves on the neutrino emission, and hence also on the GW signal
of core-collapse supernovae.  A measurement of these signals may
actually provide some insight into the conditions inside proto-neutron
stars.

Because the neutrino energy flux density varies in our models both
with latitude and longitude, the observable neutrino luminosity
$L_o(t)$ is obtained by an integral over the hemisphere visible to an
observer (Eq.\,\ref{eq:app-lum_code}).  In Fig.\,\ref{fig:lobs_vs_t}
we show the observable electron neutrino and electron anti-neutrino
luminosities for one chosen viewing direction for the three models
W15-4, L15-3, and N20-2, respectively. The results for other
directions look very similar with all characteristic features being
independent of the observer position.  We provide these quantities in
addition to the total neutrino energy loss rate (Eqs.\,\ref{eq:lamt}
and \ref{eq:eloss}; Figs.\,\ref{fig:rsh+lnu_vs_t} and
\ref{fig:rsh+lnu_vs_t_a}), because their temporal evolutions are the
ones expected to be measurable in the IceCube and Super-Kamiokande
detectors.  These detectors (mainly for ${\bar \nu_e}$) will be
sensitive to a combination of the observable neutrino luminosity $L_o$
and the observable mean neutrino energy $\langle E \rangle_o$. Thus,
we also provide in Fig.\,\ref{fig:lobs_vs_t} the time evolution of the
observable mean neutrino energy and of the combination $L_o \langle E
\rangle_o^2$, which (roughly) enters the IceCube detection rate of
Cherenkov photons originating from the dominant inverse beta decay
reaction ${\bar \nu_e} + p \rightarrow n + e^+$ \citep{Lund_etal10}.
\footnote{Note that our transport approximation only provides
  luminosities and mean energies, but not the higher moments of the
  energy spectrum (see Sect.\,\ref{subsec:nuform}). }
Again one can recognize the different evolution stages, and in
particular the post-shock convection and SASI phase, during which the
quantity $L_o \langle E \rangle_o^2$ exhibits rapid low-amplitude
variations for all three models. The level of the variations is a few
percent (Fig.\,\ref{fig:lobs_vs_t}), which is considerably lower than
that of the angular fluctuation amplitudes of the flux density, which
reaches almost 100\% for the total neutrino flux density
(Fig.\,\ref{fig:W15-4_lnue}) and almost 200\% for the electron
neutrino and electron antineutrino flux densities
(Fig.\,\ref{fig:cl_spectrogram}, lower panels). However, because the
flux density variations are caused by a few individual hot spots
covering only angular areas of size $\sim (\pi/9)^2$, the observable
fluctuations (of $L_0$ and $\langle E \rangle_0$) are lower by a
factor of roughly $(\pi/9)^2 / (2\pi) \sim 1/50$.  Some of this
activity is also present at late times in the two models W15-4 and
N20-2, where Ledoux convection develops in the simulated outer parts
of the proto-neutron star (see discussion above).

From the results presented above we conclude that the signals carry
clear information about the postshock hydrodynamic activity, and about
the duration and decay of the accretion period. Composition-shell
interfaces present in the progenitor star can also leave an
imprint. In model W15-4 the transition from the Fe-core to the
Si-shell manifests itself in fast drops of the luminosities of $\nu_e$
and $\bar\nu_e$ at $\sim150\,$msec, when the mass accretion rate
decreases steeply at the time the interface between the Fe-core and
the Si-shell of the 15\,$M_\odot$ progenitor falls through the shock.

\section{Gravitational wave signature}
\label{sec:gw-signal}

Non-radial mass motions caused by gravity waves in the near-surface
layers of the PNS, which are caused by the SASI and convection in the
post-shock region as well as by convective activity inside the
proto-neutron star \citep{Murphy_etal09, Marek_etal09} (driven by
Ledoux unstable lepton or entropy gradients) result in a
time-dependent, aspherical density stratification that produces
gravitational radiation. The anisotropic emission of neutrinos
associated with the non-radial mass flow (see
Sect.\,\ref{sec:neutrino-signal}) contributes to the gravitational
wave signal, too.  We computed and analyzed the signature of this
gravitational radiation for the 3D models discussed in
Sect.\,\ref{subsec:models}.

\subsection{Formalism}
\label{subsec:gw-form}

\subsubsection{Non-radial mass flow}
\label{subsec:gw-form-flow}

If a source is of genuine three-dimensional nature, as it is the case
for our models, it is common to express the gravitational quadrupole
radiation tensor, $\vec h^{\rm TT}$, in the transverse traceless gauge
in the following tensorial form
\begin{equation}
   \vec h^{\rm TT} (\vec X,t) = \frac{1}{R}\, 
                   (A_+ \vec e_+  +  A_{\times} \vec e_{\times}) 
\label{eq:htt-general}
\end{equation}
\citep[see, \eg][]{Misner_etal73}.  $R$ denotes the distance between
the observer and the source, and the unit linear-polarization tensors
are given by
\begin{eqnarray}
   \vec e_+ &=& \vec e_{\theta} \otimes \vec e_{\theta} - 
                \vec e_{\phi}   \otimes \vec e_{\phi}\, ,   
\label{eq:pol-tensor-p}\\
   \vec e_{\times} &=& \vec e_{\theta} \otimes \vec e_{\phi}   + 
                \vec e_{\phi}   \otimes \vec e_{\theta}\, ,  
\label{eq:polt-tensor-x}
\end{eqnarray}
with $\vec e_{\theta}$ and $\vec e_{\phi}$ being the unit polarization
vectors in $\theta$ and $\phi$-direction of a spherical coordinate
system, and $\otimes$ denoting the tensor product.  

The wave amplitudes $A_+$ and $A_{\times}$ represent the only two
independent modes of polarization in the TT gauge
\citep{Misner_etal73}. In the slow-motion limit, they are obtained
from linear combinations of the second time derivatives (evaluated at
retarded time, and denoted by a double dot accent) of the components
of the transverse traceless mass quadrupole tensor
\citep{Misner_etal73}
\begin{eqnarray}
  A_+        &=&  \ddot Q_{\theta\theta} - \ddot Q_{\phi\phi} \ ,  
\label{eq:aplus-tf} \\
  A_{\times}  &=&  2 \ddot Q_{\theta\phi} \ .               
\label{eq:acros-tf}
\end{eqnarray}
We computed the latter using a post-Newtonian approach whereby the
numerically troublesome second-order time derivatives of the mass
quadrupole tensor components are transformed into much better
tractable spatial derivatives.  Following \citet{NakamuraOohara89} and
\citet{Blanchet_etal90}, the second-order time derivatives read in a
Cartesian orthonormal basis (the spatial indices $i$ and $j$ run from
1 to 3)

\begin{equation}
  \ddot Q_{ij} =  \frac{G}{c^4} \int {\rm d}^3 x \ \rho \
                  \left(  
                        2 v_i v_j - x_i \ \partial_j \Phi_{\rm eff} 
                                  - x_j \ \partial_i \Phi_{\rm eff}
                  \right) \ , 
\label{eq:qddot}
\end{equation}
where $G$ is Newton's gravitational constant, $c$ the speed of light
in vacuum, $\Phi_{\rm eff}$ the effective Newtonian gravitational
potential including the general relativistic ``case A" correction of
the monopole term according to \citet{Marek_etal06}, $\rho$ the
mass-density, $v_ i$ the Cartesian velocity components, and
$\partial_i$ the partial derivative with respect to the coordinate
$x^i$ of a Cartesian basis.

We note that the integrand in Eq.\,(\ref{eq:qddot}) has compact
support and is known to the (2nd order) accuracy level of the
numerical scheme employed in the hydrodynamics code.  It can easily be
shown that evaluating the integral of Eq.~(\ref{eq:qddot}) by an
integration scheme (of at least 2nd order) is by one order of accuracy
superior to twice applying numerical time-differentiation methods to
quadrupole data given at discrete points of time \citep{FinnEvans90,
  Moenchmeyer_etal91}.

Exploiting the coordinate transformation between the orthonormal
Cartesian basis $x^i$ and the orthonormal basis in spherical
coordinates $\hat{x}^i$ (with $\hat{x}^i \in \lbrack r, \theta, \phi
\rbrack$), the wave amplitudes $A_+$ and $A_{\times}$
(Eqs.\,(\ref{eq:aplus-tf}) \& (\ref{eq:acros-tf})) are obtained from
the following second time derivatives of the spherical components of
the mass quadrupole tensor \citep{Oohara_etal97, Scheidegger_etal08}
\begin{eqnarray}
  I^{TT}_{\theta\theta} &=& \left(  I^{TT}_{xx} \cos^2\phi 
                              + I^{TT}_{yy} \sin^2\phi 
                              + 2\, I^{TT}_{zz} \sin\phi \cos\phi
                         \right) \cos^2\theta
\nonumber\\       
        &\phantom{= }&   + I^{TT}_{yy} \sin^2\theta                          
                         -2\, \left( I^{TT}_{xz} \cos\phi +
                                     I^{TT}_{yz} \sin\phi \right) 
                         \sin\theta \cos\theta \, ,
\\
  I^{TT}_{\phi\phi}   &=& I^{TT}_{xx} \sin^2\phi + I^{TT}_{yy} \cos^2\phi
                       -2\, I^{TT}_{xy} \sin\phi \cos\phi \, ,
\\
  I^{TT}_{\theta \phi}  &=& \left( I^{TT}_{yy} - I^{TT}_{xx} \right) 
                         \cos\theta \sin\phi \cos\phi
                         + I^{TT}_{xy} \cos\theta 
                         \left( \cos^2\phi \right. 
\nonumber\\           
          &\phantom{=}& \left. - \sin^2\phi \right)
                        + I^{TT}_{xz} \sin\theta \sin\phi  
                        - I^{TT}_{yz} \sin\theta \cos\phi \, ,
\end{eqnarray}
where we used the abbreviation
\begin{equation}
  I^{TT}_{ij} \equiv \ddot Q^{TT}_{ij}\, .
\label{eq:ieqqddot}
\end{equation}
Choosing $\phi=0$ one obtains the polarization modes (see, \eg
\cite{Misner_etal73})
\begin{eqnarray}
A_+       &=&  I^{TT}_{xx} - I^{TT}_{yy} \ ,  \label{eq:aplus_p} \\
A_{\times}  &=& 2 I^{TT}_{xy} \ ,             \label{eq:acros_p}
\end{eqnarray}
for $\theta=0$, and
\begin{eqnarray}
A_+       &=&  I^{TT}_{zz} - I^{TT}_{yy} \ ,   \label{eq:aplus_e} \\
A_{\times}  &=& -2 I^{TT}_{yz}  \, ,           \label{eq:acros_e} 
\end{eqnarray}
for $\theta = \pi/2$, respectively. These expressions were already
discussed in earlier investigations concerned with the evaluation of
the gravitational wave signature of 3D core-collapse supernova models
\citep{MuellerJanka97, Fryer_etal04, Scheidegger_etal08,
  Scheidegger_etal10}.

The total energy radiated in the form of gravitational waves due to
nonspherical mass flow is given in the quadrupole approximation by
\citep[see, \eg][]{Misner_etal73}
\begin{eqnarray}
  E_{\rm M} &=& \frac{c^3}{5G} \int_0^\infty \sum_{ij} \left\lbrack
          \frac{{\rm d}}{{\rm d}t} \left( I^{TT}_{ij} - \frac{1}{3} 
                                          \delta_{ij} \sum_l I^{TT}_{ll} \right)
         \right\rbrack^2  {\rm d}t
\nonumber \\ 
     &=& \frac{c^3}{15G} \int_0^\infty {\rm d}t \left\lbrack
         (\dot{I}^{\ TT}_{xx} - \dot{I}^{\ TT}_{yy})^2 + 
         (\dot{I}^{\ TT}_{xx} - \dot{I}^{\ TT}_{zz})^2   \right. 
\label{eq:egw} \\
 &\phantom{=}& \phantom{ \frac{c^3}{15G}}  \left. 
         + (\dot{I}^{\ TT}_{yy} - \dot{I}^{\ TT}_{zz})^2  
         + 6\, \left( (\dot{I}^{\ TT}_{xy})^2 + (\dot{I}^{\ TT}_{xz})^2 + 
                      (\dot{I}^{\ TT}_{yz})^2 \right)
         \right\rbrack  \, ,
\nonumber 
\end{eqnarray}
with $\dot{I}^{\ TT}_{ij} \equiv \partial I^{\ TT}_{ij}/ \partial t$,
and the corresponding GW spectral energy density is given by (where
$\nu$ denotes the frequency)
\begin{eqnarray}
  \frac{{\rm d}E_{\rm M}}{{\rm d} \nu} &=& \frac{2c^3}{15G} (2\pi \nu)^2
         \left\lbrack
         \left|\widetilde{I}^{\ TT}_{xx} - \widetilde{I}^{\ TT}_{yy}\right|^2 + 
         \left|\widetilde{I}^{\ TT}_{xx} - \widetilde{I}^{\ TT}_{zz}\right|^2 \right. 
\label{eq:egwnu}  \\
 &\phantom{=}& \phantom{ \frac{2}{15} } \left. 
         + \left|\widetilde{I}^{\ TT}_{yy} - \widetilde{I}^{\ TT}_{zz}\right|^2  
         + 6\, \left( \left|\widetilde{I}^{\ TT}_{xy}\right|^2 + 
                      \left|\widetilde{I}^{\ TT}_{xz}\right|^2 + 
                      \left|\widetilde{I}^{\ TT}_{yz}\right|^2  \right)
         \right\rbrack  \, ,
\nonumber
\end{eqnarray}
where 
\begin{equation}
  \widetilde{I}^{\ TT}_{ij}(\nu) = 
     \int_{-\infty}^\infty I^{TT}_{ij}(t)\, e^{-2\pi i \nu t}\, {\rm d}t
\end{equation}
is the Fourier transform of $I^{TT}_{ij} (t)$.

\subsubsection{Anisotropic neutrino emission}
\label{subsec:gw-form-nu}

To determine the gravitational wave signal associated with the
anisotropic emission of neutrinos, we follow \citet{MuellerJanka97}
and use Eq.\,(16) of \citet{Epstein78} in the limit of a distant
source, $R \to \infty$, together with the approximation that the
gravitational wave signal measured by an observer at time $t$ is
caused only by radiation emitted at time $t' = t - R/c$. Hence, we
take $t-t' = {\rm const} = R/c$, \ie we assume that only the neutrino
pulse itself causes a gravitational wave signal, whereas memory
effects, which prevail after the pulse has passed the observer, are
disregarded.
%

With these simplifications, the dimensionless gravitational wave
amplitudes of the two polarization modes are given in the
transverse-traceless gauge for an observer located at a distance $R$
along the $z$-axis of the observer frame by \citet{MuellerJanka97}
\begin{equation}
h_{+}(t)  = \frac{2G}{c^4R} \int_0^t {\rm d}t' 
                   \int_{4\pi} {\rm d}\Omega' \,
                    ( 1 + \cos\theta ) \cos 2\phi \,
                   \frac{ {\rm d}\Lambda }{ {\rm d}\Omega'} ({\vec \Omega'},t')
\label{eq:hplus_nu}
\end{equation}
and
\begin{equation}
h_{\times}(t) = \frac{2G}{c^4R} \int_0^t {\rm d}t' 
                   \int_{4\pi} {\rm d}\Omega' \,
                   ( 1 + \cos\theta ) \sin 2\phi \,
                   \frac{ {\rm d}\Lambda }{ {\rm d}\Omega'} ({\vec \Omega'},t')\ ,
\label{eq:hcros_nu}
\end{equation}
respectively. Here ${\rm d} \Lambda (\Omega', t')/{\rm d}\Omega'$ is
given by Eq.\,(\ref{eq:dlamdom}) and denotes the total neutrino energy
radiated at time $t'$ per unit of time into a solid angle ${\rm
  d}\Omega'$ in direction $(\theta', \phi')$. Except for
position-dependent factors the gravitational wave amplitudes are
simply a function of this quantity provided by the ray-by-ray
transport approximation (note that in M\"uller \& Janka (1997) we used
the symbol $L_{\nu}$ instead of $\Lambda$).
 
The angular integration, ${\rm d}\Omega' = -{\rm d} (\cos\theta')\,
{\rm d}\phi'$, in Eqs.\,(\ref{eq:hplus_nu}) and (\ref{eq:hcros_nu})
extends over all angles $\theta'$ and $\phi'$ in the coordinate frame
of the source $(x',y',z')$ that we identify with the (arbitrarily
chosen) spherical polar coordinate frame to which the hydrodynamic
results were mapped from the Yin-Yang grid employed in the
simulations.  For the evaluation of the polarisation modes we used the
(asymptotic) values of ${\rm d} \Lambda(\Omega, t)/{\rm d}\Omega$
extracted at a radius of 500\,km from our 3D models.

%
\begin{figure}[!]
\centering
\resizebox{1.0\hsize}{!}{\includegraphics{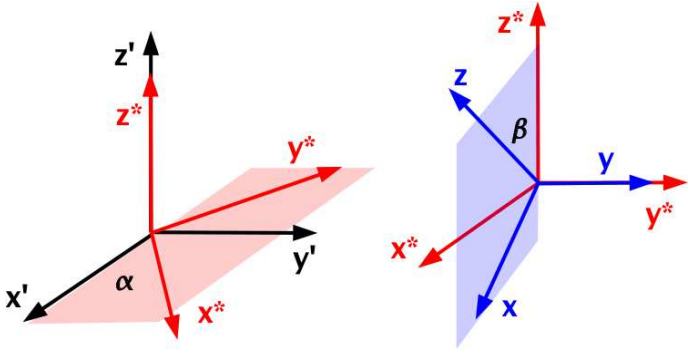}}
\caption{Relation between the source coordinate system $(x^\prime,
  y^\prime, z^\prime)$ and the observer coordinate system $(x, y,
  z)$. Changing from the observer system to the source system involves
  a rotation by an angle $\alpha$ about the $z'$-axis to an
  intermediate coordinate system $(x^\ast, y^\ast, z^\ast)$, followed
  by a rotation by an angle $\beta$ about the $y^\ast$-axis (which
  thus is also the $y$-axis). }
\label{fig:obs_src}
\end{figure}
%

The angles $\theta$ and $\phi$ in Eqs.\,(\ref{eq:hplus_nu}) and
(\ref{eq:hcros_nu}) are measured in the observer frame $(x,y,z)$,
while the neutrino luminosity is measured in the source frame
$(x',y',z')$. To allow for an arbitrary orientation of the observer
relative to the source, we introduce two viewing angles $\alpha \in
[-\pi,+\pi]$ and $\beta \in [0,\pi]$ (see Fig.\,\ref{fig:obs_src}).
The coordinates measured in the observer frame are then related to the
coordinates in the source frame by the following coordinate
transformations
\begin{eqnarray}
 x^\ast &=&  x' \cos\alpha + y' \sin\alpha\, , \\
 y^\ast &=& -x' \sin\alpha + y' \cos\alpha\, , \\
 z^\ast &=&  z'\, ,
\end{eqnarray}
and
\begin{eqnarray}
 x &=& x^\ast \cos\beta - z^\ast \sin\beta\, , \\
 y &=& y^\ast \, , \\
 z &=& x^\ast \sin\beta + z^\ast \cos\beta \, .
\end{eqnarray}
With these coordinate transformations and the relations 
\begin{eqnarray}
 x &=& r \sin\theta \cos\phi
\label{eq:csx}\, , \\
 y &=& r \sin\theta \sin\phi\, , \\
 z &=& r \cos\theta\, 
\label{eq:csz}
\end{eqnarray}
between Cartesian coordinates $(x,y,z)$ and spherical polar
coordinates $(r,\theta,\phi)$, we obtain
\begin{eqnarray}
 \sin\theta \cos\phi &=&  (\cos\phi' \cos\alpha + \sin\phi' \sin\alpha)  
                          \sin\theta' \cos\beta\, - 
\nonumber\\  
                     & &  \cos\theta' \sin\beta\, ,           
\\               
 \sin\theta \sin\phi &=&  (\sin\phi' \cos\alpha -
                           \cos\phi' \sin\alpha) \sin\theta'\, ,
 \\
 \cos\theta          &=&  (\cos\phi' \cos\alpha + \sin\phi' \sin\alpha) 
                           \sin\theta' \sin\beta\, + 
\nonumber\\
                     & &  \cos\theta' \cos\beta \, .
\label{eq:angle_trans}
\end{eqnarray}
These expressions relate the angular coordinates in the observer frame
$(\theta, \phi)$ to those in the source frame $(\theta',\phi')$. For
the special case $\alpha =0$ they were already presented by
\citet{Kotake_etal09a}.  Using Eq.\,(\ref{eq:angle_trans}) and the
equalities
\begin{eqnarray}
 \sin 2\phi &=& \frac{2xy}{x^2+y^2}\, , 
\label{eq:sintwophi}\\
 \cos 2\phi &=& \frac{x^2-y^2}{x^2+y^2}
\label{eq:costwophi}
\end{eqnarray}
derived from Eqs.\,(\ref{eq:csx}) to (\ref{eq:csz}), the two
polarization modes (Eqs.\,\ref{eq:hplus_nu} and \ref{eq:hcros_nu}) are
given by
\begin{equation}
 h_{\rm S} (t,\alpha,\beta) = \frac{2G}{c^4R} \int_0^t {\rm d}t' 
                             \Lambda(t')\, \alpha_{\rm S}(t',\alpha,\beta)\,, 
\label{eq:hnue}
\end{equation}
where ${\rm S} \in (+,\times)$ and $\Lambda(t)$ is the angular
integral of the neutrino energy radiated at time $t$ per unit of time
given in Eqs.\,(\ref{eq:lamt}) and (\ref{eq:eloss}).
\begin{equation}
  \alpha_{\mathrm S}(t,\alpha,\beta) = \frac{1}{\Lambda(t)} 
         \int_{4\pi}{\rm d}\Omega'\, W_{\rm S}(\Omega',\alpha,\beta)\,
         \frac{ {\rm d}\Lambda }{ {\rm d}\Omega'} ({\vec \Omega'}, t) \, ,
\label{eq:aniso}
\end{equation}
are anisotropy parameters, which provide a quantitative measure of the
time-dependent anisotropy of the emission in both polarization
modes.  Note that the evaluation of the anisotropy parameter
$\alpha(t)$ defined in Eq.\,(29) of \citet{MuellerJanka97}, which
should not be confused with the observer angle $\alpha$ introduced in
Fig.\,\ref{fig:obs_src}, does neither involve a dependence on observer
angles $(\alpha,\beta)$ nor on the polarization mode.

The angular weight functions appearing in the above expression for the
anisotropy parameters are given by
\begin{equation}
 W_{\rm S} (\theta',\phi',\alpha,\beta) = 
 \frac{D_{\rm S} (\theta',\phi',\alpha,\beta)}{N (\theta',\phi',\alpha,\beta)} \ ,
\end{equation}
where 
\begin{eqnarray}
 D_+ &=&  \left[ 1 + (\cos\phi' \cos\alpha + \sin\phi' \sin\alpha) 
                 \sin\theta' \sin\beta + \cos\theta' \right.
\nonumber\\
     & & \left.  \cos\beta \right] \, \left\lbrace 
         \left[ (\cos\phi' \cos\alpha + \sin\phi' \sin\alpha) 
                \sin\theta' \cos\beta\, - \right. \right.
\nonumber\\
     & &  \left.\left. \cos\theta' \sin\beta \right]^2 -
           \sin^2\theta' \left( \sin\phi' \cos\alpha -
                                \cos\phi' \sin\alpha \right)^2
                                \right\rbrace\, ,
\\
 D_\times &=& \left[ 1 + ( \cos\phi' \cos\alpha + \sin\phi' \sin\alpha) 
                  \sin\theta' \sin\beta + \cos\theta' \right.
\nonumber\\
     & &   \left.  \cos\beta \right] \, 
           2 \left[ (\cos\phi' \cos\alpha + \sin\phi' \sin\alpha) 
                  \sin\theta' \cos\beta\, - \right.
\nonumber\\
     & &  \left.  \cos\theta' \sin\beta \right]\,  
          \sin\theta'  \left( \sin\phi' \cos\alpha - 
                              \cos\phi' \sin\alpha \right)\, ,
\\
 N   &=&  \left[ (\cos\phi' \cos\alpha + 
                  \sin\phi' \sin\alpha )  
                  \sin\theta' \cos\beta\, - \, \cos\theta'\right.
\nonumber\\
     & &  \left. \sin\beta \right]^2 +
           \sin^2\theta' \left( \sin\phi' \cos\alpha - 
                                \cos\phi' \sin\alpha \right)^2 \, . 
\end{eqnarray}
Choosing $\alpha = 0$ and $\beta = \pi/2$ the observer is located in
the equatorial plane of the source (\ie perpendicular to the source's
$z'$-axis) at the azimuthal position $\phi' = 0$.  In that case one
obtains simpler expressions for the angular functions \citep[see
  also][]{Kotake_etal09a}
\begin{eqnarray}
 \left. W_+     \right|_e &=& (\cos^2\theta' - \sin^2\theta' \sin^2\phi) \,
                               \frac{1 + \sin\theta' \cos\phi'}{
                               \cos^2\theta' + \sin^2\theta' 
                               \sin^2\phi'}\, ,
\\
 \left. W_\times \right|_e &=& -2 \cos\theta' \sin\theta' \sin\phi'\,
                                 \frac{1 + \sin\theta' \cos\phi'}{
                                 \cos^2\theta' + \sin^2\theta'
                                 \sin^2\phi'}\, .
\end{eqnarray}
Note that for axisymmetric sources $h_{\times} = 0$.

In general, the total energy $E_{\rm GW}(t)$ radiated to infinity by a
source in form of gravitational waves until time $t$ is given by (see,
\eg \cite{Misner_etal73}; Greek indices run from 0 to 3, and repeated
indices are summed over)
\begin{equation}
  E_{\rm GW}(t) = \int_0^t {\rm d}t' 
                  \int_{S_\infty^2} \tau_{0\nu}\ n^\nu r^2 
                  {\rm d} \Omega \ ,
\label{eq:enue_1}
\end{equation}
where the angular integration is performed over a two-sphere at
spatial infinity $S_\infty^2$, and $n^\mu = (0,1,0,0)$ is a unit
spacelike vector in polar coordinates $\lbrace ct, r, \theta, \phi
\rbrace$ normal to $S_\infty^2$.  Denoting by $\langle \dots \rangle$
an average over several wavelengths, the gravitational-wave
energy-momentum tensor $\tau_{\mu\nu}$ is given in
transverse-traceless gauge by
\begin{equation}
  \tau_{\mu\nu} = \frac{c^5}{32\pi G} \left\langle
                (\partial_\mu h^{\rm TT}_{\rho\sigma})\
                (\partial_\nu h_{\rm TT}^{\rho\sigma}) \right\rangle  \ .
\label{eq:taunue}
\end{equation}
Thus, Eq.\,(\ref{eq:enue_1}) can be rewritten as
\begin{equation}
  E_{\rm GW}(t) = \frac{c^3}{32\pi G} \int_0^t {\rm d}t' 
                \int_{S_\infty^2} r^2 {\rm d} \Omega \left\langle
                (\partial_t h^{\rm TT}_{ik})
                (\partial_r h_{\rm TT}^{ik}) \right\rangle \ ,
\label{eq:enue_2}
\end{equation}
where we have used the facts that $h^{\rm TT}_{0\nu} = 0$, $h^{\rm
  TT}_{i,r} = 0$, and $c\ \partial_r h^{\rm TT}_{ik} = - \partial_t
h^{\rm TT}_{ik}$ for radially outgoing gravitational
radiation. Evaluating the double sum in Eq.\,(\ref{eq:enue_2}) and
using the relations $h^{\rm TT}_{\theta\theta} = - h^{\rm
  TT}_{\phi\phi} = h_+$ and $h^{\rm TT}_{\theta\phi} = h^{\rm
  TT}_{\phi\theta} = h_\times$ \citep[see, \eg][]{Misner_etal73}, we
finally find
\begin{equation}
  E_{\rm GW}(t) = \frac{c^3}{16\pi G} \int_0^t {\rm d}t' 
                \int_{S_\infty^2} r^2 {\rm d} \Omega \left\langle
                (\partial_t h_+)^2 + (\partial_t h_\times)^2
                \right\rangle \ . 
\label{eq:hnue_alpha}
\end{equation}

Inserting the expressions for $h_+$ and $h_\times$ given in
Eq.\,(\ref{eq:hnue}) into Eq.\,(\ref{eq:hnue_alpha}), we obtain for the
energy $E_{\rm N}(t)$ radiated in form of gravitational waves until
time $t$ due to anisotropic neutrino emission
\begin{equation}
  E_{\rm N}(t) = \frac{G}{4\pi c^5}   
                \int_0^t {\rm d}t' \int_{4\pi}{\rm d} \Omega_{\alpha\beta} 
                \left\lbrack  l^2_+     (t',\alpha,\beta) + 
                              l^2_\times (t',\alpha,\beta) \right\rbrack
\label{eq:enue}
\end{equation}
with ${\rm d} \Omega_{\alpha\beta} = \sin\beta\, {\rm d}\beta\, {\rm
  d} \alpha$ and
\begin{equation}
  l_{\rm S} (t,\alpha,\beta) =  \Lambda(t)\, \alpha_{\rm S}(t,\alpha,\beta) \,.
\label{eq:lnue}
\end{equation}
The corresponding spectral energy density is given by
\begin{equation}
  \frac{{\rm d} E_{\rm N}}{{\rm d} \nu} = 
     \frac{G}{2\pi c^5}\ \left| {\tilde l}(\nu) \right|^2 \, ,
\label{eq:de_ndnu}
\end{equation}
where ${\tilde l} (\nu)$ is the Fourier transform of
\begin{equation}
  l(t) = \left\lbrace 
              \int_{4\pi} {\rm d} \Omega_{\alpha\beta}\
                 \left\lbrack  l^2_+     (t,\alpha,\beta) + 
                               l^2_\times (t,\alpha,\beta) 
                 \right\rbrack
         \right\rbrace^{1/2} \, .
\label{eq:lambda}
\end{equation}

For completeness we also provide an expression for the total energy
radiated in form of gravitational waves until time $t$, \ie due to
anisotropic mass flow {\bf and} neutrino emission. It is obtained by
inserting the total GW amplitude, \ie the sum of the amplitudes given
by Eqs.\,(\ref{eq:htt-general}) and (\ref{eq:hnue}) into
Eq.\,(\ref{eq:hnue_alpha}), which leads to
\begin{equation}
  E_{\rm GW} = \frac{c^3}{16\pi G} \int_0^t {\rm d}t' 
              \int_{4\pi}{\rm d} \Omega  
              \left\lbrack
              \left( \frac{2G}{c^4} l_+     +  \partial_t A_+     \right)^2
            + \left( \frac{2G}{c^4} l_\times +  \partial_t A_\times \right)^2
              \right\rbrack \, .
\label{eq:gwetot}
\end{equation}

\subsection{Results}
\label{subsec:gw-results}

%
\begin{figure*}[t]
\centering
\resizebox{0.47\hsize}{!}{\includegraphics*{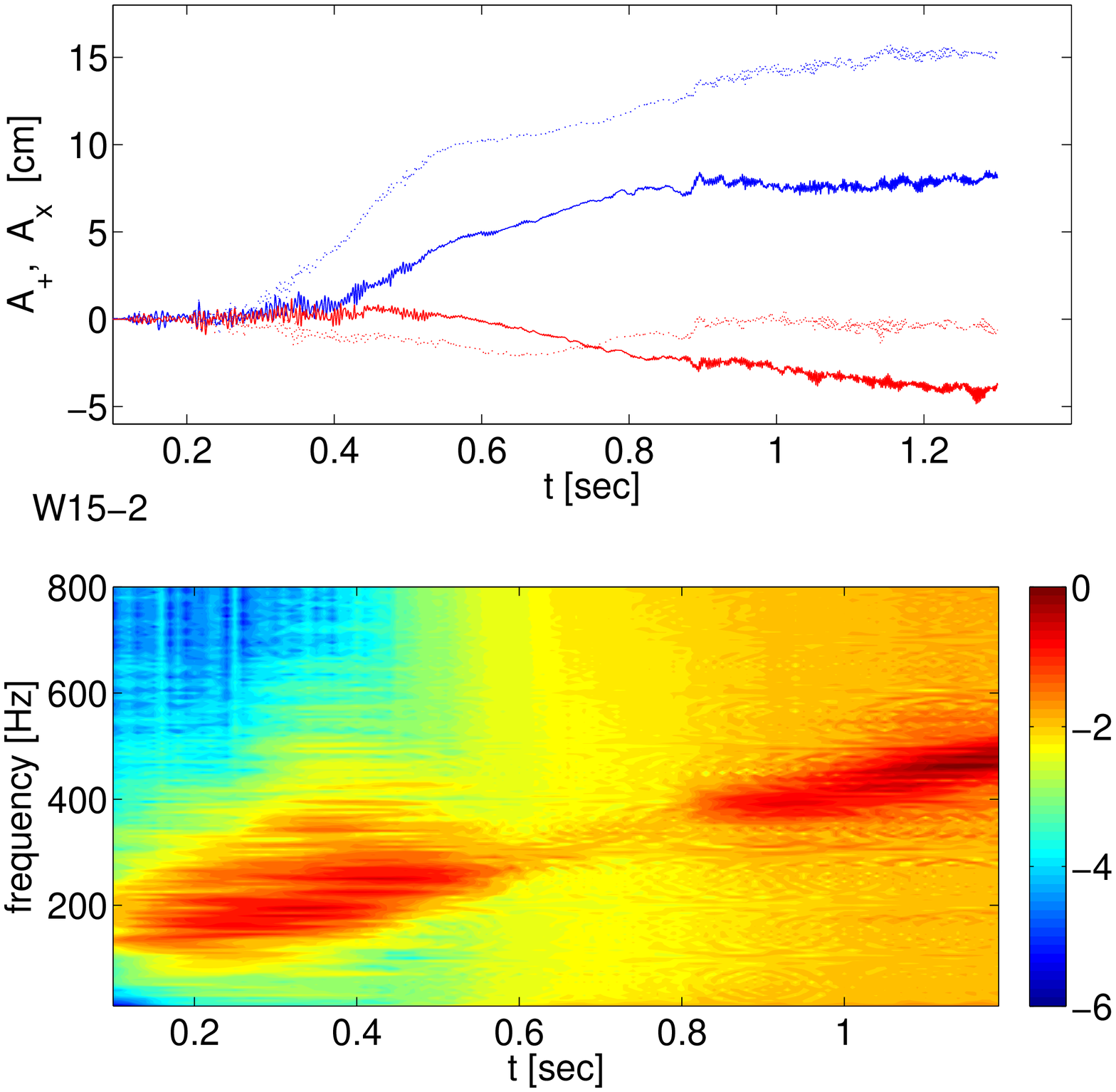}}\hspace{1cm}
\resizebox{0.47\hsize}{!}{\includegraphics*{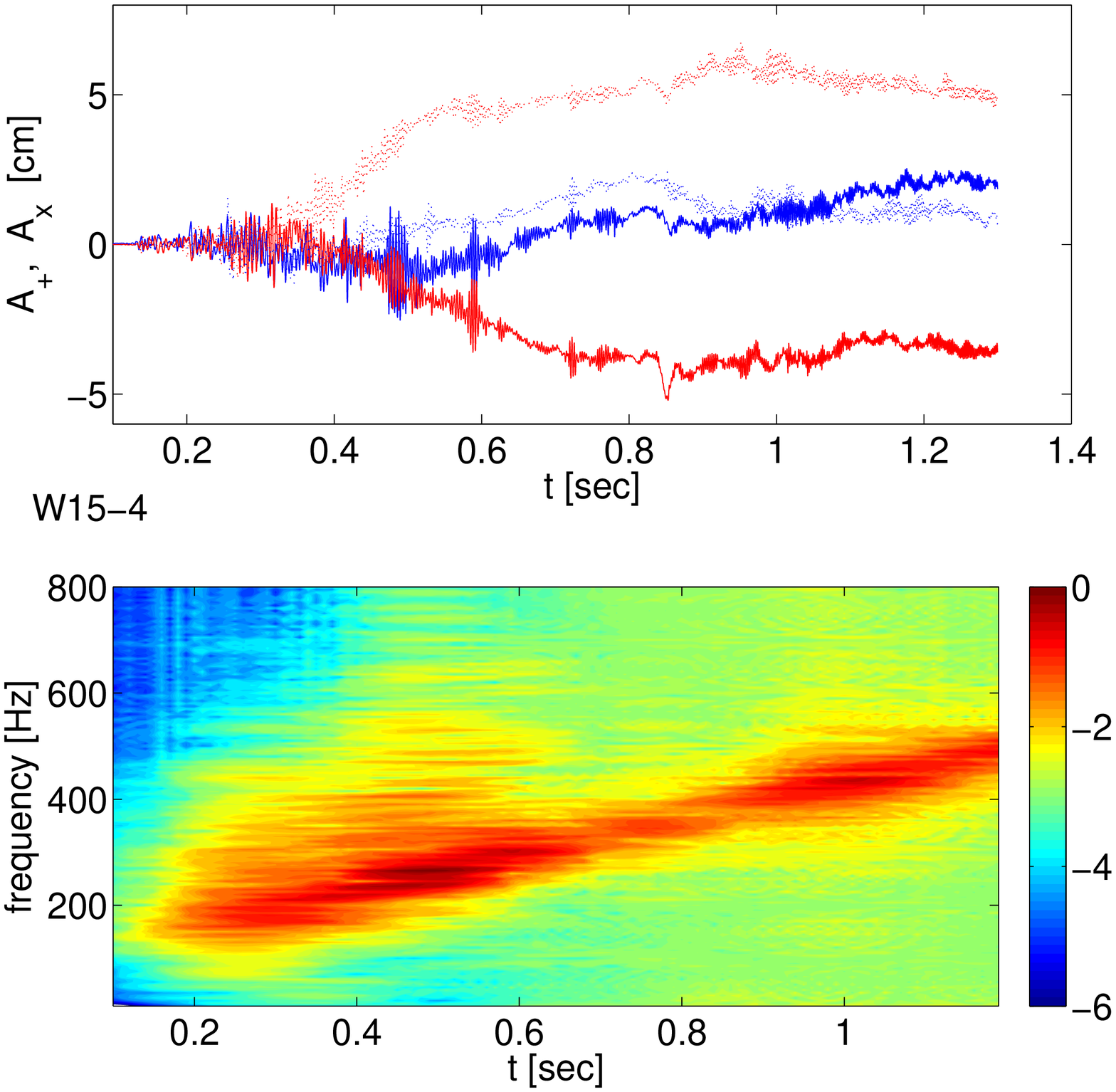}}\\
\vspace{0.5cm}
\resizebox{0.47\hsize}{!}{\includegraphics*{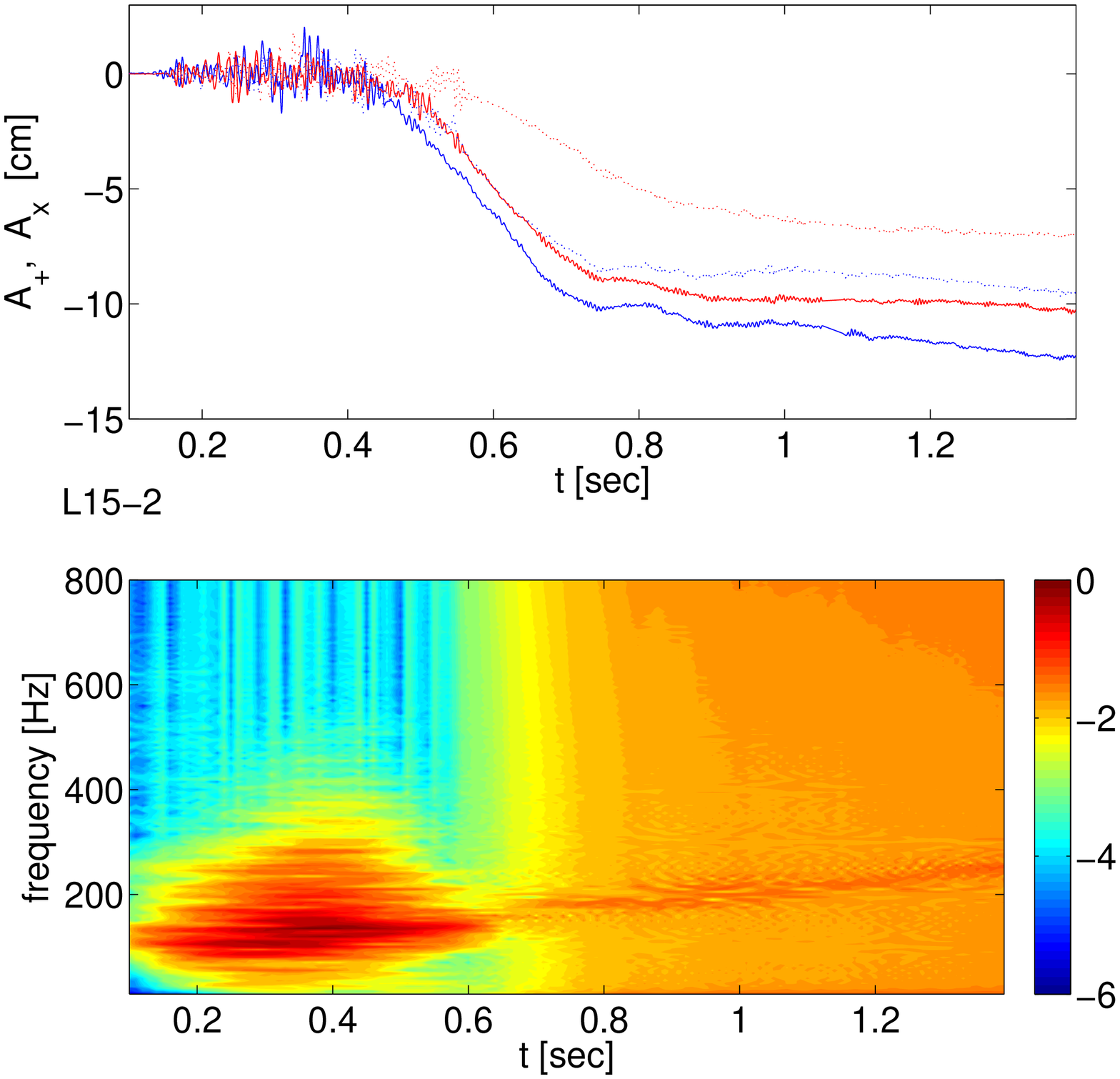}}\hspace{1cm}
\resizebox{0.47\hsize}{!}{\includegraphics*{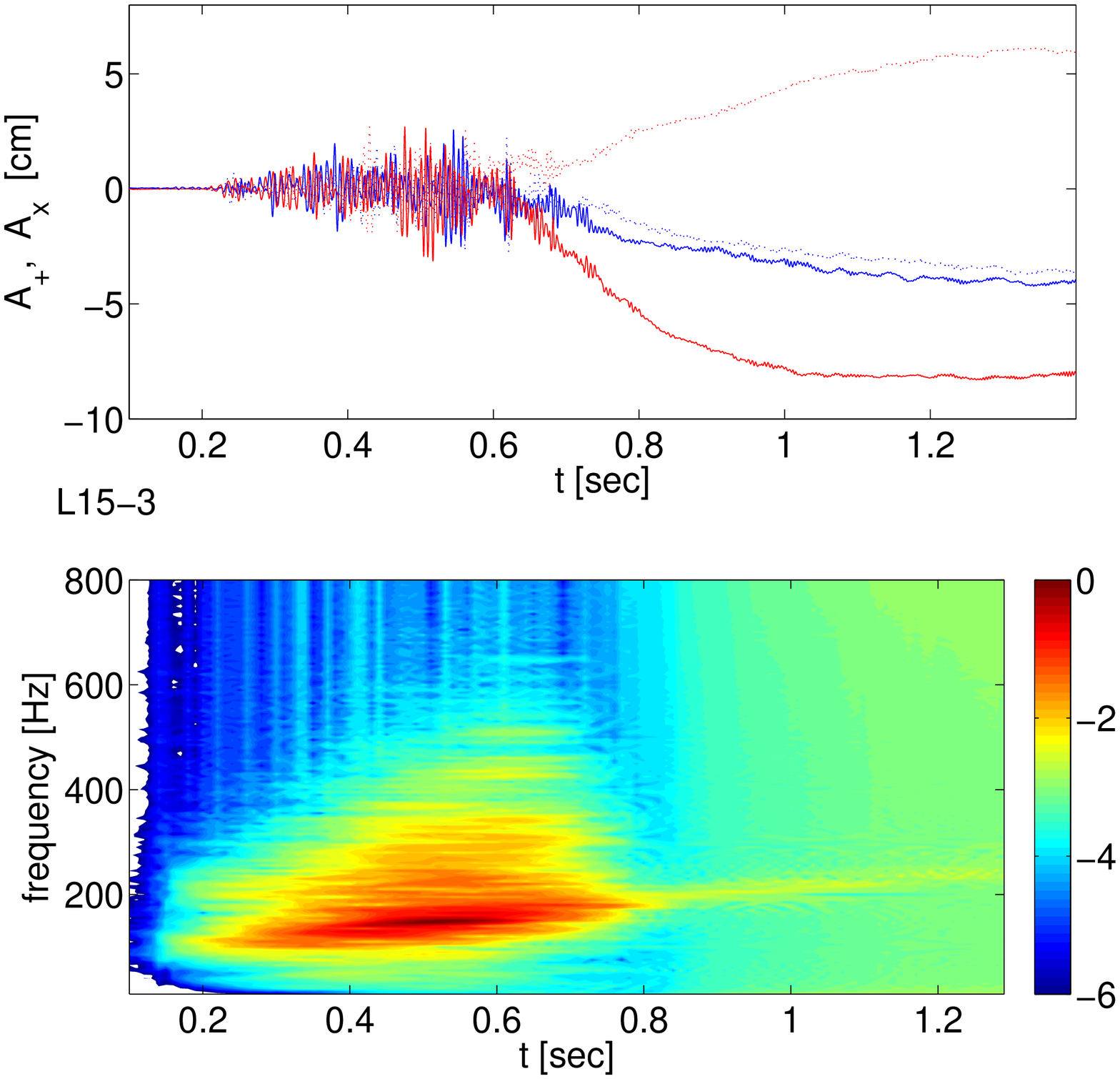}}\\
\caption{The four panels show the gravitational wave amplitudes (top)
  and spectrograms of ${\rm d} E_{\rm M} / {\rm d} \nu$ (bottom;
  normalized to the absolute maximum) arising from non-spherical mass
  flow of models W15-2 (top left), W15-4 (top right), L15-2 (bottom
  left), and L15-3 (bottom right), respectively.  Blue curves give the
  amplitude $A_+$ at the pole (solid) and the equator (dotted), while
  red curves show the other independent mode of polarization
  $A_\times$ from the same directions. }
\label{fig:W15-as}
\end{figure*}
%

%
\begin{figure*}
\centering
\resizebox{0.33\hsize}{!}{\includegraphics*{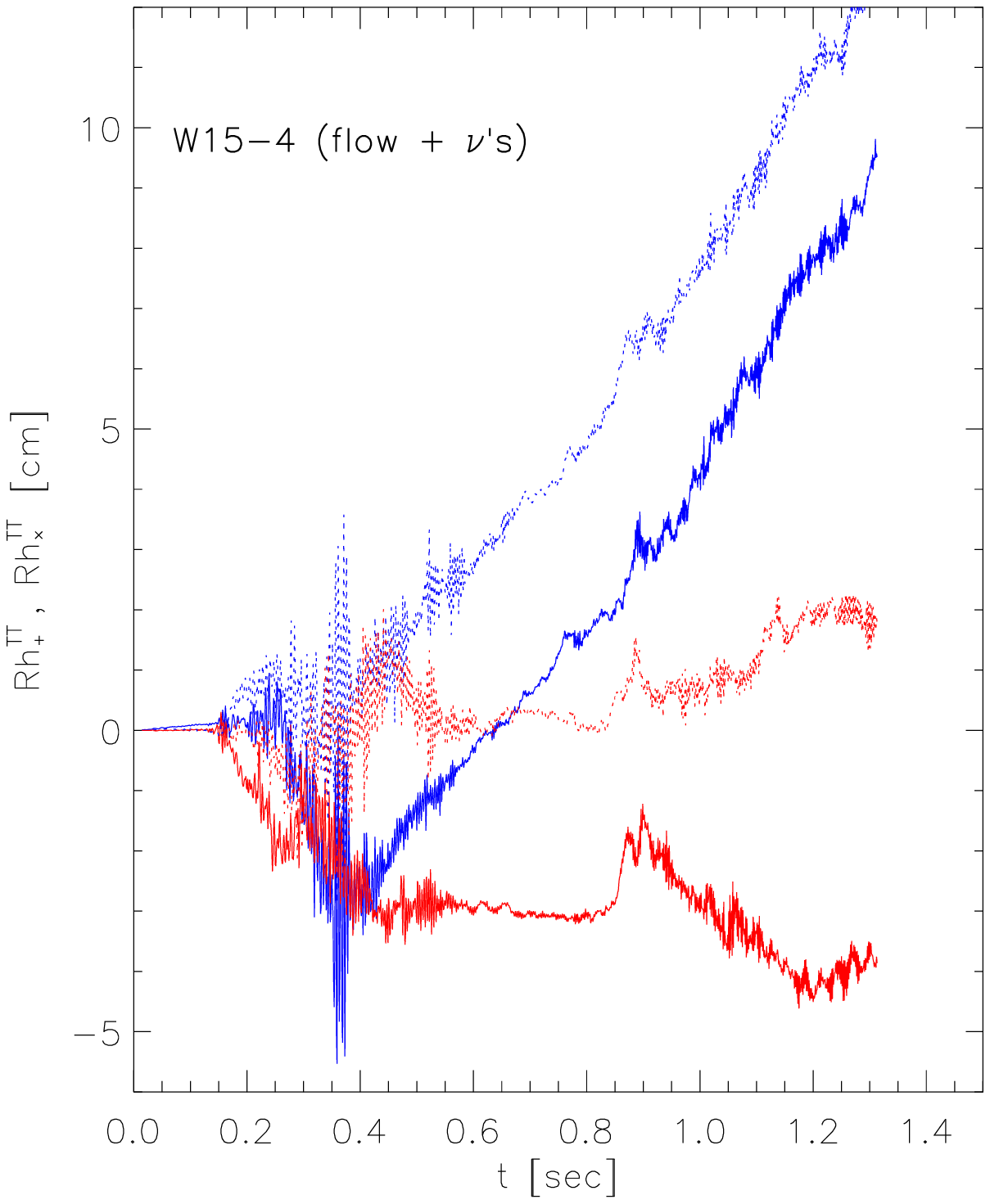}}
\resizebox{0.33\hsize}{!}{\includegraphics*{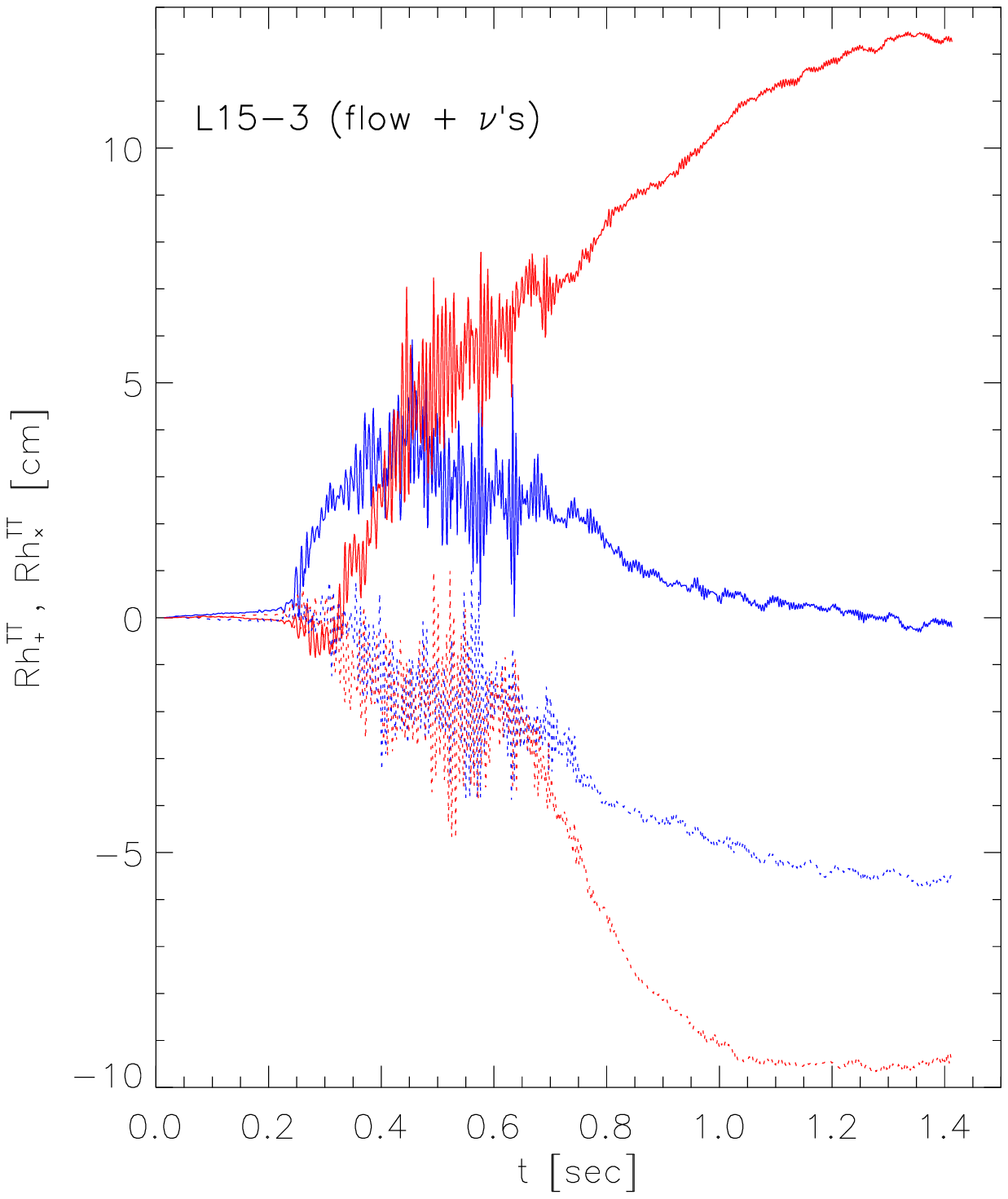}}
\resizebox{0.33\hsize}{!}{\includegraphics*{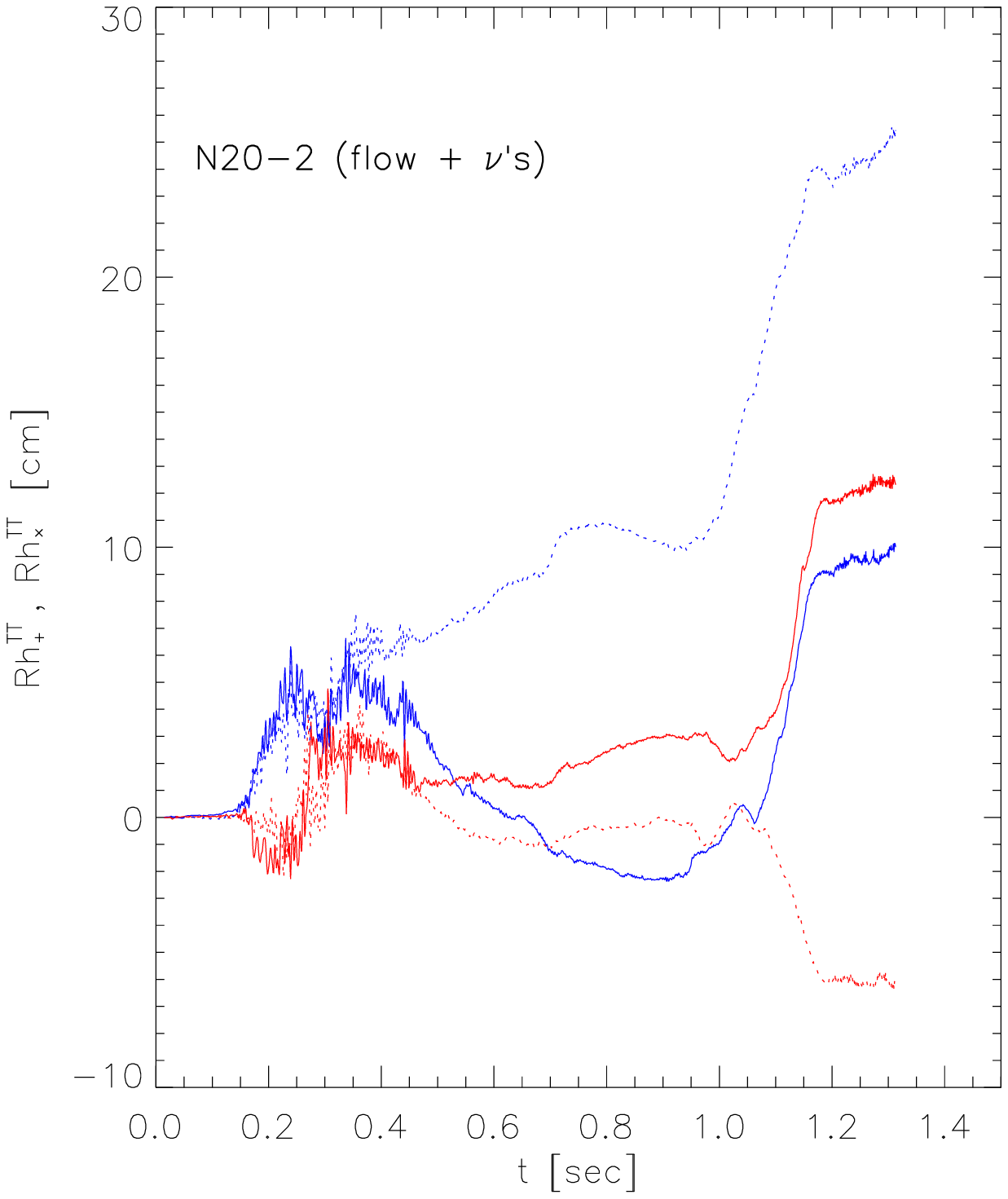}}
\caption{Gravitational wave amplitudes $R h^{TT}_{+}$ (blue) and
  $R h^{TT}_{\times}$ (red) due to anisotropic mass flow and neutrino
  emission as a function of time for models W15-4 (left), L15-3
  (middle), and N20-3 (right) , respectively. The solid curves show
  the amplitudes for an observer located above the north pole ($\alpha
  = \beta = 0$; see Fig.\,\ref{fig:obs_src}) of the source, while the
  other curves give the amplitudes at the equator ($\alpha = 0$,
  $\beta = \pi/2$). }
\label{fig:htot_vs_t}
\end{figure*}
%

%
\begin{figure*}[!]
\centering
\resizebox{0.33\hsize}{!}{\includegraphics*{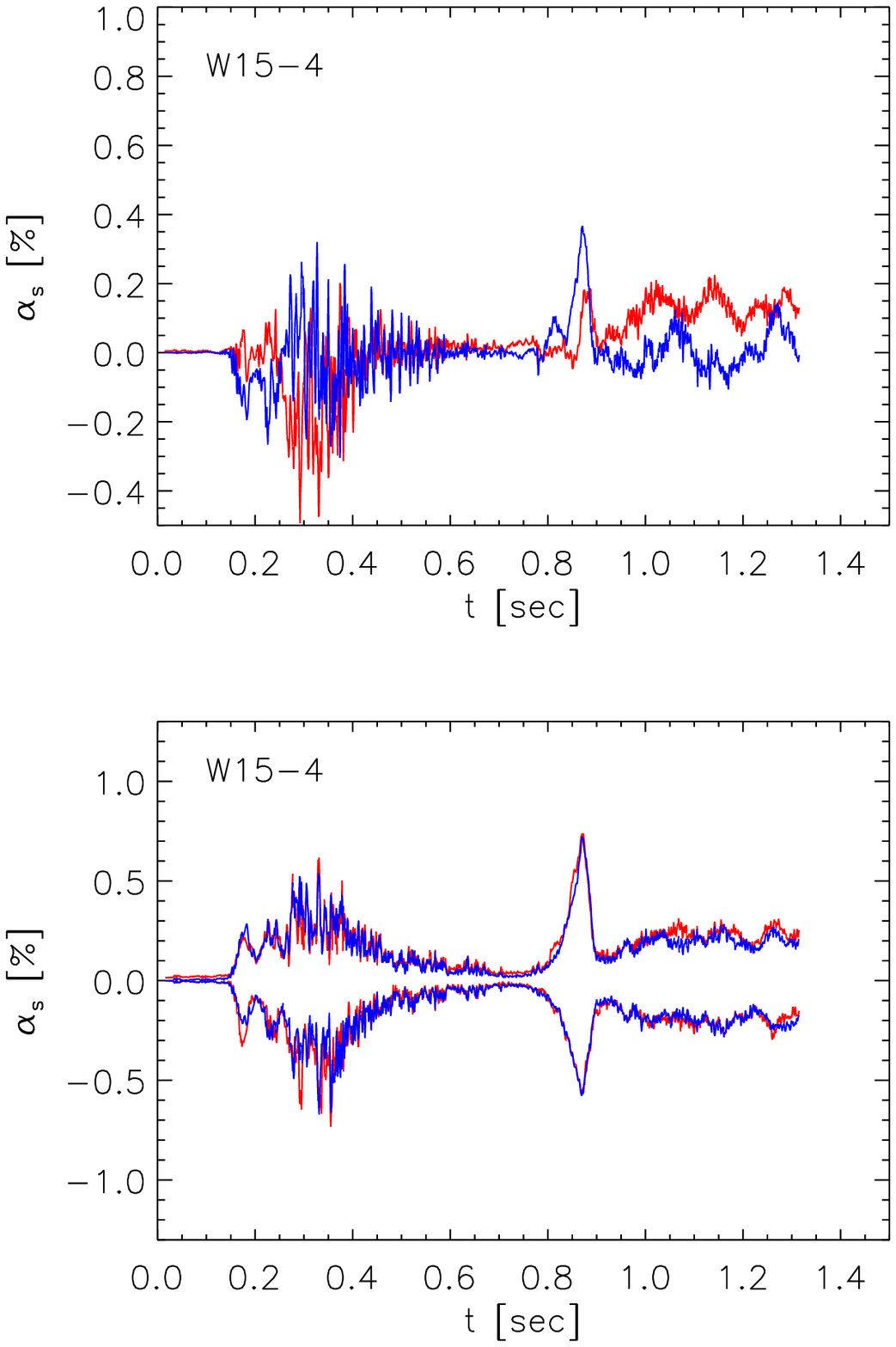}}
\resizebox{0.33\hsize}{!}{\includegraphics*{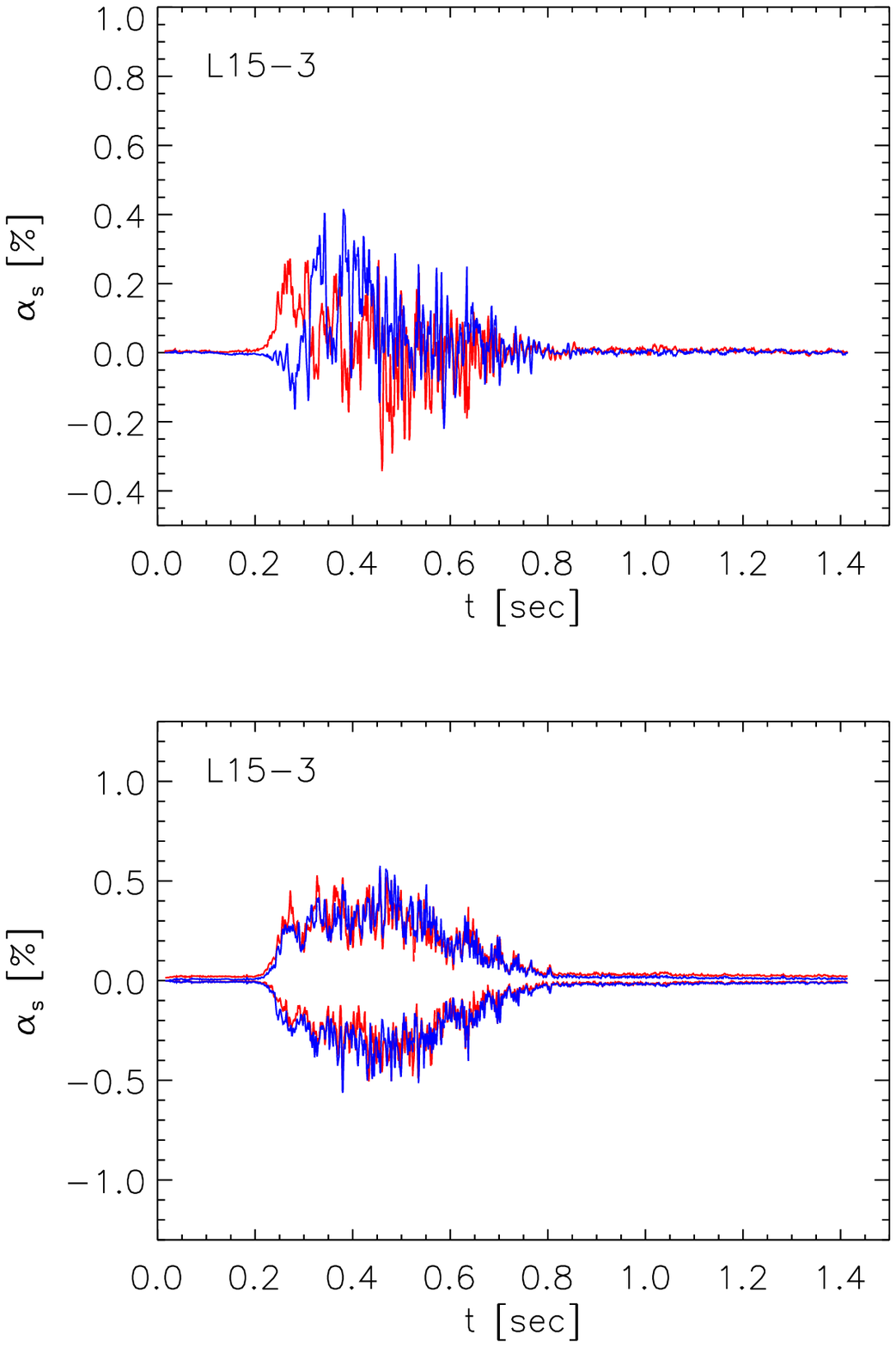}}
\resizebox{0.33\hsize}{!}{\includegraphics*{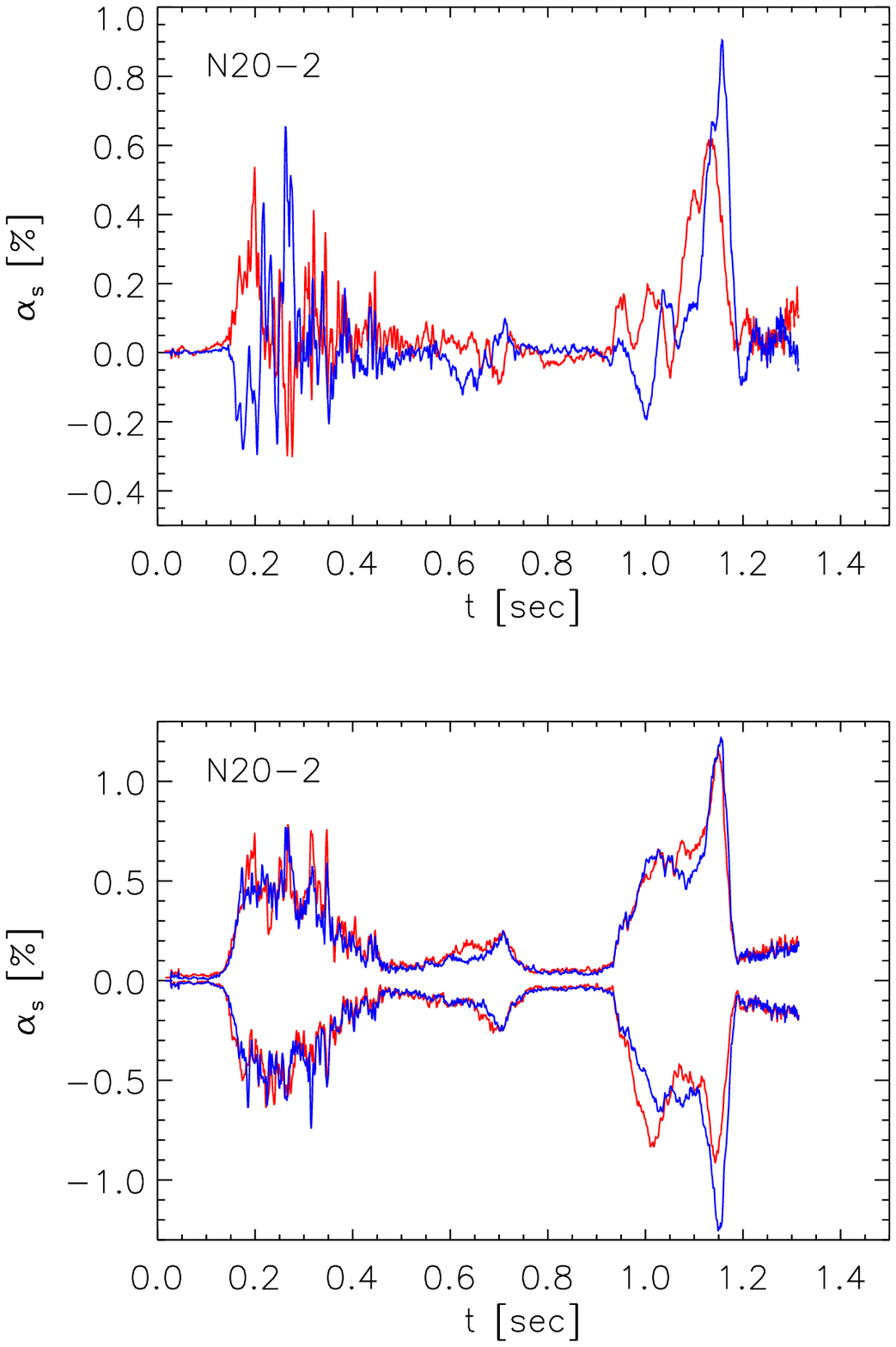}}
\caption{Asymmetry parameter $\alpha_S$ of the neutrino emission
  (Eq.\,\ref{eq:aniso}) as a function of time for models W15-4 (left),
  L15-3 (middle), and N20-2 (right), respectively. The panels in the
  upper row show $\alpha_+$ (blue) and $\alpha_\times$ (red) for a
  particular observer direction, while the panels in the lower row
  give for both parameters the maximum and minimum values in all
  directions. }
\label{fig:aniso_vs_t}
\end{figure*}

Although an observer can only measure the total gravitational wave
amplitude, \ie that due to the combined effect of non-radial flow and
anisotropic neutrino emission, we will first discuss the GW signal of
non-radial mass flow only, because it reflects the various phases of
the post-bounce evolution already introduced in the discussion of the
neutrino signal above.

Until post-shock convection and the SASI are eventually mature at
around 150\,msec, the GW signal is very small
(Fig.\,\ref{fig:W15-as}).  Note that our models are not able to follow
the GW emission that is caused by prompt post-shock convection because
of the excised inner region of the PNS \citep{Marek_etal09}.  Later
on, sizable g-mode activity is instigated in the outer layers of the
proto-neutron star by convective overturn and the SASI during the
hydrodynamically vigorous pre-explosion phase, and by the impact of
anisotropic accretion flows during the subsequent post-explosion
accretion phase \citep{Marek_etal09}. This g-mode activity is the
cause of GW signals \citep{Marek_etal09, Murphy_etal09,
  Yakunin_etal10}, whose maximum amplitudes are on the order of a few
centimeters centered around zero.

The GW frequency distribution possesses a very broad maximum in the
range of 100\,Hz to 500\,Hz, and the frequency corresponding to this
maximum slowly increases with time (Fig.\,\ref{fig:W15-as}). Partially
already during the post-explosion accretion phase, but at latest when
the shock wave starts to rapidly propagate to large radii between
$\sim\,$0.4\,sec and $\sim\,$0.7\,sec (see
Figs.\,\ref{fig:rsh+lnu_vs_t} and \ref{fig:rsh+lnu_vs_t_a}), the GW
amplitudes start to grow by about a factor of ten until approximately
asymptoting at $\sim\,$0.9\,sec in the case of the models based on the
progenitor W15, and at $\sim\,$1\,sec in the case of models based on
the progenitor L15, respectively (Fig.\,\ref{fig:W15-as}).  This
growth of the amplitude is associated with the anisotropic expansion
of the shock wave, and a positive/negative wave amplitude indicates a
prolate/oblate explosion, respectively \citep{Murphy_etal09}.
 
While the GW amplitudes grow, the GW energy distribution $\mathrm{d}
E_\mathrm{M} / \mathrm{d} \nu$ becomes narrower and dimmer, and the
frequency at maximum power continues to increase.  The latter effect
was also observed in the 2D models of \citet{Murphy_etal09}.  At late
times, the GW signal of the W15 models clearly signifies the
convective activity inside the proto-neutron star through
low-amplitude, high-frequency fluctuations around the asymptotically
roughly constant mean GW amplitudes, while no such fluctuations are
present in the case of the L15 models (see discussion of the neutrino
signal above). This model discrepancy is also evident from the energy
spectrograms, which do exhibit a pronounced broad maximum (between
$\sim\,350\,$Hz and $\sim\,550\,$Hz) at $t > 0.8\,$sec in the case of
the W15 models, but none for the L15 ones. Furthermore note that until
the end of the simulations the frequency of the maximum of $\mathrm{d}
E_\mathrm{M} / \mathrm{d} \nu$ has increased from around 100\,Hz to
almost 500\,Hz for the former models (owing to the increasing speeds
of mass motions in the postshock region at times $\la 0.5\,$sec and
because of the increased compactness of the proto-neutron star at
times $\ga 0.6\,$sec, respectively).

The behavior of the total (matter plus neutrinos) GW amplitudes is
significantly different from that of the flow-only GW amplitudes for
models that exhibit PNS convection below the neutrinosphere, \ie for
the models based on the progenitors W15 and N20.  Particularly at late
times, anisotropic neutrino emission causes a continuing growth of the
GW amplitudes (instead of a saturation) in these models, while this is
not the case for the L15 models (see Fig.\,\ref{fig:htot_vs_t}, and
compare with Fig.\,\ref{fig:W15-as}).  The latter behavior is also
reflected in the time evolution of the asymmetry parameter
$\alpha_{\mathrm S}$ (Eq.\,\ref{eq:aniso}) of the neutrino emission
(Fig.\,\ref{fig:aniso_vs_t}). The asymmetry parameter is practically
zero in model L15-3 at late times, while it remains, after having
temporarily grown to values beyond about 0.4 - 0.5\%, at the level of
$\sim 0.3\,$\% until the end of the simulations in models W15-4 and
N20-2.

\begin{figure}[!]
\centering
\resizebox{0.95\hsize}{!}{\includegraphics{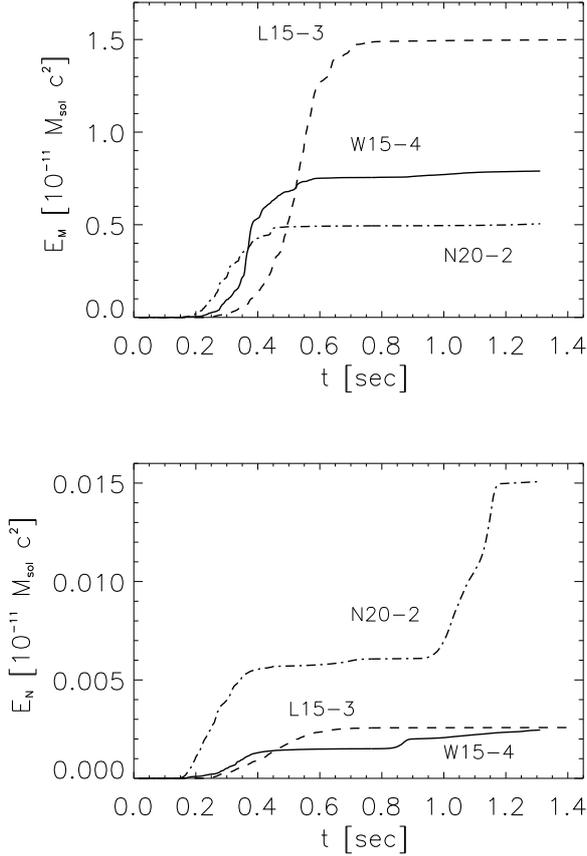}}
\caption{Energy emitted in form of gravitational waves due to
  anisotropic mass flow (top panel) and due to anisotropic neutrino
  emission (bottom panel) as a function of time for models W15-4
  (solid), L15-3 (dashed), and N20-2 (dash-dotted), respectively. }
\label{fig:etot_vs_t}
\end{figure}

The final GW amplitudes are up to a factor of two to three higher when
taking the contribution of anisotropic neutrino emission into account.
In contrast, the amount of energy radiated in the form of GW, which is
proportional to the third time-derivative of the quadrupole moment and
hence proportional to the time derivative of the GW amplitude is only
insignificantly changed, and is practically constant for all simulated
models after the onset of the explosion (see
Fig.\,\ref{fig:etot_vs_t}).  The integral value of the GW energy
radiated by neutrinos is low ($\la 1\%$) compared to that emitted by
matter through the slow variation of the GW neutrino amplitude with
time, \ie its time derivative is much smaller than that of the GW
matter amplitude.  For this reason we also abstained from evaluating
the total energy radiated in form of GW (Eq.\,\ref{eq:gwetot}). It
differs little from that caused by anisotropic matter flow alone
(Eq.\,\ref{eq:egw}), because the mixed term in Eq.\,(\ref{eq:gwetot}),
resulting from the square of the sum of the matter and neutrino parts,
contributes $\la 10\%$ to the total radiated GW energy, and the pure
neutrino term $\la 1\%$.  Figure \ref{fig:etot_vs_t} also shows that
the (small) contribution of anisotropic neutrino emission to the
radiated GW energy is enhanced at late times when proto-neutron star
convection occurs below the neutrinosphere, as it is the case for
models W15-4 and in particular N20-2.

%
%

%
\begin{figure}
\centering
\resizebox{0.95\hsize}{!}{\includegraphics*{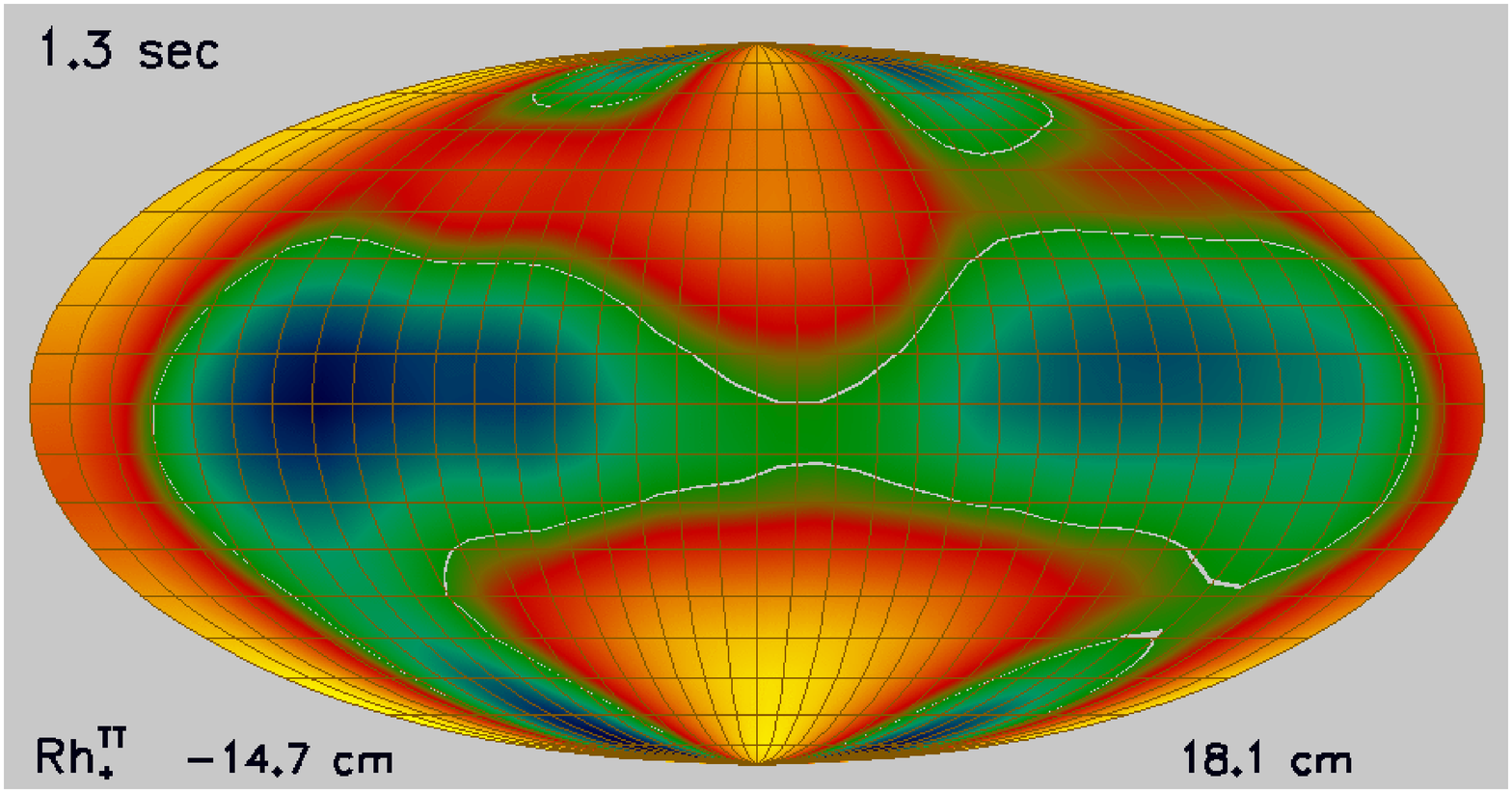}}\\
\resizebox{0.95\hsize}{!}{\includegraphics*{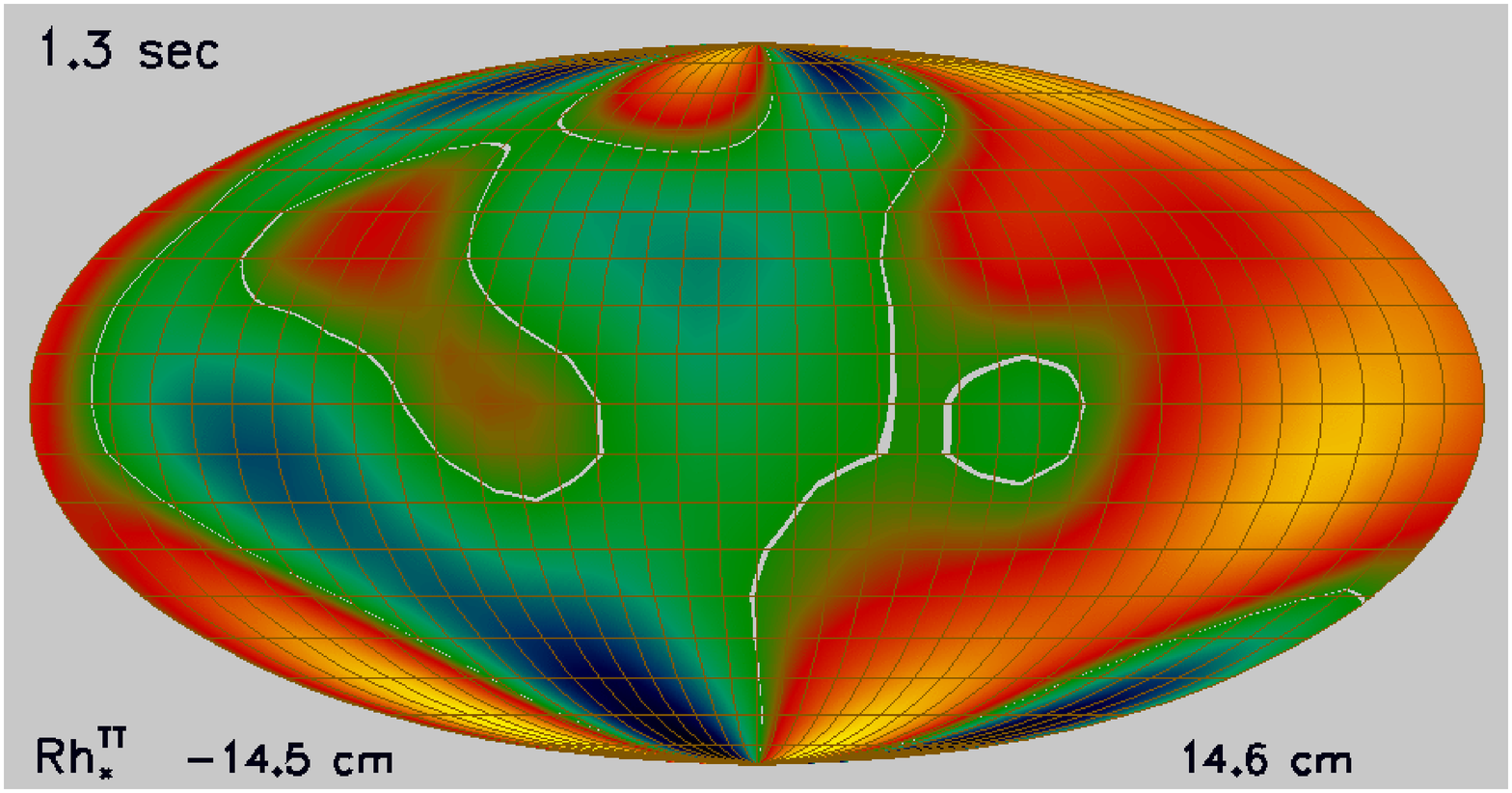}}
\caption{Gravitational wave amplitudes due to anisotropic mass flow
  and neutrino emission, $R h^{TT}_{+}$ (top) and $R h^{TT}_{\times}$
  (bottom), as functions of the observer angles (see
  Fig.\,\ref{fig:obs_src}) for model W15-4 at 1.3\,sec past
  bounce. The white contours give the locations, where the amplitudes
  are zero. Yellow and red areas indicate positive amplitudes, green
  and blue negative ones.}
\label{fig:W15-4_hnue}
\end{figure}
%

The variation of the total GW amplitudes with observer angle is
illustrated in Fig.\,\ref{fig:W15-4_hnue} for model W15-4 at 1.3\,sec
(when the simulation was stopped). Both the amplitude variations and
the typical angular size of the speckled GW emission are similar for
all other simulated models.  The model-independent level of the
amplitude variations is also supported by Fig.\,\ref{fig:htot_vs_t}
when comparing various amplitudes at any given (late) time.

\begin{figure*}
\centering
\resizebox{0.495\hsize}{!}{\includegraphics*{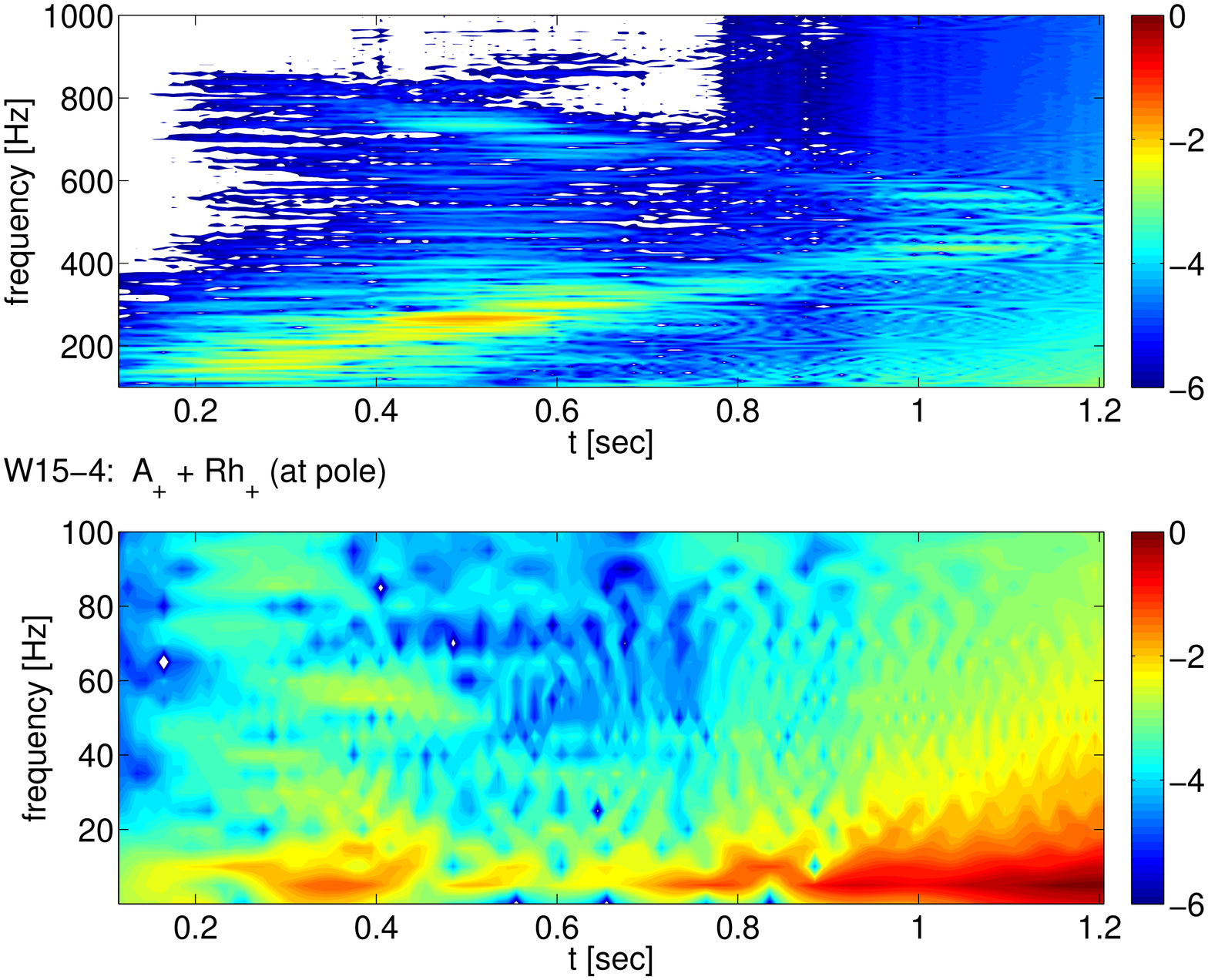}}
\resizebox{0.495\hsize}{!}{\includegraphics*{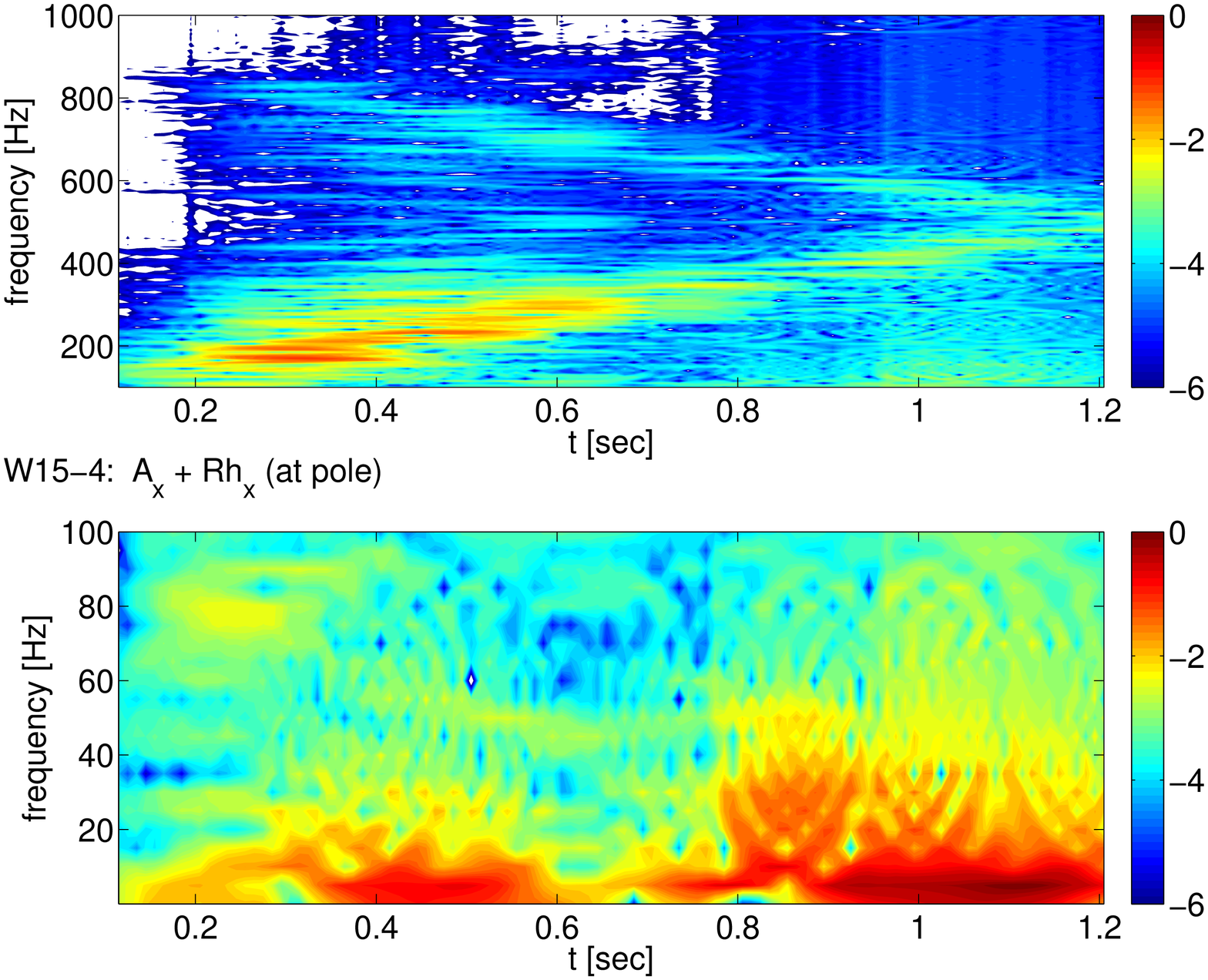}}
\caption{Normalized (to the absolute maximum) amplitude spectrograms
  of the total gravitational wave amplitudes $A_+ + Rh_+$ (left
  panels) and $A_\times + Rh_\times$ (right panels) at the pole for
  model W15-4. The lower panels show the spectrograms in the frequency
  range 5\,Hz to 100\,Hz, and the upper ones in the frequency range
  100\,Hz to 1\,kHz.}
\label{fig:W15-4_h_pol}
\end{figure*}

\begin{figure*}
\centering
\resizebox{0.495\hsize}{!}{\includegraphics*{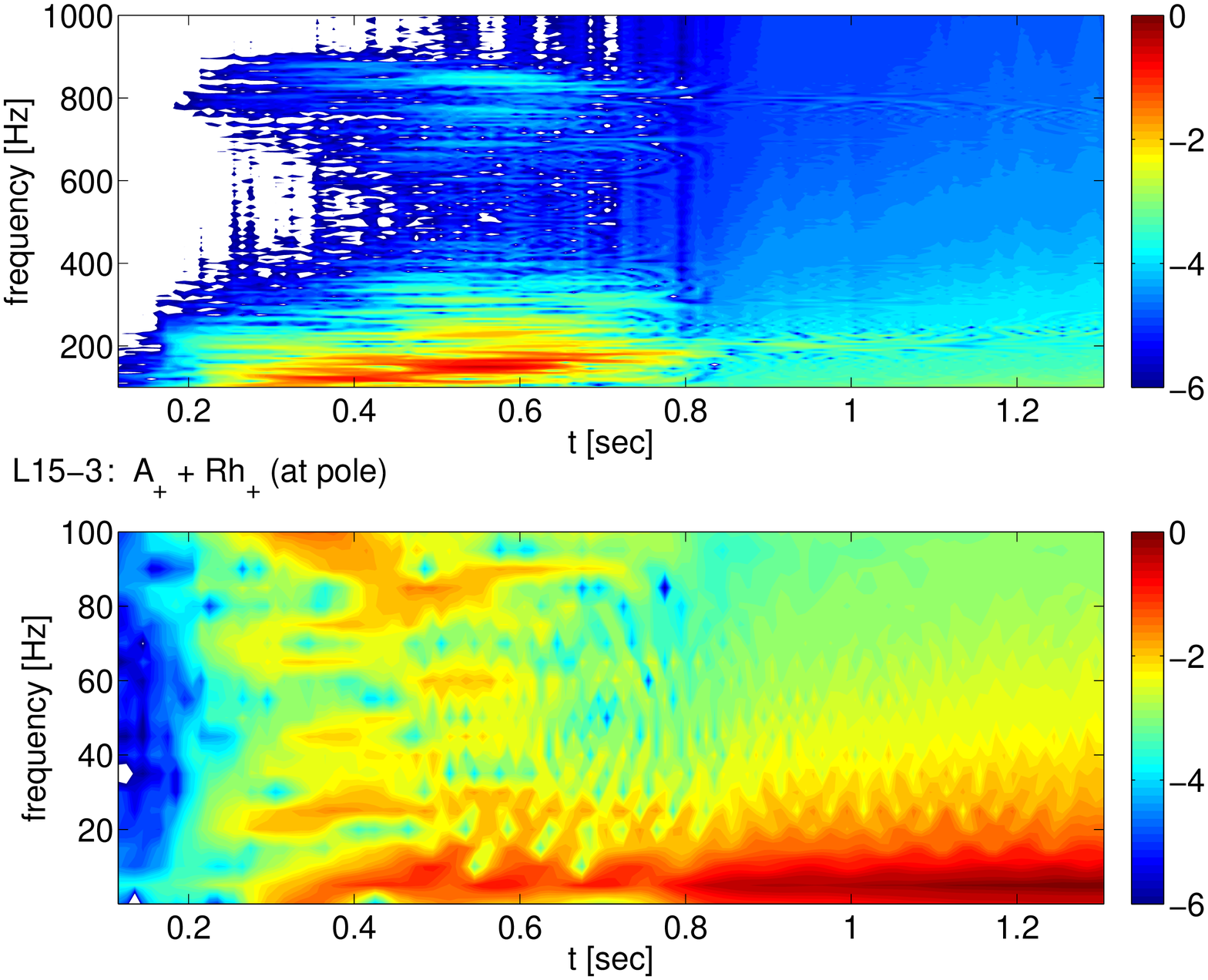}}
\resizebox{0.495\hsize}{!}{\includegraphics*{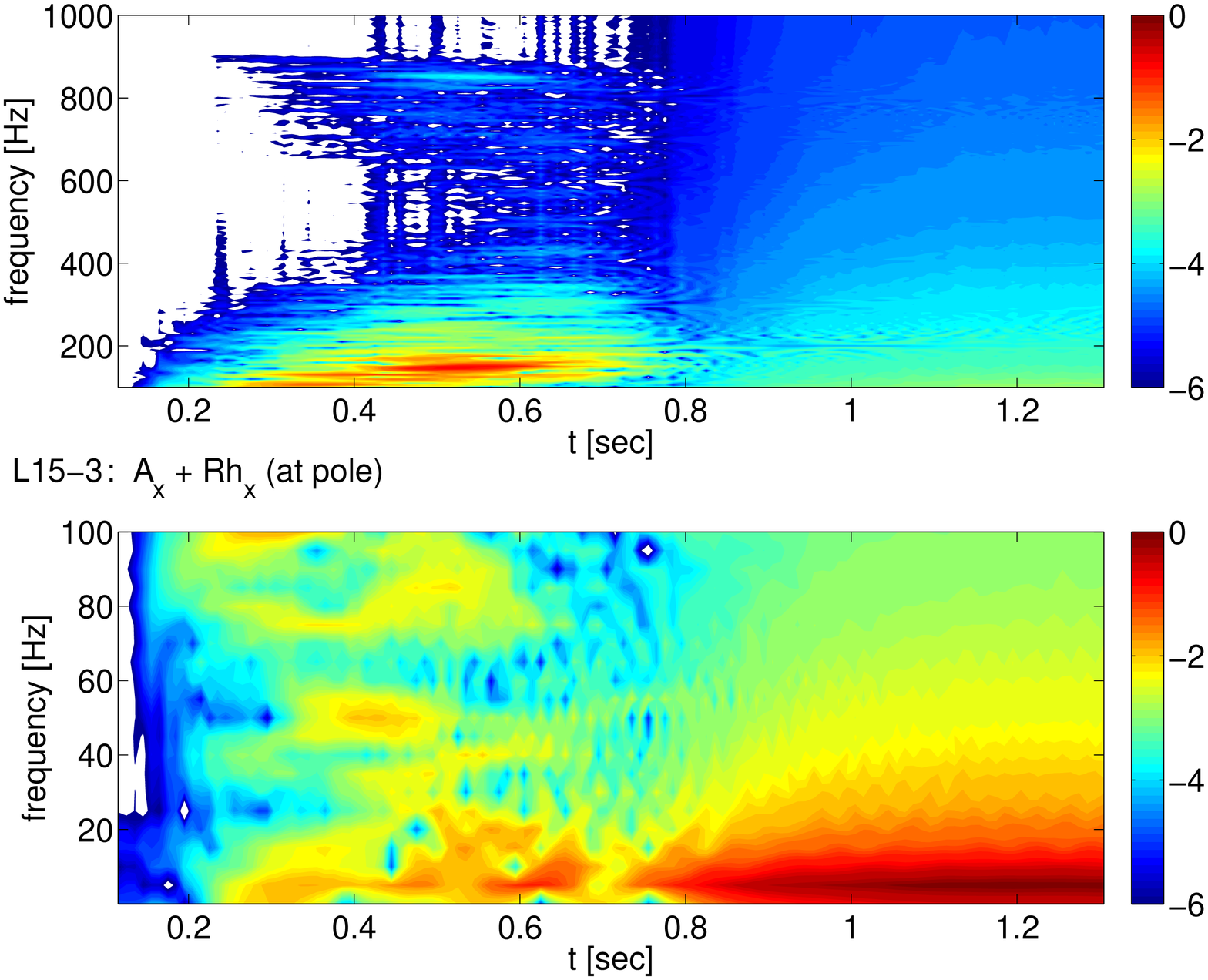}}
\caption{Same as Fig.\,\ref{fig:W15-4_h_pol}, but for model L15-3.}
\label{fig:L15-3_h_pol}
\end{figure*}

The (normalized) amplitude spectrograms of the total gravitational
wave amplitudes (${\rm d} (A_{+, \times} + R h_{+, \times})/ \rm{d}
\nu$; Figs.\,\ref{fig:W15-4_h_pol} and \ref{fig:L15-3_h_pol})
illustrate two model-independent findings.  Firstly, during the
hydrodynamically vigorous pre-explosion and post-explosion accretion
phases ($0.2 \la t \la 0.5-0.7\,$sec) the spectra of all models are
characterized by some power at low frequencies ($\la 100\,$Hz) and a
broad power maximum at frequencies $\sim 200\,$Hz and another weak one
at $\sim 800\,$Hz. The latter broad maximum at high frequency is more
pronounced in the models based on the W15 and N20 progenitors and in
the cross polarization GW mode. Secondly, during the post-accretion
phase ($t \ga 0.7\,$sec) the spectra of all models are dominated by a
low-frequency ($\la 40\,$Hz) contribution peaked toward the lower end
of the spectrogram. In the models where PNS convection occurs below
the neutrinosphere (models W15 and N20) we also find a double-peaked
high-frequency contribution decreasing/increasing from $\sim 700\,$Hz
(400\,Hz) at $t \sim 0.8\,$sec, and eventually merging into a single
power maximum at $\sim 500\,$Hz at $t \sim 1.2\,$sec. Again, this
contribution is more pronounced for the cross polarization GW mode.

The spectra of the total GW amplitudes are dominated by the
contribution from non-isotropic neutrino emission at low ($\le
100\,$Hz) frequencies (Figs.\,\ref{fig:W15-4_h_pol} and
\ref{fig:L15-3_h_pol}). At higher frequencies ($\ge 100\,$Hz) the
spectra of model W15-4 show two pronounced maxima (at 100 - 200\,Hz
and $600-800\,$Hz, respectively) at all times. These maxima are also
present in model L15-3 at times $\le 0.7\,$s, the high-frequency one
being, however, much less pronounced.  The lower maximum (at
$100-200\,$Hz) results from g-mode activity in the PNS surface
instigated by non-radial flow (SASI, accretion) in the post-shock
region until $\sim 0.5 -0.7\,$sec. At later times PNS convection is
responsible for the peak between 300 and 500\,Hz. We have proposed
this explanation already for the corresponding maxima present in the
GW energy spectrograms arising from non-spherical mass flow
(Fig.\,\ref{fig:W15-as}), and discussed why the frequencies of these
maxima increase with time. The source of the high-frequency maximum
(600 - 800\,Hz) is unclear, but a further detailed analysis shows that
(i) the maximum is solely caused by non-radial gas flow, \ie it is not
connected to neutrinos, (ii) it does not result from stellar layers
below the neutrino sphere but from those close to or slightly above
it, and (iii) does not depend on the position of the observer.

Note that the high-frequency maximum present in the amplitude
spectrograms is strongly suppressed in the corresponding energy
spectrograms (Fig.\,\ref{fig:W15-as}), because the latter involve the
squared time derivatives of the amplitudes. Thus, the already high
ratio of the low- and high-frequency maxima in the amplitude
spectrograms (about two orders of magnitude) translates into an even
higher ratio for the energy spectrogram maxima, rendering the
high-frequency maximum practically invisible.

%
%

%
%

\section{Discussion and conclusions}

Based on a set of three-dimensional (3D) parametrized neutrino-driven
supernova explosion models of non-rotating 15 and 20\,$M_\odot$ stars,
employing a neutrino transport description with a gray spectral
treatment and a ray-by-ray approximation of multi-dimensional effects
(the scheme is applicable in the regime outside the dense neutron star
core, \ie around and outside the neutrinosphere), we evaluated both
the time-dependent and direction-dependent neutrino and
gravitational-wave emission of these models.  To this end we presented
the formalism necessary to compute both the observable neutrino and
gravitational wave signals for a three-dimensional, spherical
source. For the neutrino signal we presented formulas that allow one
to estimate the apparent luminosity when the local flux density $F$ on
a sphere is known and the ansatz of Eq.\,(4) about the angular
distribution of the intensity can be made.  While in general the
location of the neutrino-decoupling surface has to be suitably defined
(\eg by a criterion based on the optical depth), our ray-by-ray
transport scheme implies that $r^2 F =$const in the free-streaming
limit in every direction. Thus, it spares us such a definition of the
neutrinosphere for evaluating the angular integration for the
observable luminosity.  Concerning the gravitational-wave analysis, we
extended and generalized previous studies, where the source was either
assumed to be axisymmetric or where the formulas for the signals of a
3D source were only given for special observer directions.

Our models followed the evolution from shortly after core bounce up to
more than one second into the early cooling evolution of the PNS
without imposing any symmetry restrictions and covering a full sphere.
The extension over such a relatively long evolution time in 3D was
possible through the usage of an axis-free overset grid (the Yin-Yang
grid) in spherical polar coordinates, which considerably eases the CFL
time-step restriction and avoids axis artifacts.  A central region,
the dense inner core of the proto-neutron star, was excised from the
computational domain and replaced by an inner, time-dependent radial
boundary condition and a gravitating point mass at the coordinate
origin.  Explosions in the models were initiated by neutrino heating
at a rate that depends on suitably chosen values of the neutrino
luminosities imposed at the inner radial boundary.

The post-bounce evolution of our models can be divided into four
distinct phases (Fig.\,\ref{fig:rsh+lnu_vs_t}).  The first phase, the
\emph{quasi-spherical shock-expansion phase}, lasts from shock
formation shortly after core bounce to $80\,-\,150$\,msec, when
convection sets in. The second phase, the hydrodynamically vigorous
\emph{pre-explosion phase}, comprises the growth of post-shock
convection and of the standing accretion shock instability (SASI).
The \emph{post-explosion accretion phase} begins when energy
deposition by $\nu$-heating in the post-shock layers becomes
sufficiently strong so that the total energy in the post-shock region
ultimately becomes positive. During this phase the shock accelerates
outward while gas is still accreted onto the PNS.  This process is
commonly called ``shock revival".  The duration of the latter two
phases depends on the progenitor.  During the \emph{post-accretion
  phase}, the fourth and final phase characterizing the evolution of
our models, accretion ends and the proto-neutron star develops a
nearly spherical neutrino-driven wind.

The neutrino emission properties (fluxes and effective spectral
temperatures) of our 3D models exhibit the generic time-dependent
features already known from 2D (axisymmetric) models \citep[\eg][]{
  Buras_etal06, Scheck_etal06, Marek_etal09, Brandt_etal11}, showing
fluctuations over the neutron star surface on different spatial and
temporal scales.  We found that non-radial mass motions caused by the
SASI and convection in the neutrino-heated hot-bubble region as well
as by PNS convection below the neutrinosphere give rise to a
time-dependent, anisotropic emission of neutrinos, particularly of
electron neutrinos and anti-neutrinos, and thus also to the emission
of gravitational waves.  Because very prominent, quasi-periodic
sloshing motions of the shock due to the standing accretion-shock
instability as visible in 2D simulations are absent and the emission
from different surface areas facing an observer adds up incoherently,
the modulation amplitudes of the measurable neutrino luminosities and
mean energies are significantly lower than predicted by 2D models
\citep[for 2D results see][]{Marek_etal09, Brandt_etal11}.

During the quasi-spherical shock expansion phase shortly after bounce
the level of temporal and angular fluctuations of the neutrino
emission is low ($\la 10^{-2}$). In contrast, the fluctuation
amplitudes reach a level of several 10\% of the average values during
the hydrodynamically vigorous pre-explosion phase and the
post-explosion accretion phase, where a few distinct, highly
time-variable regions or even short-lived single spots with an angular
size of 10$^\circ$ to 20$^\circ$ are responsible for the brightest
emission maxima.  As the outward shock expansion is well on its way in
the post-explosion accretion phase, still existing accretion
downdrafts can be responsible for similar fluctuations in the neutrino
emission, though the number of corresponding hot spots decreases with
diminishing accretion. When accretion has ended and the post-accretion
phase has started, directional variations can be caused by the
occurrence of Ledoux convection in the outer layers of the
proto-neutron star, which we indeed observe in models based on two of
our three progenitors (see also the discussion of the influence of the
inner radial boundary condition below).  The temporal and angular
variations of the emission in different directions are even more
pronounced when considering the energy flux of the electron neutrinos
or electron anti-neutrinos alone (instead of the emission in all
neutrino flavors). In that case the angular variations of local flux
densities can exceed 100\% in all models during the pre-explosion and
post-explosion accretion phases, and the peak values can be close to
200\% during short episodes. The total energy loss rates in neutrinos
and the observable luminosities as surface-integrated quantities,
however, are much smoother in time during all phases, showing
fluctuation amplitudes of at most several percent.

The gravitational wave emission also exhibits the generic
time-dependent features already known from 2D (axisymmetric) models,
but the 3D wave amplitudes are considerably lower (by a factor of
$2-3$) than those predicted by 2D models \citep{Mueller_etal04,
  Marek_etal09, Murphy_etal09, Yakunin_etal10} owing to less coherent
mass motions and neutrino emission.  Note in this respect that the GW
quadrupole amplitudes, which are usually quoted for 2D models
($A^{E2}_{20}$), have to be multiplied by a geometric factor
$\sin^2\theta\, \sqrt{15/\pi}/8$ (which is equal to $\approx 0.27$ for
$\theta = 90^o$).  Violent, non-radial hydrodynamic mass motions in
the accretion layer and their interaction with the outer layers of the
proto-neutron star give rise to a GW signal with an amplitude of $\sim
2-4\,$cm in the frequency range of $\sim\,$100\,Hz to $\sim\,$400\,Hz,
while anisotropic neutrino emission is responsible for a superimposed
low-frequency evolution of the wave amplitude, which thus can grow to
maximum values of $10-20\,$cm.  Variations of the mass-quadrupole
moment caused by convective activity inside the nascent neutron star
contribute a high-frequency component ($300\,-\,600\,$Hz) to the GW
signal during the post-accretion phase.  The GW signals exhibit strong
variability between the two polarizations, different explosion
simulations and different observer directions, and besides common
basic features do not possess any template character.

Finally we would like to reflect on some of the deficiencies of the
presented 3D models. Because of the ray-by-ray treatment of the
$\nu$-transport, the directional variations of the neutrino emission
in response to local inhomogeneities in the star may be overestimated
\citep{Ott_etal08, Brandt_etal11}.  However, when we evaluate
observable signals, these artificial effects are mostly compensated
for by integrations of the neutrino flux densities over the surface
areas visible to observers from different viewing directions (see
Eqs.\,\ref{eq:app-lum}, \ref{eq:lnum}, \ref{eq:app-lum_code},
\ref{eq:lnum_code}), or by the integration of the neutrino energy loss
in all directions (Eqs.\,\ref{eq:lamt}, \ref{eq:aniso}).

Another deficiency concerns the usage of the inner radial grid
boundary, because of which our simulations do not fully include
(either in space or time) the convective flow occurring in the PNS
interior after core bounce \citep{Keil_etal96, Buras_etal06,
  Dessart_etal06}.  Moreover, convective activity in the simulated
outer layers of the PNS occurs for special conditions: It is triggered
only when the artificially imposed inflow of neutrino energy and
lepton number through the inner radial boundary into the adjacent
layers is faster than the neutrinos can carry away this energy or
lepton number.  Whether this is the case sensitively depends on the
employed neutrino-transport approximation, but also on the location
and contraction of the grid boundary, the chosen values of the
boundary luminosities, and on the stellar progenitor. Its mass-infall
rate decides how much mass accumulates in the near-surface layers of
the PNS outside the inner grid boundary.  Because the position of and
the conditions imposed at the inner boundary can thus influence the
neutrino emission properties, in particular during the post-accretion
phase, our respective model predictions must be considered with
care. While they do not allow us to make any definite statements
concerning the detailed neutrino signal of a particular progenitor
model due to the neglected treatment of the inner parts of the
proto-neutron star, the models nevertheless show that convective flows
below the neutrinosphere are likely to imprint themselves on the
neutrino emission, and hence also on the GW signal of core-collapse
supernovae.  A measurement of these signals may actually provide some
insight into the conditions inside proto-neutron stars.

\acknowledgements 
This work was supported by the Deutsche Forschungsgemeinschaft through
the Transregional Collaborative Research Centers SFB/TR~27 ``Neutrinos
and Beyond'' and SFB/TR~7 ``Gravitational Wave Astronomy'' and the
Cluster of Excellence EXC~153 ``Origin and Structure of the Universe''
(http://www.universe-cluster.de). 

\bibliographystyle{aa}
\bibliography{3Dnugrav}
\end{document}